\documentclass[prd,preprint,eqsecnum,amsfonts,amssymb,amsmath,floatfix,nofootinbib]{revtex4}

\usepackage{bm}
\usepackage{slashed}
\usepackage{graphicx}
\usepackage{MnSymbol}
\usepackage{braket}
\usepackage{appendix}

\makeatletter
\newcommand{\raisemath}[1]{\mathpalette{\raisem@th{#1}}}
\newcommand{\raisem@th}[3]{\raisebox{#1}{$#2#3$}}
\makeatother

\begin{document}

\title{Predictions of a fundamental statistical picture}

\author{Roland E. Allen} 

\affiliation{Department of Physics and Astronomy \\ 
Texas A\&M University, College Station, Texas 77843, USA}

\date{\today}

\begin{abstract}
The discovery of a Higgs boson at the electroweak scale appears to point toward supersymmetry, as the most likely mechanism for protecting a scalar boson mass from enormous radiative corrections. The earlier discovery of neutrino masses similarly appears to point toward grand unification of nongravitational forces, which permits (for neutrinos) Majorana masses, Dirac masses, and a seesaw mechanism to drive the observed masses down to low values. A third major discovery, cosmic acceleration suggesting a relatively tiny cosmological constant, appears to point toward truly revolutionary new physics. Many other problems and mysteries also indicate a need for fresh ideas at the most fundamental level. Here a picture is proposed in which standard physics and its extensions are obtained (through a nontrivial set of arguments) from statistical counting and the local geography of our universe. The unavoidable predictions include supersymmetry (at some energy scale), $SO(N)$ grand unification, a drastic diminishing of the usual cosmological constant, and a nonsupersymmetric dark matter WIMP which should be detectable within the next several years.
\end{abstract}

\maketitle

\section{\label{sec:sec1}Introduction}

Complexity can emerge from simplicity in amazing ways, as when most of our observed world is attributed to two quarks and two leptons (plus gauge bosons and gravity). It is worthwhile to consider the possibility that all the complexities of the Standard Model and its extensions might similarly emerge from a very simple underlying description. Here we explore the results that follow from what appears to be the simplest imaginable picture, introduced in Section \ref{sec:sec2}. 

One motivation for a fresh perspective on fundamental physics is the remarkable mix of clarity and confusion that currently exists.   The situation in the early 21st century is, in fact, similar to what it was in the late 19th century. Then most physicists were generally  satisfied with the successful paradigm of classical mechanics and electrodynamics, but there were some conflicting experimental data and theoretical puzzles. Now most physicists are generally satisfied with the successful paradigm of quantum fields and gauge theories (plus Einstein gravity), but there are again mysteries that suggest the need for a deeper theory. Many excellent reviews have been given of the current situation in physics and astronomy~\cite{pdg}, but it may be worthwhile to begin with a brief summary.

Most recently, the particle discovered by the ATLAS and CMS collaborations at the LHC is now known to be a Higgs boson~\cite{pdg,ATLAS,CMS}. A naive conclusion is that the Standard Model of particle physics is now complete. But the more profound interpretation is that the discovery of a scalar boson immediately points to physics beyond the Standard Model, since otherwise radiative corrections should push the mass of this particle up to an absurdly large value. The most natural candidate for such new physics is supersymmetry (susy)~\cite{baer,nath,mohapatra,kane-susy,martin,ellis,murayama,lykken,bilal,haber-kane,drees,binetruy,terning,aitchison,weinberg-susy,wess,ramond,raby}, for which there is already indirect experimental evidence: The coupling constants of the 3 nongravitational forces are found to converge to a common value, as they are run up to high energy in a grand unified theory, only if the calculation includes susy. So, instead of acting as an endpoint for physics, and a mere capstone of the Standard Model, the observation of a Higgs boson opens the door to a plethora of new particles and effects.

Another major advance has been the discovery and exploration of neutrino masses~\cite{nu-2021}, which appear to open the door to a more fundamental understanding of forces and matter via grand unification~\cite{nath,mohapatra,raby,cheng,kounnas,ross,georgi,collins,barger,pati,baez,ozer}. There are two possibilities for a neutrino mass, either of which is inconsistent with the requirements of the Standard Model. For a Dirac mass, an extra field has to be added for each generation of fermions. For a Majorana mass, lepton number conservation has to be violated. But either or both types of mass are natural with grand unification, and in addition a seesaw mechanism can explain the small observed values of neutrino masses. At the moment, it is not known whether neutrinos have Majorana masses or Dirac masses or both. This is currently an intense area of research, and any outcome will again involve rich new physics and better understanding of Nature.

There are many other mysteries and gaps in fundamental understanding. For example, the discovery and exploration of cosmic acceleration~\cite{riess,perlmutter} has suggested the need for truly revolutionary new physics. The cause of this acceleration has increasingly been found to resemble a cosmological constant $\Lambda$, and has therefore been a strong reminder of the original cosmological constant problem~\cite{weinberg-1989}: Because of the various contributions to the vacuum energy, conventional general relativity predicts that $\Lambda$ should be vastly larger than permitted by observation. A parallel astronomical mystery is the origin of dark matter~\cite{hooper}.

Still another and even older theoretical problem is the difficulty of reconciling general relativity with quantum mechanics~\cite{rovelli}: A new fundamental theory must somehow regularize quantum gravity near the Planck scale, in addition to reducing the value of $\Lambda$ by many orders of magnitude.

The next level of theoretical understanding is not likely to be a ``theory of everything'', since ``everything'' surely transcends our current observational capabilities and imagination.  But the most ambitious version of a more fundamental theory might hope to include and explain the following: the absence of an enormous cosmological constant, the origin of dark matter, the origin of gravitational and gauge interactions, the origin of Lorentz invariance, the gravitational metric and its signature (which distinguishes time from space and characterizes spacetime as 4-dimensional),  the action for fermionic and bosonic fields, the action for gravitational and gauge fields, the regularization of quantum gravity near the Planck scale, the origin of quantum fields, and the origin of spacetime coordinates. As will be seen below, the present theory addresses all of these issues and leads to a substantial number of predictions. These are largely qualitative, because quantitative treatments in many cases would require a detailed understanding of the very complex vacuum fields after multiple symmetry breakings at various energy scales. However, there are some specific new features, with quite quantitative predictions, which should be testable in the near future. For example, the theory predicts new fundamental particles which can be produced in pairs through their couplings to vector bosons. The lowest-mass of these is a dark matter candidate, with precisely defined couplings and mass, which is consistent with all current experiments and which should be detectable within the next several years.

\section{\label{sec:sec1a}Overview}

DeWitt has provided an elegant survey of contemporary fundamental physics~\cite{dewitt}, which is based on path-integral quantization over classical
trajectories in the combined space of coordinates and fields:
``A classical dynamical system is described globally by a 
\textit{trajectory} or \textit{history}. A history is a section of a fibre
bundle $E$ having the manifold $M$ of spacetime as its base space. 
The
typical fibre is known as configuration space and will be denoted by $C$....
Denote by $\Phi $ the set, or space, of all possible
field histories, both those that do and those that do not satisfy the
dynamical equations .... The nature and dynamical properties of a classical
dynamical system are completely determined by specifying an \textit{action
functional} $S$ for it.''

This is the basic picture used in all the versions of fundamental physics
that are investigated by sizable communities of physicists. Notice that
coordinates and fields have essentially the same status. This is consistent
with the way they are defined in the present theory, starting near the
beginning of the next section. In standard field theory, the spacetime
coordinates $x^{\mu }$ correspond to $M$ and the fields to $C$. In
conventional string (or $p$-brane) theory, $M$ is a $2-$dimensional
worldsheet (or  $\left( p+1\right) -$dimensional worldvolume) and $C$ is
specified by bosonic and fermionic coordinates. 

The arguments below involve many unfamiliar definitions and redefinitions of fields, with 
physically accessible fields emerging from more primitive 
 ``hidden'' degrees of freedom. But this kind of treatment is already necessary and familiar in standard physics -- 
 for example, when the fundamental gauge fields are redefined after Higgs condensation, 
 and when many kinds of elementary excitations are defined in condensed matter physics. 
 More broadly, the deeper hidden origin
of observed phenomena is a common theme in science. 

The physical fields that have emerged by the end of this paper are interpreted as
those chosen by Nature to yield a stable vacuum $\left\vert 0\right\rangle $ satisfying
$a\left\vert 0\right\rangle =0$, where $a$ is a typical destruction operator
for one of these fields. For
example, the excitations $a^{\dag}\left\vert 0\right\rangle $ must have positive rather
than negative energy, and this is why the transformed bosonic fields 
of Section \ref{sec:DM} are physically acceptable, whereas the
initial fields $\psi _{b}$ are not.

The principal ideas and results of this paper are as follows.

(1) In standard physics, all the events of the world are simply a progression through 
the states of fields on a spacetime manifold. In the present picture, they are a progression 
through the states of a single fundamental system. This second unified picture leads back to (and beyond) the 
first via the arguments below.

The most primitive microstates $\ket{m}$ of the fundamental system are taken to have equal amplitudes, so we are beginning with 
the simplest imaginable picture, in which all possibilities are realized and have equal weights.

To avoid confusion, it should be emphasized that time, quantum mechanics, etc. are 
yet to be defined in this initial picture, but one may still write 
\begin{equation}
\ket{\mathrm{Nature}}= \sum_m a_m \ket{m} \quad , \quad a_m=1\;  \mathrm{for} \; \mathrm{all} \; m \; 
\label{a1}
\end{equation}
in the same spirit as in ordinary quantum mechanics, where this global state, expanded in ``basis states'' $\ket{m}$, is postulated to describe all of physical reality.

In a sense, Nature is viewed as existing in all these states ``simultaneously'', just as an electron in state $\ket{\psi}$ 
simultaneously exists in all position states $\ket{\vec{r}}$ with amplitudes $\langle \vec{r} | \psi \rangle $:
\begin{equation}
\ket{\psi}= \int{d^3 r}  \ket{\vec{r}} \langle \vec{r} | \psi \rangle \; .
\label{a2}
\end{equation} 

Time will be defined as a parameter describing a trajectory through the space of states. 
There are, of course, precedents for defining time internally, within a stationary state, using some internal parameter. 
For example, in minisuperspace models based on the Wheeler-DeWitt equation~\cite{DeWitt-1967,Hartle-Hawking}, this parameter is the cosmic scale factor, and time is defined by the expansion of the universe. In the present picture, \textit{all} coordinates and fields are defined internally, with the progression of events in Nature regarded as progression through a space of states. In the most primitive space, each point is defined by a microstate 
$\ket{m}$. In the emergent space corresponding to DeWitt's $\Phi $ (see above), each point is defined by a macrostate, as described above (\ref{eq2.1}).

For a concrete characterization of both microstates and macrostates, 
we adopt the picture that the fundamental
system is composed of discrete distinguishable constituents that are called 
``dits'' because each can exist in any of $d$ states labeled by $i=1,2,...,d$. 
The number of dits in the $i$th state 
determines the size of the field or coordinate labeled by $i$.

The fact that fields and coordinates have the same basic status in standard
physics, as noted above, is then explained by the fact that they have the
same fundamental origin.

(2) The action is defined to be essentially the negative of the entropy, in
(\ref{eq2.25}). Action conventionally has the units of $\hbar $, and entropy the units
of the Boltzmann constant $k$, but here we use natural units, with 
$\hbar =k=c=1$.

(3) Much later in the paper, beginning with (\ref{eq6.21}), the resulting path integrals with Euclidean form are 
transformed to equivalent Lorentzian path integrals, with the action left
unchanged. 

One has then regained the standard formulation of quantum field theory. One
can subsequently transform from path-integral quantization to canonical
quantization in the usual way (since the action has a standard form).

(4) The assumption that all states of the fundamental system are realized
leads inevitably to a multiverse picture. There are still many who reject the
possibility of a multiverse, but one should recall that most people at the
time of Galileo would have rejected the possibility of  hundreds of billions
of galaxies, or a single galaxy, or even a heliocentric Solar System, and that 
the history of physics shows a steady progression toward more expansive 
views of Nature.

Among the vast number of states of the fundamental system (in the full path
integral), there are some trajectories through these states which can
legitimately be assigned to universes, in the sense that the states or configurations can be
coherently connected with high probability. This will be the case if there is a 
path through the space of these configurations along which a local 
minimum is stably maintained in the action.

In the present picture our own universe is stabilized \textit{topologically} 
-- specifically, it has an extremely stable geometry determined by 
two topological defects in a primordial condensate, 
one in $4$-dimensional external
spacetime and the other in a $(D-4)$-dimensional internal space. 
Each of these is ``vortex-like'' (or ``instanton-like''), in the sense that they involve 
 circulation of the primordial condensate around a central point. 
 The stability of our universe is then analogous to the stability 
 of a vortex in a 2-dimensional fluid.

There is an additional nuance, in that a universe can
be made stable through an effect which is exhibited in the behavior  of the
quartic self-coupling of the recently discovered Higgs boson: Within the
Standard Model, the unrenormalized value of this coupling appears to be very
nearly equal to zero~\cite{elias-miro,alekhin,degrassi,buttazzo}, but at low energies it is made appreciably finite by
radiative corrections, so that a stable Higgs condensate forms. This
suggests that more generally there will be configurations, within the
complete path integral of all possibilities, where a universe is
``bootstrapped'' into existence, because
(at low enough energies) radiative corrections will similarly yield a
nonzero quartic self-coupling for a primordial condensate which allows it to
form. We are thus envisioning a self-consistent universe, or trajectory
through the space of all possibilities, in which a condensate is stably
maintained by this effect. 
In order to search for such a possible solution (i.e. a persistent local minimum in the
action), we adopt the artifice in (\ref{eq2.26})-(\ref{eq2.25c}) of adding an imaginary random potential
to the action, proportional to a parameter $b$ which is ultimately taken
to go to zero: $b\rightarrow 0+$.

Known physics is regained in a very simple picture based on a
primordial condensate with an $SU(2)\times U(1)$ order parameter in external
spacetime and an $SO(D-4)\times U(1)$ (or more precisely $Spin(D-4)\times U(1)$) 
order parameter in the internal space.
Each of these factors in the overall order parameter has a vortex-like
topological defect at the origin. The external topological defect is
interpreted as the Big Bang, and the internal topological defect gives rise
to an $SO(N)$ grand-unified gauge group, with $N=D-4\geq 10$. 

The gravitational vierbein and the gauge fields of other forces are
interpreted as ``superfluid velocities'',
with arbitrary curvatures permitted by a background of ``rapidly fluctuating'' topological
defects that are analogous to vortices and vortex rings (or extended and closed flux tubes) -- in roughly the same way that, as shown by Feynman and Onsager, 
vortices permit rotation of a superfluid. See the discussion below involving (\ref{curv-1})-(\ref{curv-6}).

(5) In Sections \ref{sec:sec4} and \ref{sec:sec5}, the fermionic fields and scalar bosonic
fields are found to automatically couple in the correct way to both the
gravitational field and the gauge fields of the other forces. 

(6) At the same time, local Lorentz invariance automatically emerges (rather
than being postulated), and external spacetime is automatically  $(3+1)-$dimensional.

(7) The present theory unavoidably predicts $SO(N)$ grand unification.
This is consistent with the fact that many regard $SO(10)$ as the most
appealing gauge group for unifying Standard Model forces. Family replication can result from  $D-4>10$, via a horizontal group.

Gauge symmetry results from rotational symmetry in the internal space, 
and this explains why forces are described by a basic gauge symmetry 
which is so similar to rotational symmetry.

(8) The present theory also unavoidably predicts supersymmetry, beginning
with the unphysical supersymmetry of Section \ref{sec:sec3} but ending with the standard
supersymmetric action of (\ref{eq8.3}), which automatically includes the auxiliary fields $F$, 
after transformation to the physical fields. But, as noted below (\ref{eq8.3}), there is a kind of ``F-term susy breaking''.

The lightest supersymmetric particle (as a subdominant component) can stably coexist with the dark matter candidate of the present theory~\cite{DM2021a,DM2021b}.

(9) The usual cosmological constant (regarded as one of the deepest
problems in standard physics) automatically vanishes for two independent
reasons, according to the arguments in Section \ref{sec:sec8+}: 

(i) For fermion fields and scalar boson fields, there is no factor of 
$e=\sqrt{-g}$ 
in the integrals giving their action.

(ii) When the gauge-field action is quantized, the operators must be
normal-ordered, in accordance with the interpretation of the origin of this
action in Section \ref{sec:sec8+}: It arises from the response of vacuum fields
to the curvature of the external gauge fields, and it must therefore vanish
when there are no external fields. The vacuum stress-energy  tensor for the
gauge fields then also vanishes.

Standard physics is regained in each case:

(i) As shown in (\ref{eq8.7x}), classical matter (which follows the on-shell classical
equations of motion) acts as a source for Einstein gravity in the same way
as in standard physics, and all matter and fields move in the same way.

(ii) The results are consistent with experiment and observation, even though
there is no vacuum zero-point energy for the gauge fields.

Many people (including some who are otherwise expert in this area) will
naively object that the Casimir effect, as verified experimentally,
demonstrates that the electromagnetic field does have a zero-point energy in
the vacuum. 

This belief is common but incorrect~\cite{Jaffe}. The experimentally-observed Casimir effect demonstrates only that the static
electromagnetic field energy is \textit{changed} by the modification of
boundary conditions~\cite{casimir1,casimir2}. In the simplest model, two metal plates are inserted
and the force between them calculated. There are two ways to do the
calculation: The first is indeed to assume zero-point vibrations of the
electromagnetic field, whose energy is modified when the boundary conditions
are modified. The second approach is instead to consider the processes
involving virtual photons which mediate the interaction of the plates, with
no reference to zero-point vibrations and no need for a vacuum energy. The
first method is more popular because it is easier. But the two
methods give the same answer, as they should. (The second method regards the force as 
mediated by virtual photons, and the first obtains the force from the derivative of the 
energy with respect to a displacement.) The second method is more
difficult, but is consistent with the way other virtual processes are
calculated in e.g. quantum electrodynamics. Of course, the second method
also implies a change in the static electromagnetic field energy 
(interpreted as a van der Waals interaction), but this 
\textit{change} does not imply an initially nonzero vacuum zero-point energy. 

In summary, the observed Casimir effect is perfectly consistent with the
present theory, in which there is no vacuum zero-point energy due to the
electromagnetic field or other gauge fields.

(10) Although the usual cosmological constant vanishes, there will
still be a weaker response of the vacuum to the imposition of external
fields. In Appendix \ref{sec:appG}, it is shown how the Einstein-Hilbert  
action (\ref{eq8.7}) for the gravitational field can arise from a vacuum response within 
the present picture. 

Quantum gravity is regularized by an energy and momentum cutoff $a_0^{-1}$, where $a_0$ is the minimum length of (\ref{eq2.1}), which can be regarded as comparable to the Planck length $\ell_P$.

(11) The Bekenstein-Hawking entropy of black holes automatically emerges in the present picture, for the reason given in Section \ref{sec:sec8+}: Gibbons and Hawking~\cite{Gibbons-Hawking} have shown that the Euclidean action $S_{E}$  of a black hole is equal to an expression which has the right form to be interpreted as the  Bekenstein-Hawking entropy plus a contribution from angular momentum. Until now no convincing reason has been given for the $S$ in this expression to be identified 
as a true entropy derived from microstates, but in the present picture -- before the effect of rotation is added -- (\ref{Euclid}) implies that the Euclidean action of a black hole \textit{is} its entropy. This entropy ultimately originates from the microstates of the dits that are consistent with the gravitational field configuration that  comprises the black hole.

(12) As discussed in Sections \ref{sec:DM} and \ref{sec:sec8+}, the present theory predicts new particles, including a new dark matter WIMP which is consistent with experiment and observation because it has no couplings other than its second-order gauge couplings to $W$ and $Z$ bosons. It should be observable in future colliders within roughly the next 12-20 years, and in direct-detection experiments within roughly the next 2-5 years, and it may already have been detected via the gamma rays observed by Fermi-LAT and antiprotons observed by AMS-02~\cite{DM2021a,DM2021b}. This particle is unique among viable dark matter candidates in that both its couplings and its mass are precisely determined, making clean experimental tests possible in the near future. The favorable characteristics of this candidate are summarized at the end of Section \ref{sec:sec8+}.

Let us now proceed to the detailed arguments behind the above claims.

\section{\label{sec:sec2}Statistical origin of the initial action}

For a theory to be viable, it must be mathematically (and philosophically) consistent, its premises must lead to testable predictions, and these predictions must be consistent with experiment and observation. The theory presented here appears to satisfy these requirements, but it starts with an extremely unfamiliar point of view: There are initially no laws, and instead all possibilities are realized with equal weight. The observed laws of nature are emergent phenomena, which result from statistical counting and the geography (i.e. specific features) of our particular universe in $D$ dimensions. In other words, standard physics (including familiar extensions such as grand unification and supersymmetry) emerges as an effective field theory at relatively low energies.

Our starting point is a single fundamental system which consists of
identical (but distinguishable) irreducible objects, which we will call
``dits''. Each dit can exist in any of $d$ states, with the number
of dits in the $i$th state represented by $n_{i}$. An unobservable
microstate of the fundamental system is specified by the number of dits and
the state of each dit. An observable macrostate is specified by only the
occupancies $n_{i}$ of the states. 

As discussed below, $D$ of the states are
used to define $D$ spacetime coordinates $x^{M}$, and $N_{F}$ of the states
are used to define fields $\phi _{k}$. 

Let us begin with the coordinates: 
\begin{equation}
x^M=\Delta n_M  a_0  \
\quad , \quad  \Delta n_M = n_M - \overline{n}   
\quad , \quad M=0,1,...,D-1 
\label{eq2.1}
\end{equation}
with $\overline{n}$, which is defined below, specifying the initial origin of 
coordinates. It is convenient to include a (very small) fundamental length 
$a_{0}$ in this definition, so that we can later express the coordinates in
conventional units. One can think of $a_{0}$  as being comparable to the Planck length $\ell_{P}$. 

As discussed below, we will eventually take the limit 
$\overline{n}  \rightarrow \infty$, with $\Delta n_M$ finite, and there will then be no lower bound 
to negative coordinates. I.e., $\Delta n_M$ can have any integer value. (A central feature of the present theory is that both 
coordinates and physical fields are defined by relatively 
small perturbations $\Delta n_i = n_i - \overline{n} $, analogous to waves 
on a deep ocean.)

Now define a set of initial fields $\phi _{k}$ by 
\begin{equation}
\phi _{k}^{2}\left( x\right) =\rho _{k}\left( x\right) 
\quad , \quad k=1,2,..., N_{F}  
\label{eq2.2}
\end{equation}
where 
\begin{equation}
\rho _{k}\left( x\right) =n_{k}\left( x\right) /a_{0}^{D}  \label{eq2.3}
\end{equation}
and $x$ represents all the coordinates. 
(To avoid awkward notation, we write $n_k$ for $n_{i=D+k}$.)
These primitive bosonic fields 
$\phi _{k}$ are then real, and defined only up to a phase factor $\pm 1$.

We now set out to calculate the entropy $S$ for a given configuration of the fields 
$\phi _{k}$ at all points in spacetime. This will essentially become the action for a 
given path (i.e.  specific classical field configuration) in the quantum path integral, 
beginning with the identification (\ref{eq2.25}). 

Let $S\left( x\right) $ be the entropy at a fixed point $x$, as defined by 
$S\left( x\right) =\log \,W\left( x\right) $. Here $W\left( x\right) $ is the total
number of microstates for fixed occupation numbers $n_{i}$: $W\left(
x\right) =N\left( x\right) !/\Pi _{i}\,n_{i}\left( x\right) !$, with 
\begin{equation}
N\left( x\right) =\sum_{i}n_{i}\left( x\right) \quad ,\quad i=1,2,..., d
\; .
\label{eq2.4}
\end{equation}
The total number of available microstates for all points $x$ is $W=\Pi
_{x}\,W\left( x\right) $, so the total entropy is 
\begin{eqnarray}
S = \sum_{x}\,S\left( x\right) \quad , \quad
S\left( x\right) = \log \Gamma \left( N\left( x\right) +1\right)
-\sum_{i}\log \Gamma \left( n_{i}\left( x\right) +1\right) \; .
\label{eq2.5}
\end{eqnarray}

We will see below that $n_{k}\left( x\right) $ can be approximately treated
as a continuous variable when it is extremely large, with 
\begin{eqnarray}
\frac{\partial S}{\partial n_{k}\left( x\right) } &=&\psi \,\left( N\left(
x\right) +1\right) -\psi \left( n_{k}\left( x\right) +1\right) 
\label{eq2.6}\\
\frac{\partial ^{2}S}{\partial n_{k^{\prime }}\left( x\right) \partial
n_{k}\left( x\right) } &=&\psi \,^{\left( 1\right) }\left( N\left( x\right)
+1\right) -\psi ^{\left( 1\right) }\left( n_{k}\left( x\right) +1\right)
\delta _{k^{\prime }k}\; .  \label{eq2.7}
\end{eqnarray}
The functions $\psi \,\left( z\right) =d\log \Gamma \left( z\right) /dz$ and 
$\psi ^{\left( 1\right) }\,\left( z\right) =d^{2}\log \Gamma \left( z\right)
/dz^{2}$ have the asymptotic expansions
\begin{eqnarray}
\psi \,\left( z\right) = \log z-\frac{1}{2z}-\sum_{l=1}^{\infty }
\frac{B_{2l}}{2l\,z^{2l}} \quad , \quad 
\psi ^{\left( 1\right) }\,\left( z\right) = \frac{1}{z}+\frac{1}{2z^{2}}
+\sum_{l=1}^{\infty }\frac{B_{2l}}{\,z^{2l+1}}  \label{eq2.8}
\end{eqnarray}
as $z\rightarrow \infty $. It will be assumed that 
each $n_{k}\left( x\right) $ has some characteristic value $\overline{n}
_{k}\left( x\right) $ which is vastly larger than nearby values:
\begin{equation}
n_{k}\left( x\right) =\overline{n}_{k}\left( x\right) +\Delta n_{k}
\left( x \right) \quad ,\quad \overline{n}_{k}\left( x\right) \ggg 
\left | \Delta n_{k}\left( x\right) \right | \label{eq2.9}
\end{equation}
where ``$\ggg$'' means ``is vastly greater than'', as in $10^{1000} \ggg 1$. 
This assumption is consistent with the fact that the initial action 
of (\ref{eq2.24}) and (\ref{eq2.25}) has no lower bound as $n_{k}\left( x\right) \rightarrow \infty $ before the extra stochastic term 
involving (\ref{eq2.26}) is added. 
(To state the reasoning more cleanly, but slightly out of the order of presentation, the limit (\ref{eq2.26xx}) implies 
the limit $\overline{n}_{k} \rightarrow \infty$.) 
Then it is an extremely good approximation to use the asymptotic formulas 
above and write 
\begin{eqnarray}
\hspace{-1cm}
S=S_{0}+\sum_{x,k}a_{k}\left( x\right) \Delta n_{k}\left( x\right)
-\sum_{x,k}a_{k}^{\prime }\left( x\right) \left[ \Delta n_{k}\left( x\right) 
\right] ^{2} + \sum_{x,k,k^{\prime }\ne k}
a_{kk^{\prime }}^{\prime }\left( x\right) 
\Delta n_{k}\left( x\right) \Delta n_{k^{\prime}}\left( x\right)
\label{eq2.10}
\end{eqnarray}
\begin{eqnarray}
a_{k}\left( x\right) &=& \log \overline{N}\left( x\right) -\log \overline{n}
_{k}\left( x\right) \label{eq2.11a} \\
a_{k}^{\prime }\left( x\right) &=& \left( 2
\overline{n}_{k}\left( x\right) \right) ^{-1}
- \left( 2\overline{N}\left( x \right) \right) ^{-1}  \quad , \quad
a_{kk^{\prime }}^{\prime }\left( x\right) 
= \left( 2\overline{N}\left( x \right) \right) ^{-1}  
\label{eq2.11b}
\end{eqnarray}
where $\overline{N}\left( x\right) $ is the value of $N\left( x\right) $
when $n_{k}\left( x\right) =$ $\overline{n}_{k}\left( x\right) $ for all $k$, 
and the higher-order terms have been separately neglected in $a_{k}\left(
x\right) $ and $a_{k}^{\prime }\left( x\right) $. (The above results then also follow immediately 
from Stirling's approximation for factorials.) For simplicity, we will also 
neglect the terms involving 
$\left( 2\overline{N}\left( x \right) \right) ^{-1}$. (If these small terms are retained, 
the conclusions below still hold with some trivial redefinitions, but the notation 
and algebra become much more tedious.) Since there is initially
no distinction between the fields labeled by $k$, it is consistent to assume
that they all have the same $\overline{n}_{k}\left( x\right) =\overline{n}
\left( x\right) $, and that $\overline{n}\left(
x\right) $ is independent of $x$: 
$\overline{n}\left( x\right) =\overline{n}$ and 
$\overline{N}\left( x\right) =\overline{N}$, so that 
\begin{eqnarray}
a_{k}\left( x \right) &=& a = \log \left( \overline{N}/\overline{n} \right)
\label{eq2.11c} \\
a_{k}^{\prime }\left( x\right) &=& a^{\prime } = 
\left( 2\overline{n}\right) ^{-1} \; .  \label{eq2.11d}
\end{eqnarray}
(The above assumptions are actually needed only to simplify the presentation, 
and they have no effect on the final results below as $\overline{n} \longrightarrow \infty$.)

It is not conventional or convenient to deal with $\Delta n_{k}\left(
x\right) $ and $\left[ \Delta n_{k}\left( x\right) \right] ^{2}$, so let us
instead write $S$ in terms of the fields $\phi _{k}$ and their derivatives 
$\partial \phi _{k}/\partial x^{M}$ via the following procedure: First, we
can switch to a new set of points $\overline{x}$, defined to be the corners
of the $D$-dimensional hypercubes centered on the original points $x$. It is
easy to see that 
\begin{equation}
S=S_{0}+\sum_{\overline{x},k}a\left\langle \Delta n_{k}\left( x\right)
\right\rangle -\sum_{\overline{x},k}a^{\prime }\left\langle \left[ \Delta
n_{k}\left( x\right) \right] ^{2}\right\rangle  \label{eq2.12}
\end{equation}
where $\left\langle \cdots \right\rangle $ in the present context indicates
an average over the $2^{D}$ boxes labeled by $x$ which have the common
corner $\overline{x}$. Second, at each point $x$ we can write $\Delta
n_{k}=\Delta \rho _{k}a_{0}^{D}=\left( \left\langle \Delta \rho
_{k}\right\rangle +\delta \rho _{k}\right) a_{0}^{D}$, with $\left\langle
\delta \rho _{k}\right\rangle =0$: 
\begin{eqnarray}
S &=&S_{0}+\sum_{\overline{x},k}a\left\langle \left\langle
\Delta \rho _{k}\right\rangle +\delta \rho _{k}\right\rangle a_{0}^{D}-\sum_
{\overline{x},k}a^{\prime }\left\langle \left( \left\langle \Delta \rho
_{k}\right\rangle +\delta \rho _{k}\right) ^{2}\right\rangle \left(
a_{0}^{D}\right) ^{2} \label{eq2.13} \\
&=&S_{0}+\sum_{\overline{x},k}a\left\langle \Delta \rho
_{k}\right\rangle a_{0}^{D}-\sum_{\overline{x},k}a^{\prime }\left[
\left\langle \Delta \rho _{k}\right\rangle ^{2}+\left\langle \left( \delta
\rho _{k}\right) ^{2}\right\rangle \right] \left( a_{0}^{D}\right) ^{2}
\; .
\label{eq2.14}
\end{eqnarray}
Each of the points $x$ surrounding $\overline{x}$ is displaced by 
$\delta x^{M}=$ $\pm a_{0}/2$ along each of the $x^{M}$ axes, so
\begin{eqnarray}
\hspace{-1cm}
\left\langle \left( \delta \rho _{k}\right) ^{2}\right\rangle
&=&\left\langle \left( \delta \phi _{k}^{2}\right) ^{2}\right\rangle 
\label{eq2.15} \\
&=&\left\langle \sum_{M}\left( \frac{\partial \phi _{k}^{2}}{\partial x^{M}}
\delta x^{M}+\frac{1}{2}\frac{\partial ^{2}\phi _{k}^{2}}{\partial \left(
x^{M}\right) ^{2}}\left( \delta x^{M}\right) ^{2}\right) ^{2}\right\rangle 
\label{eq2.16} \\
&=&\left\langle \sum_{M}\left( 2\phi _{k}\frac{\partial \phi _{k}}{\partial
x^{M}}\delta x^{M}+\left( \frac{\partial \phi _{k}}{\partial x^{M}}\right)
^{2}\left( \delta x^{M}\right) ^{2}+\phi _{k}\frac{\partial ^{2}\phi _{k}}
{\partial \left( x^{M}\right) ^{2}}\left( \delta x^{M}\right) ^{2}\right)
^{2}\right\rangle  \label{eq2.17}
\end{eqnarray}
to lowest order, where it is now assumed that 
at normal energies the fields are slowly varying
over the extremely small distance $a_{0}$. This assumption is justified by
the prior assumption that $\overline{n}$ is extremely large: $\phi
_{k}^{2}\left( x\right) =\rho _{k}\left( x\right) =n_{k}\left( x\right)
/a_{0}^{D}$ implies that $2\delta \phi _{k}/\phi _{k}\approx \delta
n_{k}/n_{k}$ and $\phi _{k}=n_{k}^{1/2}a_{0}^{-D/2}$, so that $\delta \phi
_{k}\sim \delta n_{k}\,n_{k}^{-1/2}a_{0}^{-D/2}$. The minimum change in 
$\phi _{k}$ is given by $\delta n_{k}=1$:
\begin{eqnarray}
\delta \phi _{k}^{\min } \sim n_{k}^{-1/2} a_{0}^{-D/2}
\label{eq2.17xx}
\end{eqnarray}
which means that $\delta \phi _{k}^{\min }$ is
extremely small if $n_{k} $ is extremely large. 

In other words, the fields $\phi _{k}$ have effectively continuous values as 
$\overline{n}\rightarrow \infty $.

For extremely large $\overline{n}$ it is an extremely good approximation to
neglect the middle term in (\ref{eq2.17}), and to replace $\phi _{k}^{2}$ by 
\begin{equation}
\overline{\phi }^{2}=\overline{\rho }=\overline{n}/a_{0}^{D}
\label{eq2.17a}
\end{equation}
giving 
\begin{equation}
a^{\prime }\left\langle \left( \delta \rho _{k}\right) ^{2}\right\rangle =
\frac{1}{2a_{0}^{D}}\sum_{M}\left[ \left( \frac{\partial \phi _{k}}{\partial
x^{M}}\right) ^{2}a_{0}^{2}+\left( \frac{\partial ^{2}\phi _{k}}{\partial
\left( x^{M}\right) ^{2}}\right) ^{2}\frac{a_{0}^{4}}{16}\right] \; .
\label{eq2.18}
\end{equation}
It is similarly an extremely good approximation to neglect the term in
(\ref{eq2.14}) 
involving $a^{\prime }\left( a_{0}^{D}\right) ^{2}
\left\langle \Delta \rho _{k}\right\rangle ^{2}
= \left\langle \Delta n_{k}\right\rangle ^{2}/2\overline{n}$ in
comparison to that involving $\left\langle \Delta \rho _{k}\right\rangle
a_{0}^{D}=\left\langle \Delta n_{k}\right\rangle $, so that 
\begin{equation}
S=S_{0}+\sum_{\overline{x},k}a_{0}^{D}\frac{\mu_{0}}{m_{0}} 
\left( \phi _{k}^{2} - \overline{\phi }^{2}\right) 
-\sum_ {\overline{x},k}\sum_{M}a_{0}^{D}
\frac{1}{2m_{0}^{2}}\left[ \left( \frac{\partial
\phi _{k}}{\partial x^{M}}\right) ^{2}+\frac{a_{0}^{2}}{16}\left( \frac
{\partial ^{2}\phi _{k}}{\partial \left( x^{M}\right) ^{2}}\right) ^{2}\right]
\label{eq2.19}
\end{equation}
where 
\begin{equation}
m_{0} = a_{0}^{-1} \quad ,\quad \mu_{0} = 
m_{0} \log \left( \overline{N}/\overline{n} \right) \; .  \label{eq2.21}
\end{equation}

The philosophy behind the above treatment is simple: We essentially wish to
replace $\left\langle f^{2}\right\rangle $ by $\left( \partial f/\partial
x\right) ^{2}$, and this can be accomplished because 
\begin{equation}
\left\langle f^{2}\right\rangle -\left\langle f\right\rangle
^{2}=\left\langle \left( \delta f\right) ^{2}\right\rangle \approx
\left\langle \left( \partial f/\partial x\right) ^{2}\left( \delta x\right)
^{2}\right\rangle =\left( \partial f/\partial x\right) ^{2}
\left( a_{0}/2\right)^{2} \; . \label{eq2.22}
\end{equation}
The form of (\ref{eq2.19}) also has a simple interpretation: The entropy 
$S$ increases with the number of dits, but decreases when the dits are 
not uniformly distributed.

In the continuum limit, 
\begin{equation}
\sum_{\overline{x}}a_{0}^{D}\rightarrow \int d^{D}x   \label{eq2.23}
\end{equation}
(\ref{eq2.19}) becomes 
\begin{equation}
S=S_{0}+\int d^{D}x\,\,\sum_{k}
\left\{ \frac{\mu_{0}}{m_{0}}
\left( \phi _{k}^{2} - \overline{\phi }^{2}\right) 
-\frac{1}{2m_{0}^{2}}\sum_{M}\left[ \left( \frac{\partial \phi _{k}}{\partial
x^{M}}\right) ^{2}+\frac{a_{0}^{2}}{16}\left( \frac{\partial ^{2}\phi _{k}}
{\partial \left( x^{M}\right) ^{2}}\right) ^{2}\right] \right\} \; .
 \label{eq2.24}
\end{equation}

A physical configuration of all the fields $\phi _{k}\left( x\right) $
corresponds to a specification of all the density variations $\Delta \rho
_{k}\left( x\right) $. In the present picture, the probability of such a
configuration is proportional to $W=e^{S}$. In a path integral with Euclidean form,
the probability is proportional to $e^{-\overline{S}_{b}}$, where 
$\overline{S}_{b}$ is the action for these bosonic fields. 
We conclude that 
\begin{equation}
\overline{S}_{b}=-S+\mathrm{constant} \label{eq2.25}
\end{equation}
and we will choose the constant to equal $S_{0}$.

In the following it will
be convenient to write the action in terms of $\widetilde{\phi }_{k}
= m_{0}^{-1/2}\phi _{k}$. For simplicity, we assume that the number of
relevant $\widetilde{\phi }_{k}$ is even, so that we can group these real
fields in pairs to form $N_{f}$ complex fields $\widetilde{\Psi }_{b,k}$. 
It is also convenient to subtract out the enormous contribution of 
$\overline{\phi }$ by defining
\begin{eqnarray}
\Psi _{b}=\widetilde{\Psi }_{b}-\overline{\Psi }_{b}
\label{eq2.25a}
\end{eqnarray}
where $\widetilde{\Psi }_{b}$ is the vector with components $\widetilde{\Psi 
}_{b,k}$ and $\overline{\Psi }_{b}$ is similarly defined with 
$\phi_{k}\rightarrow \overline{\phi }$. Then the action can be written
\begin{eqnarray}
\hspace{-0.8cm}\overline{S}_{b}=\int d^{D}x\,\Bigg\{\frac{1}{2m_{0}}\left[ 
\frac{\partial \Psi _{b}^{\dagger }}{\partial x^{M}}\frac{\partial \Psi _{b}
}{\partial x^{M}}+\frac{a_{0}^{2}}{16}\frac{\partial ^{2}\Psi _{b}^{\dagger }
}{\partial \left( x^{M}\right) ^{2}}\frac{\partial ^{2}\Psi _{b}}{\partial
\left( x^{M}\right) ^{2}}\right] -\mu _{0}\,\left( \widetilde{\Psi }
_{b}^{\dagger }\widetilde{\Psi }_{b}-\overline{\Psi }_{b}^{\dagger }
\overline{\Psi }_{b}\right) \Bigg\}
\label{eq2.25b}
\end{eqnarray}
since $\overline{\Psi }_{b}$ is constant, with summation now implied over
repeated indices like $M$.

As described above, in Section \ref{sec:sec1a}, we now add an extra imaginary term 
$i\widetilde{V}\,\Psi _{b}^{\dagger }\Psi _{b}$ in the integral giving the action. 
Here $\widetilde{V}$ is a potential which has a Gaussian distribution, with 
\begin{equation}
\left\langle \widetilde{V}\,\right\rangle =0\quad ,\quad \left\langle 
\widetilde{V}\left( x\right) \widetilde{V}\left( x^{\prime }\right)
\right\rangle =b\,\delta \left( x-x^{\prime }\right)   
\label{eq2.26}
\end{equation}
where $b$ is a constant, with
\begin{eqnarray}
b \rightarrow 0+
\label{eq2.26xx}
\end{eqnarray}
at the end of the calculations.

Then the complete action has the form 
\begin{align}
\widetilde{S}_{B}\left[ \Psi _{b}^{\dagger },\Psi _{b}\right]
=\int d^{D}x\,\Bigg\{\frac{1}{2m_{0}}\left[ \frac{\partial \Psi
_{b}^{\dagger }}{\partial x^{M}}\frac{\partial \Psi _{b}}{\partial x^{M}}+
\frac{a_{0}^{2}}{16} \frac{\partial ^{2}\Psi _{b}^{\dagger }}{\partial \left(
x^{M}\right) ^{2}}\frac{\partial ^{2}\Psi _{b}}{\partial \left( x^{M}\right)
^{2}}\right] 
-\mu _{0}\,\left( \widetilde{\Psi }_{b}^{\dagger }\widetilde{\Psi }_{b}-
\overline{\Psi }_{b}^{\dagger }\overline{\Psi }_{b}\right)  +i\widetilde{V}
\,\Psi _{b}^{\dagger }\Psi _{b}\Bigg\} \; .
\label{eq2.25c}
\end{align}
In the following we will assume that the only fields which make an
appreciable contribution in (\ref{eq2.25c}) are those for which 
$\int d^{D}x\, \overline{\Psi } _{b}^{\dagger }\Psi _{b} = 
\overline{\Psi } _{b}^{\dagger } \int d^{D}x\, \Psi _{b} = 0$. 
This assumption is justified by the fact that $\overline{\Psi } _{b}$ 
is constant with respect to all the coordinates and, in the present
picture, fields $\Psi _{b}$ corresponding to  physical gauge representations 
have nonzero angular momenta in the internal space of Section \ref{sec:sec5} 
and  Appendices \ref{sec:appA} and \ref{sec:appB}. Then (\ref{eq2.25c})
simplifies to 
\begin{eqnarray}
\hspace{-0.7cm} \widetilde{S}_{B}\left[ \Psi _{b}^{\dagger },\Psi _{b}\right]
=\int d^{D}x\,\Bigg\{\frac{1}{2m_{0}}\left[ \frac{\partial \Psi
_{b}^{\dagger }}{\partial x^{M}}\frac{\partial \Psi _{b}}{\partial x^{M}}+
\frac{a_{0}^{2}}{16}\frac{\partial ^{2}\Psi _{b}^{\dagger }}{\partial \left(
x^{M}\right) ^{2}}\frac{\partial ^{2}\Psi _{b}}{\partial \left( x^{M}\right)
^{2}}\right]  
-\mu _{0}\,\Psi_{b}^{\dagger }\Psi_{b}
 +i\widetilde{V}\,\Psi _{b}^{\dagger }\Psi _{b}\Bigg\} \; .
\label{eq2.27}
\end{eqnarray}

\section{\label{sec:sec3}Primitive supersymmetry}

If $F$ is any functional of the fundamental fields $\Psi _{b}$, 
its average value is given by 
\begin{equation}
\left\langle F\right\rangle =\left\langle \frac{\int \mathcal{D}\,\Psi
_{b}^{\dagger }\,\mathcal{D}\,\Psi _{b}\,F\left[ \Psi _{b}^{\dagger },\Psi _{b}
\right] \,e^{-\widetilde{S}_{B}\left[ \Psi _{b}^{\dagger },\Psi _{b}\right] }}
{\int \mathcal{D}\,\underline{\Psi } _{\,b}^{\dagger }\,\mathcal{D}
\,\underline{\Psi } _{\,b}
\,e^{-\widetilde{S}_{B}\left[ \underline{\Psi } _{\,b}^{\dagger },
\underline{\Psi } _{\,b}\right] }}\right\rangle   \label{eq3.1}
\end{equation}
where $\left\langle \cdots \right\rangle $ now represents an average over
the perturbing potential $i\widetilde{V} \,$ and $\int \mathcal{D}\,\Psi
_{b}^{\dagger }\,\mathcal{D}\,\Psi _{b}$ is to be interpreted as 
$\prod\nolimits_{x,k}\int_{-\infty }^{\infty }d \, 
\mathrm{Re} \, \Psi _{b,k}\left(
x\right) \,\int_{-\infty }^{\infty }d  \, 
\mathrm{Im} \, \Psi _{b,k}\left( x\right) $.
The transition from the original summation over $n_{k}\left( x\right) $ to
this Euclidean path integral has the form (with $\Delta n=1$ here) 
\begin{eqnarray}
\hspace{-0.5cm}\sum_{n=0}^{\infty }f\left( n\right) \Delta n\rightarrow
\int_{0}^{\infty }fdn\rightarrow \int_{0}^{\infty }f\;d\left( a_{0}^{D}\phi
^{2}\right) \rightarrow 2\overline{\phi }a_{0}^{D}\int_{0}^{\infty }f\;d\phi 
\rightarrow 2\overline{\phi }\,a_{0}^{D}m_{0}^{1/2}\int_{-\infty }^{\infty
}f\;d\phi ^{\prime }
\label{eq3.2}
\end{eqnarray}
where $\phi ^{\prime }=\widetilde{\phi }-m_{0}^{-1/2}\overline{\phi }$,
since $d\left( \phi ^{2}\right) \approx 2\overline{\phi }\,d\phi $ 
is an extremely good approximation for physically relevant fields, 
and since $\phi ^{\prime } $ effectively ranges from $-\infty $ 
to $+\infty $. Each $\phi ^{\prime }$ then
becomes a $\mathrm{Re} \,  \Psi _{b,k}\left( x\right) $ or 
$\mathrm{Im} \,  \Psi _{b,k}\left( x\right) $, and the constant factors 
cancel in the numerator and denominator of (\ref{eq3.1}).

The presence of the denominator makes it difficult to perform the average of
(\ref{eq3.1}), but there is a trick for removing the 
bosonic degrees of freedom $\underline{\Psi } _{\,b}$ 
in the denominator and replacing them with fermionic degrees
of freedom $\Psi _{f}$ in the numerator~\cite{parisi,efetov,huang}:
After integration by parts (with boundary terms usually assumed 
either to vanish or to be irrelevant in this paper), (\ref{eq2.27}) can 
be written in the form 
$\widetilde{S}_{B}=\int d^{D} x \,\Psi _{b}^{\dagger } A \Psi _{b} $. 
Then, since 
\begin{equation}
\int \mathcal{D}\,\underline{\Psi } _{\,b}^{\dagger }\,\mathcal{D}\,
\underline{\Psi } _{\,b}
\,e^{-\widetilde{S}_{B}\left[ \underline{\Psi } _{\,b}^{\dagger },
\underline{\Psi } _{\,b} \right] }= C \left( \det \, \cal A \right) ^{-1}  
\label{eq3.3}
\end{equation}
\begin{equation}
\int \mathcal{D}\,\Psi _{f}^{\dagger }\,\mathcal{D}\,\Psi _{f}\,
e^{-\widetilde{S}_{B}\left[ \Psi _{f}^{\dagger },\Psi _{f}\right] }=
\det \, \cal A
\label{eq3.4}
\end{equation}
where the matrix $\cal A$ corresponds to the operator $A$ and $C$ is a
constant, it follows that 
\begin{eqnarray}
\left\langle F\right\rangle &=& \frac{1}{C} \left\langle \int \mathcal{D}\,\Psi
_{b}^{\dagger }\,\mathcal{D}\,\Psi _{b}\,\mathcal{D}\,\Psi _{f}^{\dagger }\,
\mathcal{D}\,\Psi _{f}\,F\,e^{-\widetilde{S}_{B}\left[ \Psi _{b}^{\dagger
},\Psi _{b}\right] }e^{-\widetilde{S}_{B}\left[ \Psi _{f}^{\dagger },\Psi _{f}
\right] }\right\rangle \label{eq3.5} \\
&=&  \frac{1}{C} \left\langle \int \mathcal{D}\,\Psi ^{\dagger }\,
\mathcal{D}\,\Psi
\,F\,e^{-\widetilde{S}_{bf}\left[ \Psi ^{\dagger },\Psi \right]
}\right\rangle  \label{eq3.6}
\end{eqnarray}
where $\Psi _{b}$ and $\Psi _{f}$ have been combined into 
\begin{equation}
\Psi =\left( 
\begin{array}{c}
\Psi _{b} \\ 
\Psi _{f}
\end{array}
\right)  \label{eq3.7}
\end{equation}
and
\begin{equation}
\widetilde{S}_{bf}\left[ \Psi ^{\dagger },\Psi \right] =\int d^{D}x\,\left\{ 
\frac{1}{2m_{0}}\left[ \frac{\partial \Psi ^{\dagger }}{\partial x^{M}}
\frac{\partial \Psi }{\partial x^{M}}+\frac{a_{0}^{2}}{16}\frac{\partial ^{2}
\Psi^{\dagger }}{\partial \left( x^{M}\right) ^{2}}\frac{\partial ^{2}\Psi }
{\partial \left( x^{M}\right) ^{2}}\right] -\mu_{0} \,\Psi ^{\dagger }\Psi 
+i\widetilde{V}\,\Psi ^{\dagger }\Psi \right\} \; .  \label{eq3.8}
\end{equation}
In (\ref{eq3.7}), $\Psi _{f}$ consists of Grassmann variables $\Psi _{f,k}$, 
just as $\Psi _{b}$ consists of ordinary variables $\Psi _{b,k}$, and $\int 
\mathcal{D}\,\Psi ^{\dagger }\,\mathcal{D}\,\Psi $ is to be interpreted as
\begin{eqnarray}
\prod\nolimits_{x,k}\int_{-\infty }^{\infty }d  \, \mathrm{Re} \,  
\Psi _{b,k}\left( x\right) \,\int_{-\infty }^{\infty }d  \, \mathrm{Im} \,  
\Psi _{b,k}\,\int d\,\Psi
_{f,k}^{\, \ast }\left( x\right) \,\int d\,\Psi _{f,k}\left( x\right) 
\; . 
\label{eq3.9}
\end{eqnarray}
Recall that $\Psi _{b}$ and $\Psi _{f}$ each have $N_{f}$ components. 

In this early stage of the theory, the bosonic fields in the denominator -- which are 
of critical importance for proper normalization -- have effectively been transformed into fermionic fields in the numerator, where they perform the same function. This is the origin of supersymmetry in the present picture. 

For a Gaussian random variable $v$ whose mean is zero, the result 
\begin{equation}
\left\langle e^{-iv}\right\rangle =e^{-\frac{1}{2}\left\langle
v^{2}\right\rangle }  \label{eq3.10}
\end{equation}
implies that 
\begin{eqnarray}
\left\langle e^{-\int d^{D}x \, i  \widetilde{V}\,\Psi ^{\dagger
}\Psi }\right\rangle &=&e^{-\frac{1}{2}\int d^{D}x\,\,\,d^{D}x\,^{\prime
}\,\Psi ^{\dagger }\left( x\right) \Psi \left( x\right) \left\langle 
\widetilde{V}\left( x\right) \widetilde{V}\left( x^{\prime }\right)
\right\rangle \Psi ^{\dagger }\left( x^{\prime }\right) \Psi \left(
x^{\prime }\right) } \label{eq3.11}  \\
&=&e^{-\frac{1}{2}b\int d^{D}x\,\,\,\left[ \Psi ^{\dagger }\left( x\right)
\Psi \left( x\right) \right] ^{2}} \; . \label{eq3.12}
\end{eqnarray}
It follows that 
\begin{equation}
\left\langle F\right\rangle = \frac{1}{C} \int \mathcal{D}\,
\Psi ^{\dagger }\,\mathcal{D} \,\Psi \,F\,e^{-S_{E}}  \label{eq3.13}
\end{equation}
with
\begin{equation}
S_{E}=\int d^{D}x\,\left\{ \frac{1}{2m_{0}}\left[ \frac{\partial \Psi ^{\dagger }
}{\partial x^{M}}\frac{\partial \Psi }{\partial x^{M}}+\frac{a_{0}^{2}}{16}
\frac{\partial ^{2}\Psi ^{\dagger }}{\partial \left( x^{M}\right) ^{2}}
\frac{\partial ^{2}\Psi }{\partial \left( x^{M}\right) ^{2}}\right] -\mu_{0} \,
\Psi^{\dagger }\Psi +\frac{1}{2}b\left( \Psi ^{\dagger }\Psi \right) ^{2}
\right\} \; . \label{eq3.14}
\end{equation}
A special case (with $F=1$) is 
\begin{equation}
Z= \frac{1}{C} \int \mathcal{D}\,\Psi ^{\dagger }\,\mathcal{D}\,
\Psi e^{-S_{E}} \label{eq3.15}
\end{equation}
but according to (\ref{eq3.1}) $Z=1$, so 
$C=\int \mathcal{D}\,\Psi ^{\dagger }\,\mathcal{D}\,\Psi e^{-S_{E}} $
and
\begin{equation}
\left\langle F\right\rangle =\frac{\int \mathcal{D}\,\Psi ^{\dagger }\,
\mathcal{D}\,\Psi \,F\,e^{-S_{E}} }
{\int \mathcal{D}\,\Psi ^{\dagger }\,\mathcal{D}\,\Psi \; e^{-S_{E}}}
\; . \label{eq3.17}
\end{equation}

Again, notice that the fermionic fields $\Psi _{f}$ are effectively a transformed version of the bosonic fields $\underline{\Psi }_{\,b}$.
The coupling between the fields $\Psi _{b}$ and $\Psi _{f}$
(or $\underline{\Psi } _{\,b}$) is due to the random perturbing potential 
$i\widetilde{V}$. In the replacement of (\ref{eq3.1}) by
(\ref{eq3.17}), $F$ essentially serves as a test functional. The
meaning of this replacement is that the action (\ref{eq3.14}), with
both bosonic and fermionic fields, must be used instead of the
original action (\ref{eq2.27}), with only bosonic fields, in treating
all physical quantities and processes, if the average over random
fluctuations in (\ref{eq3.1}) is to disappear from the theory. 

Notice that the two steps above serve two independent purposes: The transformation of $\underline{\Psi }_{\,b}$ in the denominator 
to $\Psi _{f}$ in the numerator provides a more convenient formulation because all fields now have equal status in the numerator, and can be treated in the same way. The introduction of 
an infinitesimal perturbing potential is then preparation for the formation of a condensate at finite energy (in a cooling universe), as discussed in Section \ref{sec:sec1a}. Of course, it is the 
conjunction of these two steps that makes the following developments possible.

Ordinarily we can let $a_{0}\rightarrow 0$ in (\ref{eq3.14}), so that
\begin{equation}
S_{E}=\int d^{D}x\,\left[ \frac{1}{2m_{0}}\partial _{M}\Psi ^{\dagger }\partial
_{M}\Psi -\mu_{0} \,\Psi ^{\dagger }\Psi +\frac{1}{2}b\left( \Psi ^{\dagger
}\Psi \right) ^{2}\right] \; . \label{eq3.18}
\end{equation}
However, the higher-derivative term in (\ref{eq3.14}) is relevant in the
internal space defined below, and a finite $a_{0}$ also automatically provides an ultimate ultraviolet cutoff. 

\section{\label{sec:sec4}Origin of fermion action and (3+1) dimensional spacetime}

The present theory is based on
(1) statistical counting (which ultimately produced the results of the preceding two sections) and (2) the
geography (or specific features) of our universe, to which we now turn. 

The most central assumption is that 
\begin{eqnarray}
\Psi _{b} = \Psi _{0}^{\prime } +\Psi _{b}^{\prime } 
\label{eq4.2} 
\end{eqnarray}
where $\Psi _{0}^{\prime } $ contains the order parameter $\Psi _{0}$ for a primordial bosonic 
condensate which forms in the very early universe, and 
$\Psi _{b}^{\prime }$ contains all the other bosonic fields. I.e., in $\Psi _{0}^{\prime }$ only one set of components is nonzero and equal to $\Psi _{0}$, and this set of components is zero in $\Psi _{b}^{\prime } $.
The treatment of Appendix \ref{sec:appA} implies that 
\begin{eqnarray}
\Psi _{0}^{\prime \, \dagger } \Psi _{b}^{\prime } = 0
\end{eqnarray}
everywhere. (The fields in other representations do not overlap the representation containing  $\Psi _{0}$. Fields in the same representation are orthogonal according to (\ref{eq12.6}) and the comments above (\ref{eq12.2}) and
(\ref{eq12.4}).)
The action can then be written as 
\begin{eqnarray}
S_{E} &=&S_{cond}+S_{b}+S_{f}+S_{int}  \label{eq4.6} \\
S_{cond} &=&\int d^{D}x\,\left[ \frac{1}{2m_{0}}\partial _{M}
\Psi _{0}^{\dagger }\partial _{M}\Psi _{0} 
-\mu_{0} \Psi _{0}^{\dagger }\Psi _{0} +\frac{1}{2}
b\left( \Psi _{0}^{\dagger }\Psi _{0}\right) ^{2}\right]  
\label{eq4.7} \\
S_{b} &=&\int d^{D}x\,\left[ \frac{1}{2m_{0}}\partial _{M}\Psi _{b}^{\prime
\,\dagger }\partial _{M}\Psi _{b}^{\prime }\,\,+\left( V_{0}-\mu_{0} \right)
\,\Psi _{b}^{\prime \,\dagger }\Psi _{b}^{\prime }\,+\frac{1}{2}b\left( \Psi
_{b}^{\prime \,\dagger }\Psi _{b}^{\prime }\right) ^{2}\right]  
\label{eq4.8} \\
S_{f} &=&\int d^{D}x\,\left[ \frac{1}{2m_{0}}\partial _{M}\Psi _{f}^{
\dagger }\partial _{M}\Psi _{f}\,\,+\left( V_{0}-\mu_{0} \right)
\,\Psi _{f}^{\dagger }\Psi _{f}\,+\frac{1}{2}b\left( \Psi
_{f}^{\dagger }\Psi _{f}\right) ^{2}\right]  
\label{eq4.9} \\
S_{int} &=&\int d^{D}x\, b 
\left( \Psi _{f}^{\dagger }\Psi _{f} \right) 
\left( \Psi _{b}^{\prime \,\dagger }\Psi _{b}^{\prime } \right)
\label{eq4.9a} \\
V_{0} &=& b \, \Psi _{0}^{\dagger }\Psi _{0} \; . \label{eq4.10}
\end{eqnarray}
In most of the following, the last term will be neglected in 
(\ref{eq4.8}) and (\ref{eq4.9}); we are 
then considering the theory prior to formation of further
condensates beyond the primordial $\Psi _{0}$. 

For a static condensate we could write 
$\Psi _{0}=n_{0}^{1/2}\eta _{0}$, 
where $\eta _{0}$ is constant, $\eta _{0}^{\dagger }\eta _{0}=1 $, 
and $n_{0}=\Psi _{0}^{\dagger }\Psi _{0}$ is the condensate
density. This picture is too simplistic, however, since the order parameter
can exhibit rotations that are analogous to the rotations in 
the complex plane of the order parameter $\psi _{s}=e^{i\theta
_{s}}n_{s}^{1/2}$ for an ordinary superfluid: 
\begin{equation}
\Psi _{0}\left( x\right) =U_{0}\left( x\right) \,n_{0}\left( x\right)
^{1/2}\eta _{0}\quad ,\quad U_{0}^{\dagger }U_{0}=1 \; .  \label{eq4.11a}
\end{equation}
After an integration by parts in (\ref{eq4.7}) (with boundary terms usually
neglected in the present paper), extremalization of the action gives 
the classical equation of motion for the order parameter:
\begin{equation}
-\frac{1}{2m_{0}}\partial _{M}\partial _{M}\Psi _{0}+\left( V_{0}-\mu_{0} 
\right) \Psi _{0}=0  \; .\label{eq4.11b}
\end{equation}

Now an important nuance, which requires some references to the treatment below: Because the primordial condensate density is extremely large, (\ref{eq4.11b}) is assumed to always hold
(at normal energies). However, consistent with this constraint, the ``phase" and "superfluid velocities" of (\ref{eq4.17}) (\ref{eq4.40}), (\ref{eq8.3z} ), and (\ref{eq5.23}) are allowed to vary within the path integral. 
This means that the gauge potentials $A_{\mu}^i$ and metric tensor $g_{\mu \nu}$ are quantized. I.e., the path integral over the original field $\Psi_0$ is replaced by path integrals over $A_{\mu}^i$ and $g_{\mu \nu}$. Notice that the invariance of (\ref{eq4.7}) under a gauge transformation implies gauge-fixing within the path integrals for both  $A_{\mu}^i$ and $g_{\mu \nu}$, as usual. In the present treatment, we additionally require that (\ref{torsion-free}) always hold, so that the path integral is restricted to torsion-free spacetime geometries.

In specifying the geography of our universe, it will be assumed that
$\Psi _{0}$ can be written as the product of a $2$-component external 
order parameter $\Psi _{ext}$, which is a function of $4$ external coordinates 
$x^{\mu }$, and an internal order parameter $\Psi _{int}$, which is 
primarily a function of $D-4$ internal coordinates $x^{m}$, but which also 
varies with $x^{\mu }$:
\begin{eqnarray}
\Psi _{0} &=&\Psi _{ext}\left( x^{\mu }\right) \,\Psi _{int}\left(
x^{m},x^{\mu }\right)  \label{eq4.12}  \\
\Psi _{ext}\left( x^{\mu }\right) &=&U_{ext}\left( x^{\mu }\right)
\,n_{ext}\left( x^{\mu }\right) ^{1/2}\eta _{ext}\quad ,\quad \mu =0,1,2,3 
\label{eq4.13}  \\
\Psi _{int}\left( x^{m},x^{\mu }\right) &=&U_{int}\left( x^{m},x^{\mu
}\right) \,n_{int}\left( x^{m},x^{\mu }\right) ^{1/2}\eta _{int}\quad ,\quad
m=4,...,D-1  \label{eq4.14}
\end{eqnarray}
where again $\eta _{ext}$ and $\eta _{int}$ are constant, and 
$\eta _{ext}^{\dagger }\eta _{ext}=\eta _{int}^{\dagger }\eta _{int}=1 $.
Here, according
to a standard notation, $x^{\mu }$ actually represents the set of $x^{\mu }$,
and $x^{m}$ the set of $x^{m}$.

Let us define external and internal ``superfluid velocities'' by 
\begin{eqnarray}
m_{0}v_{\mu } = -iU_{ext}^{-1} \partial  _{\mu } U_{ext} \quad , \quad 
m_{0}v_{m}    = -iU_{int}^{-1}\partial _{m}   U_{int} \;  . \label{eq4.17}
\end{eqnarray}
The fact that $U_{ext}^{\dagger } U_{ext} = 1$ implies that 
$\left( \partial _{\mu } U_{ext}^{\dagger } \right) U_{ext}
= - U_{ext}^{\dagger } \left( \partial_{\mu }U_{ext} \right) $ with 
$U_{ext}^{\dagger }=U_{ext}^{-1} $, or $m_{0}v_{\mu } = 
i \left( \partial _{\mu }U_{ext}^{\dagger } \right) U_{ext} $, 
so that 
\begin{equation}
v_{\mu }^{\dagger }=v_{\mu } \; . \label{eq4.19}
\end{equation}

For simplicity, let us first consider the case 
\begin{equation}
\partial _{\mu }U_{int}=0  \label{eq4.20}
\end{equation}
for which there are separate external and internal equations of
motion: 
\begin{equation}
\left( -\frac{1}{2m_{0}}\partial_{\mu }\partial_{\mu }-\mu _{ext}\right) \Psi
_{ext}=0  \quad , \quad
\left( -\frac{1}{2m_{0}}\partial _{m}\partial _{m}-\mu _{int}+V_{0}\right) \Psi
_{int}=0  \label{eq4.21}
\end{equation}
with 
\begin{equation}
\mu _{int}=\mu_{0} -\mu _{ext} \; . 
\label{eq4.21a}
\end{equation}
The quantities $\mu _{int}$ and $V_{0}$ have a 
relatively slow parametric dependence on $x^{\mu }$.

When (\ref{eq4.13}) and (\ref{eq4.17}) are used in 
(\ref{eq4.21}), we obtain 
\begin{equation}
\eta _{ext}^{\dagger }\,n_{ext}^{1/2}\left[ \left( \frac{1}{2}
m_{0}v_{\mu }v_{\mu }-\frac{1}{2m_{0}}\partial _{\mu }\partial _{\mu }-\mu
_{ext}\right) -i\left( \frac{1}{2}\partial _{\mu }v_{\mu }+v_{\mu }\partial
_{\mu }\right) \right] n_{ext}^{1/2}\eta _{ext}=0  \label{eq4.22}
\end{equation}
and its Hermitian conjugate 
\begin{equation}
\eta _{ext}^{\dagger }\,n_{ext}^{1/2}\left[ \left( \frac{1}{2}
m_{0}v_{\mu }v_{\mu }-\frac{1}{2m_{0}}\partial _{\mu }\partial _{\mu }-\mu
_{ext}\right) +i\left( \frac{1}{2}\partial _{\mu }v_{\mu }+v_{\mu }\partial
_{\mu }\right) \right] n_{ext}^{1/2}\eta _{ext}=0 \; . \label{eq4.23}
\end{equation}
Subtraction gives the equation of continuity 
\begin{equation}
\partial_{\mu }\,j_{\mu }^{ext}=0~\quad ,\quad j_{\mu }^{ext}=n_{ext}\,\eta
_{ext}^{\dagger }\,v_{\mu }\eta _{ext}  \label{eq4.24}
\end{equation}
and addition gives the Bernoulli equation 
\begin{equation}
\frac{1}{2}m_{0}\bar{v}_{ext}^{2}+P_{ext}=\mu _{ext}  \label{eq4.25}
\end{equation}
where 
\begin{eqnarray}
\bar{v}_{ext}^{2}=\eta _{ext}^{\dagger }\,v_{\mu }v_{\mu }\,\eta _{ext}
\quad , \quad 
P_{ext}=-\frac{1}{2m_{0}}n_{ext}^{-1/2}\partial _{\mu }\partial _{\mu
}n_{ext}^{1/2}\; . \label{eq4.27}
\end{eqnarray}

Since the order parameter $\Psi _{ext}$ in external spacetime has $2$
components, its ``superfluid velocity'' 
$v_{\mu }$ can be written in terms of the identity matrix $\sigma ^{0}$ and
Pauli matrices $\sigma ^{a}$ : 
\begin{equation}
v_{\mu }=v_{\alpha \mu }\sigma ^{\alpha } \quad , \quad \alpha = 0,1,2,3 \; . \label{eq4.28}
\end{equation}
Let us now transform to a coordinate system in which 
\begin{equation}
v_{0 k}=v_{0}^{k}=v_{a 0}=v_{a}^{0}=0  \quad , \quad k = 1,2,3 \quad \mathrm{and} \quad a=1,2,3  \label{eq4.29}
\end{equation}
(with the volume element held constant) so that (\ref{eq4.25}) becomes
\begin{eqnarray}
{\frac{1}{2}}m_{0} v_{\alpha }^{\mu  } v_{\alpha \mu }+P_{ext}=\mu _{ext}
\; .
\label{eq4.30}
\end{eqnarray}
To avoid notational complexity we will still use $x^{\mu}$ to label the new coordinates. At this point a physically meaningful metric tensor has not yet been introduced, and the notation in (\ref{eq4.30}) merely indicates that  $v_{\alpha }^{\mu  } v_{\alpha \mu }$ is to be kept constant under coordinate transformations. The same is true for the other quantities in (\ref{eq4.37}). Later in the development, where Lorentz invariance and results like those in (\ref{eq8.3}) hold (i.e. at energies far below the Planck energy $\ell_P^{-1}$), the usual conventions are used for raising and lowering indices.

The transformation to (\ref{eq4.29})  is trivial in, e.g., a cosmological model in which the Big Bang is at the origin of the new coordinates, with the $U(1)$ phase of $\Psi _{0} $ varying only with respect to the radial coordinate $x^{0}$, and the ``$SU(2)$ phase'' involving the Pauli matrices varying within successive $3$-spheres with coordinates $x^{k}$, so that $v_{a k}$ has a vortex-like (or instanton-like) configuration. More generally, the time coordinate $x^{0}$ is distinguished from the spatial coordinates $x^{k}$ in (\ref{eq4.29}) because it is the direction of $U(1)$ rather than $SU(2)$ rotations of the order parameter.

As $v_{\alpha }^{\mu  } v_{\alpha \mu } $ varies, $\mu _{ext}$ varies in 
response, with $\mu _{int} $ determined by (\ref{eq4.21a}).

Now expand $\Psi _{b}^{\prime }$ in terms of a complete set of basis
functions $\widetilde{\psi }_{int}^{r}$ in the internal space:
\begin{equation}
\Psi _{b}^{\prime }\left( x^{\mu },x^{m}\right) =\widetilde{\psi }
_{b}^{r}\left( x^{\mu }\right) \widetilde{\psi }_{int}^{r}
\left( x^{m}\right)  \label{eq4.31}
\end{equation}
with 
\begin{eqnarray}
\left( -\frac{1}{2m_{0}}\partial _{m}\partial _{m}-\mu _{int}+V_{0}\right) 
\widetilde{\psi }_{int}^{r}\left( x^{m}\right) &=&\varepsilon _{r}
\widetilde{\psi }_{int}^{r}\left( x^{m}\right) \label{eq4.32}  \\
\int d^{D-4}x\,\widetilde{\psi }_{int}^{r\dag }\left( x^{m}\right) 
\widetilde{\psi }_{int}^{r^{\prime }}\left( x^{m}\right) &=&\delta
_{rr^{\prime }}  \label{eq4.33}
\end{eqnarray}
and with the usual summation over repeated indices in (\ref{eq4.31}). For reasons that will
become fully apparent below, but which are already suggested by the form of
the order parameter, each $\widetilde{\psi }_{b}^{r}\left( x^{\mu }\right) $
has two components. As usual, only the zero ($\varepsilon_{r}=0$) 
modes will be kept. (To simplify the presentation, the higher-derivative 
terms are not explicitly shown in 
the present section; they will be restored in the next section.) When 
(\ref{eq4.31})-(\ref{eq4.33}) are then used in (\ref{eq4.8}) (with the
last term neglected), the result is 
\begin{equation}
S_{b}=\int d^{4}x\,\widetilde{\psi }_{b}^{\dagger }
\left( -\frac{1}{2m_{0}}\partial ^{\mu }\partial _{\mu }-\mu _{ext}\right) 
\widetilde{\psi }_{b}  \label{eq4.34}
\end{equation}
where $\widetilde{\psi }_{b}$ is the vector with components $\widetilde{\psi 
}_{b}^{r}$.

Let $\widetilde{\psi }_{b}$ be written in the form 
\begin{equation}
\widetilde{\psi }_{b}\left( x^{\mu }\right) =U_{ext}\left( x^{\mu }\right)
\psi _{b}\left( x^{\mu }\right)  \label{eq4.35}
\end{equation}
or equivalently
\begin{eqnarray}
\widetilde{\psi }_{b}^{r}\left( x^{\mu }\right) =U_{ext}\left( x^{\mu
}\right) \psi _{b}^{r}\left( x^{\mu }\right) \; . \label{eq4.36}
\end{eqnarray}
Here $\psi _{b}$ has a simple interpretation: It is the field seen by an
observer in the frame of reference that is moving with the condensate. In
the present theory, a (very high density) condensate $\Psi _{0}$ forms in the
very early universe, and the other bosonic and fermionic fields 
are subsequently born into it. It is therefore natural to define the
fields $\psi _{b}^{r}$ in the condensate's frame of reference. 

Equation (\ref{eq4.35}) is, in fact, exactly analogous to 
rewriting the wavefunction of 
a particle in an ordinary superfluid moving with velocity $v_{s}$: 
$\widetilde{\psi }_{par}\left( x\right) = \exp \left( imv_{s}x\right) \psi
_{par}\left( x\right) .$ Here $\psi _{par}$ is the wavefunction in the 
superfluid's frame of reference.

When (\ref{eq4.35}) is substituted into (\ref{eq4.34}), the result is 
\begin{equation}
S_{b}=\int d^{4}x~\psi _{b}^{\dagger }\left[
\left( \frac{1}{2}m_{0}v^{\mu }v_{\mu }-\frac{1}{2m_{0}}\partial ^{\mu }\partial _{\mu }-\mu _{ext}\right) -i\left( \frac{1}{2}\partial _{\mu }v^{\mu}+v^{\mu }\partial _{\mu }\right) \right] \psi _{b}\; . \label{eq4.37}
\end{equation}
In the following it will be assumed that 
\begin{equation}
\partial _{\mu }v^{\mu} = 0 
\end{equation}
since, after the definition (\ref{eq4.40}) and the introduction of a covariant derivative, the more general version of this equation (which permits general coordinate transformations and local Lorentz transformations) follows from (\ref{torsion-free}). 
If the condensate density $n_{ext}$ is slowly varying, so that $P_{ext}$ 
can be neglected, 
(\ref{eq4.29}) and (\ref{eq4.30}) then lead to the simplification 
\begin{equation}
S_{b}=-\int d^{4}x~\psi _{b}^{\dagger }\left( 
\frac{1}{2m_{0}}\partial ^{\mu }\partial _{\mu }+iv_{\alpha }^{\mu  }\sigma
^{\alpha }\partial _{\mu }\right) \psi _{b} \; . \label{eq4.38}
\end{equation}
In most of the remainder of the paper it will be assumed that the first term in parentheses is negligible compared to the second for 
states $\psi $ with energies $\sim $ 1 TeV or less (as would be the case if we had, e.g., $m_{0}=a_{0}^{-1} \gtrsim 10^{15}$
TeV and $v_{\alpha }^{\mu }\sim 1$ for $\mu=\alpha$), so that (\ref{eq4.38}) reduces to just
\begin{eqnarray}
S_{b} &=& \int d^{4}x \, \psi _{b}^{\dagger }ie_{\alpha
}^{\mu }\sigma ^{\alpha }\partial _{\mu }\psi _{b}  \label{eq4.39} \\
e_{\alpha }^{\mu } &=& - v_{\alpha }^{\mu } \; . \label{eq4.40}
\end{eqnarray}
With this choice all fields are initially right-handed. With the choice 
$e_{\alpha }^{0 } = - v_{\alpha }^{0 }$, $e_{\alpha }^{k } = v_{\alpha }^{k }$ 
all fields would be initially left-handed, as they are for fermions in conventional 
$SU(5)$ and $SO(10)$ grand-unified theories~\cite{cheng,kounnas}. 
It is trivial to change from one convention to the other, of course.

A central feature of the present theory is that the \textit{primitive} bosonic fields $\psi _{b}$ are 2-component spinors, but, as mentioned in Section \ref{sec:sec1a}, the final \textit{physical} fields derived from these are the 1-component scalar boson fields of Section \ref{sec:DM} -- the Higgs and higgson fields of an extended Higgs sector. 

To permit local Lorentz transformations as well as general coordinate transformations, let us rewrite (\ref{eq4.39}) as
\begin{eqnarray}
S_{b} = \int d^{4}x \,  \overline{\mathcal {L}}_b \quad , \quad  \overline{\mathcal {L}}_b  = \psi _{b}^{\dagger }ie_{\alpha}^{\mu }\sigma ^{\alpha }\nabla _{\mu }^{R}\psi _{b}  
\label{invariant} 
\end{eqnarray}
where $\nabla _{\mu }^{R}$ gives the standard (curved spacetime) covariant derivative 
for a right-handed Weyl field. The full covariant derivative for a Dirac field, with right-handed and left-handed  Weyl components, is given by~\cite{Parker,GSW}
\begin{eqnarray}
\nabla_{\mu }&=&\partial _{\mu }+ i \omega _{\mu }^{\alpha \beta } \Sigma_{\alpha \beta } \; .
\end{eqnarray}
(The notation and conventions in this context usually follow those most common in the gravitational and string theory communities, as in 
Ref.~\cite{GSW,misner,weinberg-grav,wald,carroll}, rather than the particle physics and field theory communities, as in Ref.~\cite{Parker}.  Mainly, the metric tensor convention is (-+++) throughout this paper. However, the Dirac gamma matrices are defined as in most field theory textbooks~\cite{peskin,schwartz}.)
The origin and meaning of curvature involving the $ \omega _{\mu }^{\alpha \beta } $
will be considered below. 
For a vector field, the usual covariant derivative provides invariance under a coordinate transformation. The vierbein introduced above has both coordinate and tangent-space indices, so~\cite{GSW,Parker}
\begin{equation}
\nabla _{\mu }e^{\alpha}_{\nu} = \partial_{\mu} e^{\alpha}_{\nu}  - \Gamma^{\rho}_{\mu \nu} e^{\alpha}_{\rho}  +\omega_{\mu \; \;  \beta} ^{\; \, \alpha} e^{\beta}_{\nu} 
\end{equation}
and in the usual minimal case of a torsion-free universe
\begin{equation}
\nabla _{\mu }e^{\alpha}_{\nu} = 0 \; .
\label{torsion-free}
\end{equation}

The above arguments also hold for fermions, with (in the initial notation)
\begin{equation}
S_{f}=\int d^{D}x\,\left( -\frac{1}{2m_{0}}\Psi _{f}^{\dagger }\partial
^{M}\partial _{M}\Psi _{f} 
-\mu_{0} \Psi _{f}^{\dagger }\Psi _{f}+V_{0}\Psi
_{f}^{\dagger }\Psi _{f} \right)  \label{eq4.41}
\end{equation}
\begin{equation}
\Psi _{f}\left( x^{\mu },x^{m}\right) =
\widetilde{\psi }_{f}^{r}\left(
x^{\mu }\right) \widetilde{\psi }_{int}^{r}\left( x^{m}\right)
\label{eq4.42}
\end{equation}
leading to the final result
\begin{eqnarray}
S_{f} = \int d^{4}x \,  \overline{\mathcal {L}}_f \quad , \quad  \overline{\mathcal {L}}_f  = \psi _{f}^{\dagger }ie_{\alpha}^{\mu }\sigma ^{\alpha }\nabla _{\mu }^{R}\psi _{f} \label{eq4.43} \; .
\end{eqnarray}

The present theory thus yields the basic form of the standard Lagrangian for Weyl fermions, with
the gravitational vierbein $e_{\alpha }^{\mu }$ interpreted as
essentially a ``superfluid velocity'' associated with the
condensate $\Psi _{0}$. The path integral still has a Euclidean form, 
the action for bosons is also not yet in standard form, and there is no factor of 
\begin{equation}
e=\left\vert \det \, e_{\mu }^{\alpha }\right\vert =\left( -\det \, g_{\mu \nu} \right) ^{1/2}
\label{eq8.4b}
\end{equation}
multiplying $ \overline{\mathcal {L}}_f$ or $\overline{\mathcal {L}}_b$, but we will return
to these points below.

Consistency with (\ref{eq4.43}) is obtained if the metric tensor is related to the vierbein through
\begin{eqnarray}
g^{\mu \nu }=\eta ^{\alpha \beta }e_{\alpha }^{\mu }e_{\beta }^{\nu } \quad , \quad \eta ^{\alpha \beta } = diag \left(-1, 1, 1, 1 \right) 
\; .
\label{eq8.3z} 
\end{eqnarray}
In the present picture, spacetime is automatically 4-dimensional with one time coordinate, because there are 3 Pauli matrices and one $2 \times 2$ unit matrix.

\section{\label{sec:sec5}Gauge fields}

In this section and those following, down to and including (\ref{eq7.58x}), we will temporarily ignore the spin connection and write $\partial _{\mu }$ instead of $\nabla _{\mu } $, to avoid irrelevant complications in notation.
Let us now relax assumption (\ref{eq4.20}) and allow $U_{int}$ to vary 
with the external coordinates $x^{\mu }$.
The more general version of (\ref{eq4.11b}) is satisfied if (\ref{eq4.21}) is generalized to
\begin{equation}
\left( -\frac{1}{2m_{0}}\partial ^{\mu }\partial _{\mu }-\mu _{ext}\right) \Psi
_{ext}\left( x^{\mu }\right) \Psi _{int}\left( x^{m},x^{\mu }\right) =0
\label{eq5.2}
\end{equation}
with $\Psi _{int}$ required to satisfy the internal equation of motion 
(at each $x^{\mu }$)
\begin{equation}
\left[ \sum_{m} \frac{1}{2m_{0}}\left( -\frac{\partial ^{2}}{\partial 
\left( x^{m}\right) ^{2}}+
\frac{a_{0}^{2}}{16}\frac{\partial ^{4}}{\partial \left( x^{m}\right) ^{4}}
\right) 
+V_{0}\left( x^{m}\right) -\mu _{int}\right] \Psi _{int}\left( x^{m},x^{\mu
}\right) =0 \; .  \label{eq5.3}
\end{equation}
The higher-derivative 
term of (\ref{eq3.14}) has been retained and two integrations by 
parts have been performed. (In order to simplify the notation, we do not
explicitly show the weak parametric dependence of $\mu _{int}$, $V_{0}$, and 
$n_{int}$ on $x^{\mu }$.) This is a nonlinear equation because (at each 
$x^{\mu }$) $V_{0}\left( x^{m}\right) $ 
is mainly determined by 
$n_{int} = \Psi _{int}^{\dag } \Psi _{int} $.

The internal basis functions satisfy the more general version of 
(\ref{eq4.32}) with $\varepsilon _{r}=0$: 
\begin{equation}
\left[ \sum_{m} \frac{1}{2m_{0}} \left( -\frac{\partial ^{2}}{\partial 
\left( x^{m}\right) ^{2}}+
\frac{a_{0}^{2}}{16}\frac{\partial ^{4}}{\partial \left( x^{m}\right) ^{4}}
\right) 
+V_{0}\left( x^{m}\right) -\mu _{int}\right] \widetilde{\psi }
_{int}^{r}\left( x^{m},x^{\mu }\right) =0 \; .  \label{eq5.4}
\end{equation}
This is a linear equation because $V_{0}\left( x^{m}\right) $ is now
regarded as a known function. 

The full path integral 
involving (\ref{eq3.14}) contains all configurations of the
fields, including those with nontrivial topologies. In the present
theory, the geography of our universe
includes a topological defect in the ($D-4$)-dimensional internal space 
which is analogous to a vortex. 
(See Appendix \ref{sec:appA}.) The standard features of 
four-dimensional physics arise from the presence of this internal 
topological defect. For example, it compels the initial gauge symmetry 
to be $SO(D-4)$.

The behavior of the condensate and basis functions in the internal
space is discussed in Appendices \ref{sec:appA} and \ref{sec:appB}.
In (\ref{eq12.9}), the parameters $\overline{\phi } _{i}$ specify a rotation 
of $\Psi _{int}\left( x^{m},x^{\mu }\right)$ as the external coordinates 
$x^{\mu}$ are varied, and according to (\ref{eq12.9a}) the 
$\overline{J}_{i}$ satisfy the $SO(D-4)$ algebra
\begin{eqnarray}
\overline{J}_{i}\overline{J}_{j}-\overline{J}_{j}\overline{J}_{i}
=ic_{ij}^{k}\overline{J}_{k} \; .
\label{eq5.9}
\end{eqnarray}
For simplicity of notation, let 
\begin{equation}
\left\langle r\,|Q|r^{\prime }\right\rangle =\int d^{D-4}x\,\widetilde{\psi }
_{int}^{r\dagger }Q\,\widetilde{\psi }_{int}^{r^{\prime}}\quad \mbox{with}
\quad \left\langle r\,|r^{\prime }\right\rangle =\delta _{rr^{\prime }}
\label{eq5.16}
\end{equation}
for any operator $Q$, and in particular let 
\begin{equation}
t_{i}^{rr^{\prime }}=\left\langle r\,|\,\overline{J}_{i}|r^{\prime }
\right\rangle
\label{eq5.17}
\end{equation}
with the matrices $t_{i}^{rr^{\prime }}$ (which are constant according
to (\ref{eq12.9b})) inheriting the $SO(D-4)$ algebra: 
\begin{eqnarray}
\left( t_{i}t_{j}-t_{j}t_{i} \right) ^{r r^{\prime}}
&=&\sum_{r^{\prime \prime }}\left\langle
r\,|\,\overline{J}_{i}|r^{\prime \prime }\right\rangle 
\left\langle r^{\prime \prime }|\,\overline{J}_{j}|r^{\prime }\right\rangle 
-\sum_{r^{\prime \prime }}\left\langle r\,|\,\overline{J}_{j}|
r^{\prime \prime }\right\rangle \left\langle r^{\prime \prime }|
\,\overline{J}_{i}|r^{\prime }\right\rangle  \label{eq5.18} \\
&=&\left\langle r\,|\,\overline{J}_{i}\,\overline{J}_{j}|\,
r^{\prime} \,\right\rangle -\left\langle r\,|\,
\overline{J}_{j}\,\overline{J}_{i}|\,r^{\prime} \,\right\rangle  
\label{eq5.19}  \\
&=&ic_{ij}^{k}t_{k} ^{r r^{\prime}} \; . \label{eq5.20}
\end{eqnarray}
The $t_{i}$ are the generators in the $N_{g}$-dimensional reducible
representation determined by the physically significant solutions to 
(\ref{eq5.4}), which spans all the irreducible (physical) gauge 
representations.

When $x^{\mu }\rightarrow x^{\mu }+\delta x^{\mu }$, 
$\Psi _{int} $ and $\widetilde{\psi} _{int}^{r} $
rotate together, and (\ref{eq12.9}) implies that
\begin{eqnarray}
\partial _{\mu }\widetilde{\psi} _{int}^{r}\left( x^{m},x^{\mu }\right) &=&
\frac{\partial \overline{\phi } _{i}}{\partial x^{\mu }} \frac{\partial }
{\partial \overline{\phi } _{i}} \widetilde{\psi} _{int}^{r} 
\left( x^{m},x^{\mu }\right) 
\label{eq5.21}  \\
&=&-i\,A_{\mu }^{i}\overline{J}_{i}\, \widetilde{\psi} _{int}^{r} 
\left( x^{m},x^{\mu }\right) 
\label{eq5.22}
\end{eqnarray}
where
\begin{eqnarray}
A_{\mu }^{i}= \frac{\partial \overline{\phi } _{i}}{\partial x^{\mu }} \; . 
\label{eq5.23}
\end{eqnarray}
The $A_{\mu }^{i}$ will be interpreted below as gauge potentials. In other
words, the gauge potentials are simply the rates at which the internal 
order parameter $\Psi _{int}\left( x^{m},x^{\mu }\right) $ is rotating as a 
function of the external coordinates $x^{\mu }$.

Let us return to the fermionic action (\ref{eq4.41}). 
If (\ref{eq4.42}) is written in the more general form
\begin{eqnarray}
\hspace{-0.5cm}
\Psi _{f}\left( x^{\mu },x^{m}\right) 
=\widetilde{\psi }_{f}^{r}\left(
x^{\mu }\right) \widetilde{\psi }_{int}^{r}\left( x^{m}, x^{\mu }\right)
=U_{ext}\left( x^{\mu }\right) \psi _{f}^{r}\left( x^{\mu }\right)
\widetilde{\psi }_{int}^{r} \left( x^{m}, x^{\mu } \right) 
\label{eq5.25}
\end{eqnarray}
we have 
\begin{equation}
\partial _{\mu }\Psi _{f}
=U_{ext}\left( x^{\mu }\right) \left( \partial _{\mu }^{\, \prime}
-im_{0}e_{\alpha \mu }\sigma ^{\alpha }-i\,A_{\mu }^{i}\overline{J}_{i}
\,\right) \psi _{f}^{r}\,\widetilde{\psi }_{int}^{r}   \label{eq5.26}
\end{equation}
where the prime indicates that $\partial _{\mu }^{\, \prime}$ 
does not operate on $\widetilde{\psi }_{int}^{r}$, and 
\begin{eqnarray}
\hspace{-1cm}
&& \int d^{D-4}x \, \Psi _{f}^{\dagger }\,\partial ^{\mu
}\partial _{\mu }\,\Psi _{f} \nonumber \\
\hspace{-1cm}
&&=\int d^{D-4}x\,\widetilde{\psi }_{int}^{r\dag }\psi
_{f}^{r\dagger }\left( \partial ^{\, \prime\, \mu}
-im_{0}e_{\alpha }^{\mu }\sigma ^{\alpha }-i\,A^{\mu i}\overline{J}_{i}
\,\right) \left( \partial _{\mu }^{\, \prime} -im_{0}
e_{\alpha ^{\prime } \mu }\sigma ^{\alpha ^{\prime }}
-i\,A_{\mu }^{i^{\prime }}\overline{J}_{i^{\prime }}\,\right) 
\psi _{f}^{r^{\prime }}\widetilde{\psi }_{int}^{r^{\prime }}  \nonumber
\\
\hspace{-1cm}
&&=\psi _{f}^{r\dagger }\,\langle r|\left( \partial ^{\, \prime\, \mu}
-im_{0}e_{\alpha }^{\mu }\sigma ^{\alpha }-i\,A^{\mu i}\overline{J}_{i}
\,\right) \,\sum_{r^{\prime \prime }}|r^{\prime \prime }\rangle 
\langle r^{\prime \prime }|\,\left( \partial _{\mu }^{\, \prime} -im_{0}
e_{\alpha ^{\prime } \mu }\sigma ^{\alpha ^{\prime }}
-i\,A_{\mu }^{i^{\prime }}\overline{J}_{i^{\prime }}\,\right) 
|r^{\prime }\rangle \,\psi _{f}^{r^{\prime }} \nonumber   \\
\hspace{-1cm}
&&=\psi _{f}^{r\dagger }\,\left[ \delta _{rr^{\prime \prime }}
\left( \partial ^{\mu }-im_{0}e_{\alpha }^{\mu }\sigma ^{\alpha }\right)
-iA^{\mu i}t_{i}^{rr^{\prime \prime }}\right] \,\left[ \delta _{r^{\prime
\prime }r^{\prime }}\left( \partial _{\mu }-im_{0}e_{\alpha ^{\prime } \mu }
\sigma ^{\alpha ^{\prime }}\right) -iA_{\mu }^{i^{\prime }}
t_{i^{\prime }}^{r^{\prime \prime }r^{\prime }}\right] \,
\psi _{f}^{r^{\prime }} \nonumber   \\
\hspace{-1cm}
&&=\psi _{f}^{\dagger }\,\left[ \left( \partial ^{\mu
}-iA^{\mu i}t_{i}\right) -im_{0}e_{\alpha }^{\mu }\sigma ^{\alpha }\right] \,
\left[ \left( \partial _{\mu }-iA_{\mu }^{i^{\prime }}t_{i^{\prime }}\right)
-im_{0}e_{\alpha ^{\prime } \mu }\sigma ^{\alpha ^{\prime }}\right] \,\psi_{f} 
\; . \nonumber   \label{eq5.27}
\end{eqnarray}
Then (\ref{eq4.41}) becomes 
\begin{equation}
\hspace{-1cm} S_{f}=\int d^{4}x\,\psi _{f} ^{\dagger }\left(-\frac{1}{2m_{0}}
D^{\mu }D_{\mu }+\frac{1}{2}ie_{\alpha }^{\mu }\sigma ^{\alpha }D_{\mu }+
\frac{1}{2}D^{\mu }ie_{\alpha \mu }\sigma ^{\alpha }+\frac{1}{2}
m_{0}e_{\alpha }^{\mu }\sigma^{\alpha}e_{\alpha^{\prime} \mu }
\sigma^{\alpha^{\prime}} -\mu _{ext}\right) \psi _{f}
\nonumber
\end{equation}
where 
\begin{equation}
D_{\mu }=\partial _{\mu }-iA_{\mu }^{i}t_{i} \; . \label{eq5.29}
\end{equation}

With (\ref{eq4.29}) and the approximations above (\ref{eq4.38}), 
(\ref{eq4.30}) implies that 
\begin{equation}
S_{f}=\int d^{4}x\,\overline{\mathcal{L}}_f \quad , \quad \overline{\mathcal{L}}_f =\psi _{f} ^{\dagger }\left( -\frac{1}{2m_{0}}D^{\mu }
D_{\mu}+ie_{\alpha }^{\mu }\sigma ^{\alpha }D_{\mu }\right) \psi _{f} \; .
\label{eq5.30}
\end{equation}
This is the generalization of (\ref{eq4.43}) 
when the internal order parameter\
is permitted to vary as a function of the external coordinates $x^{\mu }$.
Again, for momenta and gauge potentials that are small compared to 
$m_{0}e_{\alpha }^{\mu }$ with $\mu = \alpha$, 
the first term may be neglected. Furthermore, the entire treatment above
can be repeated for the bosonic action, finally giving
\begin{align}S_{f}=\int d^{4}x\,\overline{\mathcal{L}}_f \quad , \quad 
\overline{\mathcal{L}}_f  &= \psi _{f} ^{\dagger }ie_{\alpha }^{\mu }\sigma
^{\alpha }D_{\mu }\psi _{f} \label{eq5.31} \\
S_{b}=\int d^{4}x\,\overline{\mathcal{L}}_b \quad , \quad 
\overline{\mathcal{L}}_b &= \psi _{b} ^{\dagger }ie_{\alpha }^{\mu }\sigma
^{\alpha }D_{\mu }\psi _{b}  \; . \label{eq5.32}
\end{align}

\section{\label{sec:sec6}Transformation to Lorentzian path integral: fermions}

All of the foregoing is within a Euclidean picture, but we will now show
that, in the case of fermions, there is a relatively trivial transformation
to the more familiar Lorentzian description. A key point is that the
low-energy \textit{operator} $ie_{\alpha }^{\mu }\sigma ^{\alpha }D_{\mu }$
in $S_{f}$ is automatically in the correct Lorentzian form, even though the
initial \textit{path integral} is in Euclidean form. It is this fact which
permits the following transformation to a Lorentzian path integral. Within
the present theory, neither the fields nor the operators (nor the meaning of
the time coordinate) need to be modified in performing this transformation.

The operator within $S_{f} $ can be diagonalized to give 
\begin{eqnarray}
S_{f} =\sum\nolimits_{s}\,\,\overline{\psi }_{f}^{\, \ast }\left( s\right)
\,a\left( s\right) \,\overline{\psi }_{f}\left( s\right)  \label{eq6.2}
\end{eqnarray}
where 
\begin{equation}
\psi _{f}\left( x\right) =\sum\limits_{s}U\left( x,s\right) \,\overline{\psi 
}_{f}\left( s\right) \quad ,\quad \overline{\psi }_{f}\left( s\right) =\int
d^{4}x\,U^{\dag }\left( x,s \right) \,\psi _{f}\left( x\right)  
\label{eq6.3}
\end{equation}
with 
\begin{eqnarray}
ie_{\alpha }^{\mu }\sigma ^{\alpha }D_{\mu }U\left( x,s\right) &=& 
a\left( s\right) U\left( x,s\right)   \label{eq6.4} \\
\int d^{4}x\,U^{\dag }\left( x,s \right) U\left( x,s^{\prime }\right)
&=&\delta _{ss^{\prime }}\;\quad ,\quad \;\sum\limits_{s}U\left( x,s\right)
U^{\dag }\left( x^{\prime },s \right) =\delta \left( x-x^{\prime }\right)
\; . \label{eq6.5}
\end{eqnarray}
Here, and in the following, $x$ represents a point in external 
spacetime, and $U\left( x,s\right) $ is a multicomponent 
eigenfunction. There is an implicit inner product in 
\begin{eqnarray}
U^{\dag }\left( x,s \right) \,\psi _{f}\left( x\right)
&=&\sum\limits_{r}U_{r}^{\dag }\left( x,s \right) \,\psi _{f}^{r}\left( x\right)
 \label{eq6.7} 
\end{eqnarray}
with the $2 N_{g}$ components of $\psi _{f}\left( x\right) $ 
labeled by $r=1,...,N_{g}$ (spanning all components of all irreducible gauge 
representations) and $a=1,2$ (labeling the components of Weyl spinors), 
and with $s$ and $\left( x,r,a \right) $ each having $N$ values. 
Also, the delta function in (\ref{eq6.5}) implicitly multiplies the $2 N_{g} \times 2 N_{g}$ 
identity matrix.

Evaluation of the present Euclidean path integral (a Gaussian integral with
Grassmann variables) is then trivial for fermions; as usual, 
\begin{eqnarray}
Z_{f} &=&\int \mathcal{D}\,\psi _{f}^{\dag }\left( x\right) \,\mathcal{D}
\,\psi _{f}\left( x\right) \,\,\,e^{-S_{f}}   \label{eq6.9}  \\
&=&\prod_{x,ra}\int d\,\psi _{f}^{ra \, \ast }\left( x\right) \int d\,\psi
_{f}^{ra}\left( x\right) \,e^{-S_{f}}   \label{eq6.10}  \\
&=&\prod_{s}z_{f}\left( s\right)  \label{eq6.11}
\end{eqnarray}
with 
\begin{eqnarray}
z_{f}\left( s\right) &=&\int d\,\overline{\psi }_{f}^{\, \ast }\left( s\right)
\,\int d\,\overline{\psi }_{f}\left( s\right) \,e^{-\overline{\psi }
_{f}^{\, \ast }\,\left( s\right) \,a\left( s\right) \,\overline{\psi }
_{f}\left( s\right) }  \label{eq6.12}  \\
&=&a\left( s\right)  \label{eq6.13}
\end{eqnarray}
since the transformation is unitary~\cite{peskin}. Now let 
\begin{eqnarray}
Z_{f}^{L} &=& \int \mathcal{D}\,\overline{\psi }_{f}^{\, \dag }\left( s\right) 
\, \mathcal{D}\,\overline{\psi }_{f}\left( s\right) \,\,e^{iS_{f}} 
\label{eq6.16} \\
&=& \prod_{s}z_{f}^{L}\left( s\right)  \label{eq6.17}
\end{eqnarray}
where 
\begin{eqnarray}
z_{f}^{L}\left( s\right) &=&\int d\,\overline{\psi }_{f}^{\, \ast }\left(
s\right) \int \,d\,\overline{\psi }_{f}\left( s\right) \,e^{i\,\overline
{\psi }_{f}^{\, \ast }\,\left( s\right) \,a\left( s\right) \,\overline{\psi }
_{f}\left( s\right) }  \label{eq6.18}  \\
&=&-ia\left( s\right)  \label{eq6.19}
\end{eqnarray}
so that 
\begin{eqnarray}
Z_{f}^{L}=c_{f}Z_{f}\;\quad ,\quad c_{f}=\prod_{s}\left( -i\right) 
\; . \label{eq6.20}
\end{eqnarray}
This result holds for the path integral over an arbitrary time interval, 
with the fields, operator, and meaning of time left unchanged.

The transition amplitude from an initial state to a final state is equal 
to the path integral between these states, so transition 
probabilities are the same with the Lorentzian and Euclidean forms of the path integral. 
This result is consistent with the fact 
that the classical equations of motion are also the same, since they 
follow from extremalization of the same action. Furthermore, using 
the method on pp. 290-291 or 302-303 of Ref.~\cite{peskin}, it is easy to 
show that the magnitude $\left| G \left( x,x^{\prime} \right) \right| $ of 
the $2$-point function is again the same, so particles propagate the 
same way in both descriptions. This result is also obtained in Appendix 
\ref{sec:appC} with a different method.

It may seem strange that the Lorentzian and Euclidean forms of the path integral yield the same physical results, 
but perusal of the standard arguments in e.g. field theory textbooks shows that the physically significant features 
of the results derive from the Lorentzian form of the \textit{action} rather than the path integral. 

When the inverse transformation from $\overline{\psi }_{f}$ to $\psi _{f}$
is performed, we obtain 
\begin{equation}
Z_{f}^{L}=\int \mathcal{D}\,\psi _{f}^{\dag }\left( x\right) \,\mathcal{D}
\,\psi _{f}\left( x\right) \,e^{iS_{f}}  \label{eq6.21}
\end{equation}
with $S_{f}$ having its form (\ref{eq5.31}) in the coordinate representation.

One may perform calculations in either the path-integral formulation or the
equivalent canonical formulation, which can now be obtained in the 
standard way: Let us use the notation $\int_{a}^{b}$ to indicate that the
fields in a path integral are specified to begin in a state $\left\vert
a\right\rangle $ at time $t_{a}$ and end in state 
$\left\vert b\right\rangle $ at time $t_{b}$, and also to indicate that 
a path integral showing these limits has its conventional definition 
(so that it may differ by a normalization constant from $Z_{f}^{L}$ 
as defined above). Then the Hamiltonian $H_{f}$ is defined by 
\begin{eqnarray}
\left\langle b\right\vert \,U_{f}\left( t_{b},t_{a}\right) \left\vert
a\right\rangle  &=&\int_{a}^{b}\mathcal{D}\,\psi _{f}^{\dag }\left( x\right)
\,\mathcal{D}\,\psi _{f}\left( x\right) \,e^{iS_{f}} \label{eq6.24} \\
i\frac{d}{dt}U_{f}\left( t,t_{a}\right)  &=&H_{f}\left( t\right)
U_{f}\left( t,t_{a}\right) \quad ,\quad U_{f}\left( t_{a},t_{a}\right) =1 \; .
\label{eq6.25} 
\end{eqnarray}
I.e., the time evolution operator 
$U_{f}\left( t_{b},t_{a}\right) $ 
is defined to have the same effect as the path 
integral over intermediate states, and it is then straightforward to 
reverse the usual logic which leads from canonical quantization to 
path-integral quantization~\cite{peskin,weinberg}.

\section{\label{sec:sec7}Transformation to Lorentzian 
path integral: bosons}

For bosons we can again perform the transformation (\ref{eq6.3}) to obtain 
\begin{equation}
S_{b}=\sum\nolimits_{s}\overline{\psi }_{b}^{\, \ast }\left( s\right)
\,a\left( s\right) \,\overline{\psi }_{b}\left( s\right) \;.
\label{eq7.1}
\end{equation}
We will now show how this action can be put into a form which corresponds
to scalar bosonic fields plus their auxiliary fields, temporarily working in a 
locally inertial coordinate system, so that 
$e_{\alpha }^{\mu }\sigma ^{\alpha} \rightarrow \sigma ^{\mu} $.
First, if the gauge potentials $A_{\mu }^{i}$ were zero, we would have 
\begin{equation}
i\sigma ^{\mu }\partial _{\mu }U^{0}\left( x,s\right) =a_{0}\left(
s\right) U^{0}\left( x,s\right) \;.  \label{eq7.6}
\end{equation}
Then 
\begin{equation}
\,U^{0}\left( x,s\right) =\mathcal{V}^{-1/2}u\left( s\right) e^{ip_{s}\cdot
x}\;\;,\;\;p_{s}\cdot x=\eta _{\mu \nu }p_{s}^{\mu }x^{\nu }
\quad , \quad \eta _{\mu \nu } = \mathrm{diag} \left( -1,1,1,1 \right) 
\label{eq7.7}
\end{equation}
(with $\mathcal{V}$ a four-dimensional normalization volume) gives 
\begin{equation}
-\eta _{\mu \nu }\sigma ^{\mu }p_{s}^{\nu }U^{0}\left( x,s\right)
=a_{0}\left( s\right) U^{0}\left( x,s\right)  \label{eq7.8}
\end{equation}
where $\sigma ^{\mu }$ implicitly multiplies the identity matrix for the
multicomponent function $U^{0}\left( x,s\right) $. A given 2-component
spinor $u_{r}\left( s\right) $ has two eigenstates of 
$p_{s}^{k}\sigma ^{k}$: 
\begin{equation}
p_{s}^{k}\sigma ^{k}u_{r}^{+}\left( s\right) =\left\vert \overrightarrow{p}
_{s}\right\vert u_{r}^{+}\left( s\right) \quad ,\quad
p_{s}^{k}\sigma ^{k}u_{r}^{-}\left( s\right) =-\left\vert \overrightarrow{p}
_{s}\right\vert u_{r}^{-}\left( s\right)  \label{eq7.9}
\end{equation}
where $\overrightarrow{p}_{s}$ is the 3-momentum, with magnitude 
$\left\vert \overrightarrow{p}_{s}\right\vert $. 
The multicomponent eigenstates of $i\sigma ^{\mu }\partial _{\mu }$ and
their eigenvalues $a_{0}\left( s\right) =p_{s}^{0}\mp $ $\left\vert 
\overrightarrow{p}_{s}\right\vert $ thus come in pairs, corresponding to
opposite helicities.

For nonzero $A_{\mu }^{i}$, the eigenvalues $a\left( s\right) $ will also
come in pairs, with one growing out of $a_{0}\left( s\right) $ and the other
out of its partner $a_{0}\left( s^{\prime }\right) $ as the $A_{\mu }^{i}$
are turned on. To see this, first write (\ref{eq6.4}) as 
\begin{equation}
\left( i\partial _{0}+A_{0}^{i}t_{i}\right) U\left( x,s\right) +\sigma
^{k}\left( i\partial _{k}+A_{k}^{i}t_{i}\right) U\left( x,s\right) =a\left(
s\right) U\left( x,s\right)  \label{eq7.10}
\end{equation}
or 
\begin{equation}
\left( i\partial _{0}\delta_{rr^{\prime }}
+A_{0}^{i}t_{i}^{rr^{\prime }}\right) U_{r^{\prime }}
\left( x,s\right) -P_{rr^{\prime }}U_{r^{\prime }}\left( x,s\right)
-a\left( s\right) \delta _{rr^{\prime }}U_{r^{\prime }}\left( x,s\right) =0
\label{eq7.11}
\end{equation}
\begin{equation}
P_{rr^{\prime }}\equiv -\sigma ^{k} \left( i\partial _{k} 
\delta_{rr^{\prime }} + A_{k}^{i}t_{i}^{rr^{\prime }}\right) \label{eq7.12}
\end{equation}
with the usual implied summations over repeated indices. 
At fixed $r$, $r^{\prime }$ (and $x,s$), 
apply a matrix $s$ which will diagonalize
the $2\times 2$ matrix $P_{rr^{\prime }}$, bringing it into the form 
$p_{rr^{\prime }}\sigma ^{3}+\overline{p}_{rr^{\prime }}\sigma ^{0}$, where 
$p_{rr^{\prime }}$ and $\overline{p}_{rr^{\prime }}$ are 1-component
operators, while at the same time rotating the 2-component spinor 
$U_{r^{\prime }}$: 
\begin{eqnarray}
sP_{rr^{\prime }}s^{-1} &=&P_{rr^{\prime }}^{\prime }=p_{rr^{\prime
}}\sigma ^{3}+\overline{p}_{rr^{\prime }}\sigma ^{0}\quad ,\quad
U_{r^{\prime }}^{\prime }=sU_{r^{\prime }} 
\label{eq7.13}  \\
\sigma ^{0} &=&\left( 
\begin{array}{cc}
1 & 0 \\ 
0 & 1
\end{array}
\right) \quad ,\quad \sigma ^{3}=\left( 
\begin{array}{cc}
1 & 0 \\ 
0 & -1  
\end{array}
\right)  \; .  \label{eq7.14}
\end{eqnarray}
But $P_{rr^{\prime }}$ is traceless, and the trace is invariant under a
similarity transformation, so $\overline{p}_{rr^{\prime }}=0$. Then the second
term in (\ref{eq7.11}) (for fixed $r$ and $r^{\prime }$) 
becomes $s^{-1 }p_{rr^{\prime }}\sigma ^{3}U_{r^{\prime }}
^{\prime }\left( x,s\right) $. The two independent choices 
\begin{eqnarray}
U_{r^{\prime }}^{\prime }\left( x,s\right) &\propto&\left( 
\begin{array}{c}
1 \\ 
0
\end{array}
\right) \quad ,\quad \sigma ^{3}U_{r^{\prime }}^{\prime }\left( x,s\right)
=+U_{r^{\prime }}^{\prime }\left( x,s\right)  \label{eq7.15}  \\
U_{r^{\prime }}^{\prime }\left( x,s\right) &\propto&\left( 
\begin{array}{c}
0 \\ 
1
\end{array}
\right) \quad ,\quad \sigma ^{3}U_{r^{\prime }}^{\prime }\left( x,s\right)
=-U_{r^{\prime }}^{\prime }\left( x,s\right)
\label{eq7.16}
\end{eqnarray}
give $\pm s^{-1 }p_{rr^{\prime }}U_{r^{\prime }}^{\prime }
\left( x,s\right) $. Now use $s^{-1 }U_{r^{\prime }}^{\prime }
=U_{r^{\prime }}$ to obtain for (\ref{eq7.11}) 
\begin{equation}
\left( i\partial _{0}\delta_{rr^{\prime }}
+A_{0}^{i}t_{i}^{rr^{\prime }}\right) U_{r^{\prime }} \left( x,s\right) 
 \mp p_{rr^{\prime }}U_{r^{\prime }}\left( x,s\right)
-a\left( s\right) \delta _{rr^{\prime }}U_{r^{\prime }}\left( x,s\right) =0
\label{eq7.17}
\end{equation}
so (\ref{eq7.10}) reduces to two sets of equations with 
different eigenvalues $a\left( s\right)$ and 
$a\left( s^{\prime }\right)$: 
\begin{eqnarray}
a\left( s\right) = a_{1}\left( s\right) +a_{2}\left( s\right)
\quad , \quad 
a\left( s^{\prime }\right) = a_{1}\left( s\right) -a_{2}\left( s\right)
\label{eq7.21}
\end{eqnarray}
where these equations define $a_{1}\left( s\right) $ and $a_{2}\left(
s\right) $. Notice that letting $\sigma ^{k}\rightarrow -\sigma ^{k}$ in 
(\ref{eq7.10}) reverses the signs in (\ref{eq7.17}), and results in 
$a\left( s \right) \rightarrow a\left( s^{\prime }\right)$:
\begin{equation}
\left( i\partial _{0}+A_{0}^{i}t_{i}\right) U\left( x,s\right) -\sigma
^{k}\left( i\partial _{k}+A_{k}^{i}t_{i}\right) U\left( x,s\right) =a\left(
s^{\prime } \right) U\left( x,s\right) \; . \label{eq7.21a}
\end{equation}

The action for a single eigenvalue $\,a\left( s\right) $ and its partner 
$a\left( s^{\prime }\right) $ is 
\begin{eqnarray}
\widetilde{s}_{b}\left( s\right) &=&\overline{\psi }_{b}^{\, \ast }\left(
s\right) \,a\left( s\right) \,\overline{\psi }_{b}\left( s\right) +\overline
{\psi }_{b}^{\, \ast }\left( s^{\prime }\right) \,a\left( s^{\prime} \right) \,
\overline{\psi }_{b}\left( s^{\prime }\right)  \label{eq7.22} \\
&=&\overline{\psi }_{b}^{\, \ast }\left( s\right) \left( a_{1}\left( s\right)
+a_{2}\left( s\right) \right) \overline{\psi }_{b}\left( s\right) +\overline
{\psi }_{b}^{\, \ast }\left( s^{\prime }\right) \left( a_{1}\left( s\right)
-a_{2}\left( s\right) \right) \overline{\psi }_{b}\left( s^{\prime }\right)
\; .  \label{eq7.23}
\end{eqnarray}
In the following we will circumvent singularities by implicitly following the standard prescription 
$a_1 \rightarrow a_1 + i \epsilon $, $\epsilon \rightarrow 0+$,
which reduces to $\omega \rightarrow \omega  + i \epsilon $
when there are no gauge fields.

For $a_{1}\left( s\right) \ge 0$, let us choose $a_{2}\left( s\right) \ge 0$ 
and define 
\begin{eqnarray}
\overline{\psi }_{b}\left( s^{\prime }\right) &=&a\left( s\right) ^{1/2}
\overline{\phi }_{b}\left( s^{\prime} \right) =\left( a_{1}\left( s\right)
+a_{2}\left( s\right) \right) ^{1/2}\overline{\phi }_{b}\left( s^{\prime} \right)
\label{eq7.24} \\
\overline{\psi }_{b}\left( s\right) &=&a\left( s\right) ^{-1/2}\overline{F}
_{b}\left( s\right) =\left( a_{1}\left( s\right) +a_{2}\left( s\right)
\right) ^{-1/2}\overline{F}_{b}\left( s\right)  \label{eq7.25}
\end{eqnarray}
so that 
\begin{eqnarray}
\widetilde{s}_{b}\left( s\right) =\overline{\phi }_{b}^{\, \ast }\left(
s^{\prime} \right) \widetilde{a}\left( s\right) \overline{\phi }_{b}\left( s^{\prime} \right) +
\overline{F}_{b}^{\, \ast }\left( s\right) \overline{F}_{b}\left( s\right)
\quad ,\quad a_{1}\left( s\right) \ge 0  \label{eq7.26}
\end{eqnarray}
where 
\begin{eqnarray}
\widetilde{a}\left( s\right) =a\left( s\right) a\left( s^{\prime}\right)
=a_{1}\left( s\right) ^{2}-a_{2}\left( s\right) ^{2} \; . \label{eq7.27}
\end{eqnarray}
For $a_{1}\left( s\right) <0$, let us choose $a_{2}\left( s\right) \le 0$ and
write 
\begin{eqnarray}
\overline{\psi }_{b}\left( s^{\prime }\right) &=&\left( -a\left( s\right)
\right) ^{1/2}\overline{\phi }_{b}\left( s^{\prime} \right) =\left( -a_{1}\left(
s\right) -a_{2}\left( s\right) \right) ^{1/2}\overline{\phi }_{b}\left(
s^{\prime} \right)  \label{eq7.28} \\
\overline{\psi }_{b}\left( s\right) &=&\left( -a\left( s\right) \right)
^{-1/2}\overline{F}_{b}\left( s\right) =\left( -a_{1}\left( s\right)
-a_{2}\left( s\right) \right) ^{-1/2}\overline{F}_{b}\left( s\right)
\label{eq7.29}
\end{eqnarray}
so that 
\begin{equation}
\widetilde{s}_{b}\left( s\right) =-\left[ \overline{\phi }_{b}^{\, \ast }\left(
s^{\prime} \right) \widetilde{a}\left( s\right) \overline{\phi }_{b}\left( s^{\prime} \right) +
\overline{F}_{b}^{\, \ast }\left( s\right) \overline{F}_{b}\left( s\right) 
\right] \quad ,\quad a_{1}\left( s\right) <0 \; .  \label{eq7.30}
\end{equation}

Then we have
\begin{align}
S_{b} &= \displaystyle\sideset{}{'}\sum\nolimits_{s}\widetilde{s}_{b}\left( s\right) 
\label{eq7.31} \\
&= \displaystyle\sideset{}{'}\sum\limits_{a_{1}\left( s\right) \ge 0}\, \left[ \overline{\phi }
_{b}^{\, \ast }\left( s^{\prime} \right) \widetilde{a}\left( s\right) \overline{\phi }
_{b}\left( s^{\prime} \right) + \overline{F}_{b}^{\, \ast }\left( s\right) \overline{F}
_{b}\left( s\right) \right] 
-\displaystyle\sideset{}{'}\sum\limits_{a_{1}\left( s\right) <0}\left[ 
\overline{\phi }_{b}^{\, \ast }\left( s^{\prime} \right) \widetilde{a}\left( s\right) 
\overline{\phi }_{b}\left( s^{\prime} \right) +\overline{F}_{b}^{\, \ast }\left( s\right) 
\overline{F}_{b}\left( s\right) \right]  
\label{minus}
\end{align}
where a prime on a summation or product over $s$ means that only one member
of an $s,s^{\prime }$ pair (as defined in (\ref{eq7.17}) and (\ref{eq7.21})) is
included. Let us separate the non-negative contribution $S_{+}$ from the anomalous negative contribution  $S_{-}$:
\begin{eqnarray}
S_{b} &=& S_{+} + S_{-} \label{eq7.32aa} \\
S_{+} &=& \displaystyle\sideset{}{'}\sum\limits_{s \ge 0}\, \overline{\phi }
_{b}^{\, \ast }\left( s^{\prime } \right) \left| \widetilde{a}\left( s\right) \right| \overline{\phi }
_{b}\left( s^{\prime } \right) 
+ \displaystyle\sideset{}{'}\sum\limits_{a_{1}\left( s\right) \ge 0}\overline{F}_{b}^{\, \ast }\left( s\right) \overline{F}
_{b}\left( s\right) \label{eq7.32a} \\
S_{-} &=& - \left[ \displaystyle\sideset{}{'}\sum\limits_{s<0}
\overline{\phi }_{b}^{\, \ast }\left( s^{\prime } \right) \left| \widetilde{a} \left( s\right) \right| 
\overline{\phi }_{b}\left( s^{\prime } \right) 
+\displaystyle\sideset{}{'} \sum\limits_{a_{1}\left( s\right) <0}\overline{F}_{b}^{\, \ast }\left( s\right) 
\overline{F}_{b}\left( s\right)  \right] 
\label{eq7.32b} 
\end{eqnarray}
where 
\begin{eqnarray}
s < 0\; &\longleftrightarrow & \; \widetilde{a}\left( s\right) 
= a_{1}\left( s \right) ^{2}-a_{2}\left( s\right) ^{2} < 0\;\;\;
\text{if}\;a_{1}\left( s\right) \ge 0 \label{eq7.37} \\
&\longleftrightarrow &  \; \widetilde{a}\left( s\right) =a_{1}\left( s\right)
^{2}-a_{2}\left( s\right) ^{2} > 0\;\;\;\text{if}\;a_{1}\left( s\right) <0
\label{eq7.38} 
\end{eqnarray}
with $s\geq 0$ otherwise. 

Recall that if the gauge potentials $A_{\mu }^{i}$ were zero, we would have 
$a_{1}=\omega $ and $a_{2}= \mp \left\vert \overrightarrow{p}\right\vert $, where 
$\omega $ is the frequency and $\overrightarrow{p}$ the 3-momentum. 

The negative-action modes of (\ref{eq7.32b}) are discussed in Appendix \ref{sec:appD}, where it is found that their \textit{excitations} can be treated in the same way as the positive-action modes. (Here and below, ``positive-action modes'' means those which do not have negative action in the present context, before mass and interaction terms are acquired. After the transformation to a Lorentzian path integral below, some fields may come to have negative action due to symmetry breaking and condensation, but this is permissible within a Lorentzian description.) It follows that, in the treatment below, no degrees of freedom are lost in the bosonic \textit{excitations}. On-shell modes already have non-negative action, so both the excitation and condensation of scalar boson fields remain exactly the same as in standard physics.

The path integral for positive-action modes in $S_{+} $ is 
\begin{eqnarray}
Z_{+} &=&\int \mathcal{D}\,\psi _{b}^{\dag }\left( x\right) \,\mathcal{D}
\,\psi _{b}\left( x\right) \,\,\,e^{-S_{+}}  \label{eq7.2} \\
&=&\prod_{x,ra}\int_{-\infty }^{\,\infty }d\,\left( \mathrm{Re}\,\psi
_{b,ra}\left( x\right) \right) \,\int_{-\infty }^{\,\infty }d\,\left( 
\mathrm{Im}\,\psi _{b,ra}\left( x\right) \right) \,e^{-S_{+}} \nonumber
\label{eq7.3}  \\
&=&\prod_{s} \int_{-\infty }^{\,\infty }d(\mathrm{Re}\,\overline
{\psi }_{b}\left( s\right) )\int_{-\infty }^{\,\infty }d(\mathrm{Im}\,
\overline{\psi }_{b}\left( s\right) )\,e^{-S_{+}}  \; . \label{eq7.5}
\end{eqnarray}
Each of the 
transformations above from $\overline{\psi }_{b}$ to $\overline{\phi }_{b}$ 
and $\overline{F}_{b}$ has the form 
\begin{equation}
\overline{\psi }_{b}\left( s^{\prime} \right) = A\left( s\right)^{1/2}
\overline{\phi }_{b}\left( s^{\prime} \right) \;\; ,\;\; \overline{\psi }_{b}
\left( s\right) = A\left( s\right) ^{-1/2}\overline{F}_{b}\left( s\right)
\label{eq7.33}
\end{equation}
so that $d\overline{\psi }_{b}\left( s^{\prime}\right) 
= A\left( s\right) ^{1/2}d\overline
{\phi }_{b}\left( s^{\prime} \right)$, $d\overline{\psi }_{b}\left(s 
\right) =A\left( s\right) ^{-1/2}d\overline{F}_{b}\left( s\right)$, 
and the Jacobian is 
$\prod\nolimits_{s}^{\prime }A\left( s\right) ^{1/2}A\left( s\right)^{-1/2}=1$.
These transformations then lead to
\begin{eqnarray}
Z_{sb}\; = \; \displaystyle\sideset{}{'}\prod_{s\geq 0}z_{\phi }\left( s \right) \cdot \displaystyle\sideset{}{'} \prod_{s < 0}z_{\phi }\left( s \right) 
\cdot  \displaystyle\sideset{}{'} \prod_{a_{1}\left( s\right) \geq 0} z_{F}\left( s\right) 
\; =\; \displaystyle\sideset{}{'}\prod_{s}z_{\phi }\left( s \right)\cdot 
\displaystyle\sideset{}{'}\prod_{a_{1}\left( s\right) \geq 0}z_{F}\left( s\right)  
\label{eq7.36}
\end{eqnarray}
where the excitations of negative-action modes have ben added and 
\begin{eqnarray}
\hspace{-1.0cm} z_{\phi }\left( s\right) &=&\int_{-\infty }^{\,\infty }d(\mathrm{Re}\,
\overline{\phi }_{b}\left( s^{\prime} \right) )\int_{-\infty }^{\,\infty }d(\mathrm{Im}\,
\overline{\phi }_{b}\left( s^{\prime} \right) ) e^{- \left|\widetilde{a}\left(
s\right) \right| \left[ \left( \mathrm{Re}\,\overline{\phi }_{b}\left( s^{\prime} \right)
\right) ^{2}+\left( \mathrm{Im}\,\overline{\phi }_{b}\left( s^{\prime} \right) \right)
^{2}\right] } \label{eq7.38b} \\
&=&\frac{\pi }{\left| \widetilde{a}\left( s\right)\right| } \\
z_{F}\left( s\right) &=&\int_{-\infty }^{\,\infty }d(\mathrm{Re}\,\overline{F}
_{b}\left( s\right) )\int_{-\infty }^{\,\infty }d(\mathrm{Im}\,\overline{F}
_{b}\left( s\right) ) e^{-\left[ \left( \mathrm{Re}\,\overline{F}
_{b}\left( s\right) \right) ^{2}+\left( \mathrm{Im}\,\overline{F}_{b}\left(
s\right) \right) ^{2}\right] } \label{eq7.38c} \\
&=&\pi \; .
\end{eqnarray}

Now let 
\begin{eqnarray}
S_{sb} &=& \displaystyle\sideset{}{'}\sum\limits_{s}\, \overline{\phi }
_{b}^{\, \ast }\left( s^{\prime} \right) \widetilde{a}\left( s\right) \overline{\phi }
_{b}\left( s^{\prime} \right) 
+ \displaystyle\sideset{}{'}\sum\limits_{a_{1}\left( s\right) >0}\overline{F}_{b}^{\, \ast }\left( s\right) \overline{F}
_{b}\left( s\right) \label{eq7.40}  \\
Z_{sb}^{L} &=&\int \mathcal{D}\,\overline{\phi }_{b}^{\dag }\,\left( s^{\prime} \right) 
\mathcal{D}\,\overline{\phi }_{b}\left( s^{\prime} \right) \,\mathcal{D}\,\overline{F}
_{b}^{\dag }\,\left( s\right) \mathcal{D}\,\overline{F}_{b}\left( s\right)
\,\,e^{iS_{sb}}  \label{eq7.41} \\
&=&\displaystyle\sideset{}{'}\prod_{s}z_{\phi }^{L}\left( s \right) \cdot 
\displaystyle\sideset{}{'}\prod_{a_{1}\left( s\right) \geq 0}z_{F}^{L}\left( s\right)  
\label{eq7.42}
\end{eqnarray}
where 
\begin{eqnarray}
\hspace{-0.3cm} z_{\phi }^{L}\left( s \right) &=& \int_{-\infty }^{\,\infty }d(\mathrm{Re}\,
\overline{\phi }_{b}\left( s^{\prime} \right) ) \int_{-\infty }^{\,\infty }d(\mathrm{Im}
\,\overline{\phi }_{b}\left( s^{\prime} \right) ) 
e^{i\widetilde{a}\left( s\right) \left[ \left( 
\mathrm{Re}\,\overline{\phi }_{b}\left( s^{\prime} \right) \right) ^{2}+\left( \mathrm{
Im}\,\overline{\phi }_{b}\left( s^{\prime} \right) \right) ^{2}\right] } 
\label{eq7.43a} \\
&=& i \frac{\pi }{\widetilde{a}\left( s\right)} \label{eq7.43b} \\
\hspace{-0.3cm} z_{F}^{L}\left( s\right) &=& 
\int_{-\infty }^{\,\infty }d(\mathrm{
Re}\,\overline{F}_{b}\left( s\right) )\int_{-\infty }^{\,\infty }d(\mathrm{Im
}\,\overline{F}_{b}\left( s\right) ) e^{i\left[
\left( \mathrm{Re}\,\overline{F}_{b}\left( s\right) \right) ^{2}+\left( 
\mathrm{Im}\,\overline{F}_{b}\left( s\right) \right) ^{2}\right] }
\label{eq7.44a} \\
&=& i \pi   \label{eq7.44b}
\end{eqnarray}
since $\int_{-\infty }^{\,\infty }dx\,\int_{-\infty }^{\,\infty }dy\,\exp
\left( \,ia\left( x^{2}+y^{2}\right) \right) = i\pi /a$. (Nuances of
Lorentzian path integrals are discussed in, e.g., Peskin and Schroeder~\cite{peskin}, 
p. 286.) We have then obtained 
\begin{equation}
Z_{sb}^{L}=c_{b}Z_{sb}
\label{eq7.45}
\end{equation}
where $c_{b} $ is a product of factors of $ i$ and $-1$.

To return to the coordinate representation, let us define physical fields 
\begin{eqnarray}
\Phi \left( x\right) =\displaystyle\sideset{}{'}\sum_sU\left( x,s^{\prime }\right) 
\,\overline{\phi }_{b}\left( s^{\prime }\right) 
\label{equ7.87}
\end{eqnarray}
and auxiliary fields 
\begin{eqnarray}
\mathcal{F}\left( x\right) =\displaystyle\sideset{}{'}\sum_s
U\left( x,s\right) \overline{F}_{b}\left( s\right) \; .
\label{equ7.80}
\end{eqnarray}
Recall that these fields include only positive-action modes and \textit{excitations} of negative-action modes.

As a reminder of the notation, recall that, according to (\ref{eq7.10}) and 
(\ref{eq7.21a}), 
\begin{eqnarray}
i\sigma ^{\mu }D_{\mu }U\left( x,s^{\prime }\right) =a\left( s^{\prime
}\right) U\left( x,s^{\prime }\right) \quad ,\quad i\overline{\sigma }^{\mu
}D_{\mu }U\left( x,s^{\prime }\right) =a\left( s\right) U\left( x,s^{\prime
}\right) 
\label{equ7.88}
\end{eqnarray}
with $\overline{\sigma }^{0}=\sigma ^{0}$, $\overline{\sigma }^{k}=-\sigma
^{k}$, $a\left( s\right) =a_{1}\left( s\right) +a_{2}\left( s\right) $, 
$a\left( s^{\prime }\right) =a_{1}\left( s\right) -a_{2}\left( s\right) $, 
$\widetilde{a}\left( s\right) =a\left( s\right) a\left( s^{\prime }\right)
=a_{1}\left( s\right) ^{2}-a_{2}\left( s\right) ^{2}$, and $s>0$ or $<0$
defined by (\ref{eq7.37})-(\ref{eq7.38}) and the line following. Again, in
the absence of gauge potentials we have $a_{1}=\omega $ and $a_{2}=\mp
\left\vert \overrightarrow{p}\right\vert $.

We could return to the original coordinate system, with the action in 
(\ref{eq7.40}) becoming 
\begin{eqnarray}
S_{sb}=S_{\Phi }+S_{\mathcal{F}}
\label{eq7.52y}
\end{eqnarray}
where 
\begin{eqnarray}
S_{\Phi }=\int d^{4}x\, \overline{\mathcal{L}}_{\Phi }\quad ,\quad S_{\mathcal{F}}=\int
d^{4}x\,\overline{\mathcal{L}}_{\mathcal{F}}\quad ,\quad \overline{\mathcal{L}}_{\mathcal{F}}=
\mathcal{F}^{\dag }\left( x\right) \mathcal{F}\left( x\right) 
\label{equ7.90}
\end{eqnarray}
and $\overline{\mathcal{L}}_{\Phi }$ is obtained via $\sigma ^{\mu }\rightarrow
e_{\alpha }^{\mu }\sigma ^{\alpha }$, as in (\ref{eq6.4}). However, in the
following it is more convenient to remain in the locally inertial coordinate
system used above, where 
\begin{eqnarray}
\overline{\mathcal{L}}_{\Phi }=\frac{1}{2}\Phi ^{\dag }\left( x\right) i
\overline{\sigma }^{\mu }D_{\mu }\,i\sigma ^{\nu }D_{\nu }\Phi \left( x\right) 
+\frac{1}{2}\Phi ^{\dag }\left( x\right) i\sigma ^{\mu }D_{\mu }\,i\overline{\sigma }
^{\nu }D_{\nu }\Phi \left( x\right) \;.  
\label{eq7.209}
\end{eqnarray}
To simplify the mathematics below, it is convenient to temporarily write 
\begin{eqnarray}
\Phi _{b}=\left( 
\begin{array}{c}
\Phi  \\ 
\Phi 
\end{array}
\right) 
\label{equ7.92}
\end{eqnarray}
and to use the same Weyl representation as is used for Dirac fermions, with 
\begin{eqnarray}
\gamma ^{\mu }=\left( 
\begin{array}{cc}
0 & \sigma ^{\mu } \\ 
\overline{\sigma }^{\mu } & 0
\end{array}
\right) 
\label{equ7.93}
\end{eqnarray}
so that (\ref{eq7.209}) can be written as 
\begin{eqnarray}
\overline{\mathcal{L}}_{\Phi } &=&-\frac{1}{2}\Phi _{b}^{\dag }\left( x\right) \gamma ^{\mu }D_{\mu }\,\gamma ^{\nu }D_{\nu }\Phi _{b}\left( x\right) \,
\label{eq7.52x} \\
&=&-\frac{1}{2}\Phi _{b}^{\dag }\left( x\right) \,\slashed{D}^{2}\,\Phi_{b}\left( x\right) \;.  \label{eq7.52q}
\end{eqnarray}

A result~\cite{peskin,schwartz} that can be extended to the nonabelian case gives
\begin{eqnarray}
- \slashed{D}^{2}=D^{\mu }D_{\mu }-S^{\mu \nu }F_{\mu \nu }  \label{eq7.205}
\end{eqnarray}
with the present convention for the metric tensor. (See pp. 173-174 of \cite{schwartz} for this result and those immediately below.) Here the field strength
tensor $F_{\mu \nu }$ spans all the irreducible (physical) gauge
representations, and the second term can be rewritten with
``magnetic'' and ``electric'' fields $B_{k}$ and $E_{k}$ defined by 
\begin{eqnarray}
F_{kk^{\prime }}=-\varepsilon _{kk^{\prime }k^{\prime \prime }}B_{k^{\prime
\prime }}\quad ,\quad F_{0k}=E_{k}
\label{equ7.96a}
\end{eqnarray}
since \cite{schwartz}
\begin{eqnarray}
- S^{\mu \nu }F_{\mu \nu }=\left( 
\begin{array}{cc}
\left( \overrightarrow{B}+i\overrightarrow{E}\right) \cdot 
\overrightarrow{\sigma } & 0 \\ 
0 & \left( \overrightarrow{B}-i\overrightarrow{E}\right) \cdot 
\overrightarrow{\sigma }
\end{array}
\right) 
\label{equ7.97}
\end{eqnarray}
where $a\cdot b=a_{k}b^{k}$. (Recall that $\mu =0,1,2,3$ and $k=1,2,3$.) We
then obtain
\begin{eqnarray}
\overline{\mathcal{L}}_{\Phi }=\Phi ^{\dag }\left( x\right) D^{\mu }D_{\mu }\Phi \left(
x\right) +\Phi ^{\dag }\left( x\right) \,\overrightarrow{B}\cdot 
\overrightarrow{\sigma }\,\Phi \left( x\right) \;.
\label{eq7.58x}
\end{eqnarray}
The second term above is invariant under a rotation, but not under a boost,
so it breaks Lorentz invariance for the primitive bosonic fields in $\Phi$. This issue is considered in the next section, where the physical scalar boson fields are defined using arguments 
that are an extension of those we have given elsewhere~\cite{DM2021a,DM2021b}.

All of the above is in a locally inertial frame of reference. Now let us return to a general coordinate system and initially assume no nongravitational gauge fields, so that $D_{\mu} \rightarrow \nabla_{\mu}$ and 
\begin{eqnarray}
\overline{\mathcal{L}}_{\Phi } &=&-\frac{1}{2}\Phi _{b}^{\dag }\left( x\right) \underline{\gamma}
^{\mu }\nabla_{\mu }\,\underline{\gamma} ^{\nu }\nabla_{\nu }\Phi _{b}\left( x\right) =
-\frac{1}{2}\Phi _{b}^{\dag }\left( x\right) \,\slashed{\nabla}^{2}\,\Phi_{b}\left( x\right)  \label{eq7.52qq}
\end{eqnarray}
where
\begin{eqnarray}
\,\slashed{\nabla} \equiv \underline{\gamma }^{\mu }\nabla _{\mu } \quad , \quad
\underline{\gamma }^{\mu } = \left( 
\begin{array}{cc}
0 & \underline{\sigma }^{\mu} \\ 
\overline{\underline{\sigma }}^{\mu } & 0
\end{array}
\right) 
\end{eqnarray}
with $\underline{\sigma }^{\mu } =
e_{\alpha }^{\mu }\,\sigma ^{\alpha }$ and 
$\overline{\underline{\sigma }}^{\alpha } = e_{\alpha }^{\mu }\,\overline{\sigma}
^{\alpha }$. But
\begin{equation}
-\slashed{\nabla}^{2} = g^{\mu \nu } \nabla_{\mu}\nabla_{\nu} - \frac{1}{4} R 
\end{equation}
follows from e.g. (5.293) of Ref.~\cite{Parker} with our $\left( -+++\right) $
convention for the metric tensor, where $R$ is the curvature scalar in 4-dimensional
spacetime, so
\begin{eqnarray}
\overline{\mathcal{L}}_{\Phi} 
=\Phi ^{\dag }\left( x\right) \left( g^{\mu \nu } \nabla_{\mu}\nabla_{\nu} - \frac{1}{4} R
\right)  \Phi  \left( x\right) \; .
\label{xi}
\end{eqnarray}
(See also (15.5.5) of Ref.~\cite{GSW}, with the present convention for the metric tensor, but the opposite convention of their p. 274 for the Dirac gamma matrices.)
It is well known that a scalar field can have a Lagrangian which includes a term $- \, \xi \, R$ (as in (2.35) of Ref.~\cite{Parker}), with $\xi$ undetermined in standard physics, but the present picture yields 
$\xi = 1/4 $.

\section{\label{sec:DM}Scalar bosons}

Let $\Phi_r$ be one of the primitive 2-component spin 1/2 bosonic fields in $\Phi $. 
We can construct physical fields, which satisfy Lorentz invariance, by eliminating 
the anomalous second term of (\ref{eq7.58x}) in either of two ways.

\textbf{Higgs fields} can be constructed by combining $2$-component 
fields $\Phi _r$ and $\Phi _{r^{\prime }}$ with the same gauge quantum numbers but opposite spins (and equal amplitudes):
\begin{align}
\Phi _{R}=\left( 
\begin{array}{c}
\Phi _{r} \\ 
\Phi _{r^{\prime }}
\end{array}
\right)  \label{e2}
\end{align}
so that 
\begin{align}
\Phi _{R}^{\dag }\left( x\right) \,\overrightarrow{\sigma }\, 
\Phi _{R}\left( x\right) &=\left( 
\begin{array}{cc}
\Phi _{r}^{\dag } & \Phi _{r^{\prime }}^{\dag }
\end{array}
\right) \left( 
\begin{array}{cc}
\overrightarrow{\sigma }\, & 0 \\ 
0 & \overrightarrow{\sigma }\,
\end{array}
\right) \left( 
\begin{array}{c}
\Phi _{r} \\ 
\Phi _{r^{\prime }}
\end{array}
\right)  \notag \\
&=\Phi _{r}^{\dag }\,\overrightarrow{\sigma }\,\Phi 
_{r}+\Phi _{r^{\prime }}^{\dag }\,\overrightarrow{\sigma }\, 
\Phi _{r^{\prime }}  \label{eq103a} \\
&=0\;.
\end{align}
In this case we write for each gauge component 
\begin{align}
\Phi _{R}\left( x\right) =\phi _{R}\left( x\right) \,\xi _{R}\quad \mathrm{\
with}\quad \xi _{R}^{\,\dag }\,\xi _{R}=1
\end{align}
where $\xi _{R}$ has $4$ constant components and $\phi _{R}\left( x\right) $
is a $1$-component complex amplitude. If this can be done for a full gauge multiplet $\widetilde{\Phi}_{R}$ containing $\Phi _{R}$, yielding a multiplet 
 $\widetilde{\phi}_{R}$ of 1-component scalar boson fields, then (\ref{eq7.58x}) implies that
\begin{align}
S_{R}=\int d^{4}x \, \overline{\mathcal{L}}_{R}\quad ,\quad \overline{\mathcal{L}}_{R}=
\widetilde{\phi }_{R}^{\dag }\left( x\right) D^{\mu }D_{\mu }\,\widetilde{\phi }
_{R}\left( x\right)
\end{align}
is the action for this multiplet (before masses and other interaction terms are added).

\textbf{Higgson fields} can be constructed by combining
a $2$-component field $\Phi _{s}$ and its charge conjugate $\Phi _{s}^{c}$ (with opposite gauge quantum numbers but the same spin):
\begin{align}
\Phi _{S}=\frac{1}{\sqrt{2}}\left( 
\begin{array}{c}
\Phi _{s} \\ 
\Phi _{s}^{c}
\end{array}
\right) \;.
\end{align}
The subscripts $s$ and $S$ are used in this context to avoid confusion.

$\Phi _{s}$ and $\Phi _{s}^{c}$ have opposite 
expectation values for the generators $t^{j}$ (which are here treated
as operators rather than matrices):
\begin{align}
\Phi _{s}^{c\,\dag }t^{j}\,\Phi _{s}^{c}=-\,\Phi _{s}^{\dag }t^{j}\,\Phi
_{s}\;.
\end{align}
Since
\begin{align}
B_{k^{\prime \prime }}=-\varepsilon _{k^{\prime \prime }kk^{\prime
}}F_{kk^{\prime }}\quad ,\quad F_{kk^{\prime }}=F_{kk^{\prime }}^{j}t^{j}
\end{align}
we have
\begin{align}
\Phi _{S}^{\dag }\left( x\right) \,\overrightarrow{B}\,\Phi _{S}\left(
x\right) & =\left( 
\begin{array}{cc}
\Phi _{s}^{\dag } & \Phi _{s}^{c\,\dag }
\end{array}
\right) \left( 
\begin{array}{cc}
\overrightarrow{B}\, & 0 \\ 
0 & \overrightarrow{B}\,
\end{array}
\right) \left( 
\begin{array}{c}
\Phi _{s} \\ 
\Phi _{s}^{c}
\end{array}
\right)  \\
& =\Phi _{s}^{\dag }\,\overrightarrow{B}\,\Phi _{s}+\Phi _{s}^{c\,\dag }\,
\overrightarrow{B}\,\Phi _{s}^{c} \\
& =0\;.
\end{align}

Before mass is acquired, the Lagrangian for
a higgson field in the electroweak sector contains the terms~\cite{DM2021a}
\begin{align}
\overline{\mathcal{L}}_{H_i}^{0} =\widetilde{H_i}^{\dag } \partial^{\mu}\partial_{\mu} \widetilde{H_i} \;\; , \;\;
\overline{\mathcal{L}}_{H_i}^{Z} =-\frac{g_{Z}^{2}}{4}\widetilde{H_i}^{\dag }Z^{\mu }Z_{\mu }\widetilde{H_i} \;\; , \;\;
\overline{\mathcal{L}}_{H_i}^{W}=-\frac{g^{2}}{2}\widetilde{H_i}^{\dag }W^{\mu +}W_{\mu }^{-}\widetilde{H_i}  
\label{equation173}
\end{align}
where we have modified the notation slightly, with $H \rightarrow \widetilde{H_i}$ for the $i$th species.
Since $\widetilde{H_i}$ has spin $0$ and is real~\cite{DM2021a}, we can write
\begin{align}
\widetilde{H_i} \left( x\right) =H_i \left( x\right) \,\zeta  \quad \mathrm{with} \quad \zeta ^{\,\dag }\,\zeta =1 \; ,
\label{equation174}
\end{align}
where $\zeta $ has $4$ constant components which incorporate the quantum numbers of $\widetilde{H_i} $, and $H_i \left( x\right)$ is again an amplitude mode. Each $\Phi_S$, with four degrees of freedom, has first become four independent 4-component real fields $ \widetilde{H_i}$, and then four 1-component real fields $H_i$. Here we define $H_i \left( x\right) $ to be a higgson field (in a slight change of nomenclature from that of previous papers). It 
has no interactions other than those in the reduced version of (\ref{equation173}):
\begin{align}
\overline{\mathcal{L}}_{H_i}^{Z} =-\frac{g_{Z}^{2}}{4} H_i Z^{\mu }Z_{\mu }H_i \quad , \quad
\overline{\mathcal{L}}_{H_i}^{W}=-\frac{g^{2}}{2} H_i W^{\mu +}W_{\mu }^{-}H_i  \; .
\label{equation175}
\end{align}
Each such higgson field can then be treated (and quantized) like a standard real scalar field, but with no quantum numbers and no interactions except those of of (\ref{equation175}).

Some further discussion of both Higgs and higgson modes is given in Refs.~\cite{DM2021a} and \cite{DM2021b}. Note that (i) the second term in  (\ref{xi}) is obtained for all scalar boson fields 
and (ii) cross terms in the full $\slashed{D}^{2}$ of (\ref{full}) below ultimately give zero in all physical fields, because the factor involving $\Sigma_{\alpha \beta }$ (containing Pauli matrices) produces cancellation in Higgs fields and the factor involving $A_{\mu}$ produces cancellation in higgson fields. For higgson fields we then have, in a general coordinate system,
\begin{align} 
\overline{\mathcal{L}}_{H_i} =  H_i  \left(  g^{\mu \nu } \nabla_{\mu}\nabla_{\nu} - \frac{1}{4} R \right) H_i  + \overline{\mathcal{L}}_{H_i}^{Z} + \overline{\mathcal{L}}_{H_i}^{W}   \; .
\label{higgson}
\end{align}

\section{\label{sec:sec8}Fundamental action for fermions, scalar bosons, and \\
auxiliary fields}

When the various components in the preceding sections are assembled, the total action for fermion fields $\psi _{f}$, Higgs fields $\phi_R$,  auxiliary fields $F_R$, and higgson fields $H_i$ is
\begin{eqnarray}
 S_{matter}  =\int d^{4}x\, \overline{\mathcal{L}}_{matter} 
 \label{eq8.3a}
 \end{eqnarray}
 \begin{align}
 \hspace{-0.25cm} 
\overline{\mathcal{L}}_{matter} = 
\psi _{f}^{\dagger } \left( x \right) 
ie_{\alpha }^{\mu }\,\sigma ^{\alpha } 
 D_{\mu }\,\psi _{f}\left( x\right) +  \sum_R  \phi_R ^{\dagger } \left( x\right)    \left( g^{\mu \nu } D_{\mu}  D_{\nu } -  \frac{1}{4} R \right)  \phi_R \left( x\right) 
 + \sum_R F_R^{\dagger }\left( x\right) F_R\left( x\right)  
  +\sum_i \overline{\mathcal{L}}_{H_i}
\label{eq8.3}
\end{align}
after transformation to a general coordinate system, and before masses and further interactions result from symmetry breakings and other effects.  $D_{\mu }$ should now be interpreted as the full covariant derivative
\begin{align} 
D_{\mu } = \nabla_{\mu} - i A_{\mu} 
\label{full}
\end{align}
where $\nabla_{\mu}$ and $A_{\mu}$ can be regarded as operators which yield the appropriate generators and potentials for each representation. 

The spin 1/2 fermion fields in $\psi _{f}$ span the various physical
representations of the most fundamental gauge group, which must be $SO(D-4)$ in
the present theory. (More precisely, the group is $Spin(D-4)$, but $SO(D-4)$
is conventional terminology.) $F_R$ is the amplitude of a component of $\mathcal{F}$, defined in the same way as $\phi _R$.  The last term in (\ref{eq8.3}) contains the action for all higgson fields, including those which lie outside the electroweak sector. We thus obtain the basic form for a Lorentz-invariant, gauge-invariant, and supersymmetric action, with the addition of a new kind of particle. Notice, however, that susy is broken by the ``condensation'' of the negative-action $F$ modes of (\ref{eq7.32b}).

The gravitational and gauge curvatures of  (\ref{eq8.7}) and (\ref{eq8.8}) below must ultimately originate from a background (in the path integral) of ``rapidly fluctuating'' $4$-dimensional topological defects (analogous to vortices and vortex rings, or extended and closed flux tubes) associated with the gauge potentials of (\ref{eq5.23}) and the vierbein of (\ref{eq4.40}). Here ``rapidly fluctuating'' means that the $A_{\mu}$ and 
$g_{\mu \nu}$ (or $e_{\alpha }^{\mu }$) are actually averages over many topological configurations of the field $\Psi_0$. As mentioned below (\ref{eq4.11b}), the path integral over all these configurations is replaced by a path integral over the $A_{\mu}$ and $g_{\mu \nu}$ with an effective action. The gauge and gravitational fields then vary over all possibilities, and this is how these force fields are quantized in the present theory. To discuss the topological defects in detail -- with examples representing various physical phenomena associated with abelian and nonabelian gauge fields --  is beyond the scope of this paper, but all that is required here is the fact that topological defects permit the curvature associated with the the gauge potentials and vierbein to be nonzero. I.e., the gauge curvature (in a locally inertial frame, and with the coupling constant no longer absorbed into $A_{\mu} $)
\begin{align}
F^i_{\mu \nu} = \partial_{\mu}A^i_{\nu} - \partial_{\nu}A^i_{\mu} +g f^i_{j \, k} A^j_{\mu} A^k_{\mu}
\label{curv-1}
\end{align}
and the gravitational spin connection and curvature (see p.~274 of \cite{GSW})
\begin{align}
& \omega_{\mu}^{\alpha \beta} = \frac{1}{2}e^{\nu \alpha} \left( \partial_{\mu} e^{\beta}_{\nu} - \partial_{\nu} e^{\beta}_{\mu} \right) - \frac{1}{2}e^{\nu \beta} \left( \partial_{\mu} e^{\alpha}_{\nu} - \partial_{\nu} e^{\alpha}_{\mu} \right)
- \frac{1}{2}e^{\rho \alpha} e^{\sigma \beta}  \left( \partial_{\rho} e_{\sigma \gamma} - \partial_{\sigma} e_{\rho \gamma}  \right) e^{\gamma}_{\mu} \\
& R_{\mu \nu \;\;  \beta}^{ \;\; \; \;  \alpha} = \partial_{\mu} \omega_{\nu \;\;  \beta}^{ \;\;   \alpha} - \partial_{\nu} \omega_{\mu \;\;  \beta}^{ \;\;   \alpha} + \left[\omega_{\mu }  , \omega_{\nu } \right]^{\alpha}_{\;\; \beta}
\quad , \quad
R_{\mu \nu \;\;  \beta}^{ \;\; \; \;  \alpha} = e^{\alpha}_{\sigma} e^{\tau}_{\beta} R_{\mu \nu \;\;  \tau}^{ \;\; \; \;  \sigma}
\label{curv-4}
\end{align}
originate from vortex-like configurations in the same way that the vorticity of a superfluid 
\begin{align}
\omega^s_{k \ell} =\partial_k v^s_{\ell} - \partial_{\ell} v^s_k \quad  \mathrm{or} \quad \omega^s_z =\partial_x v^s_y - \partial_y v^s_x
\label{curv-5}
\end{align}
originates from ordinary vortices, as pointed out by Feynman and Onsager. The simplest case is a magnetic field with
\begin{align}
B_z = \partial_x A_y - \partial_y A_x
\label{curv-6}
\end{align}
but all the force fields above have the same basic form. In each case, the curvature is nonzero in a region penetrated by flux lines that are interpreted as vortex lines.

\section{\label{sec:sec8+}Cosmological constant, Einstein-Hilbert action, \\ black hole entropy, and dark matter}

\subsection{\label{sec:cos}Cosmological constant and Einstein-Hilbert action}

In conventional physics, the contribution of fermion and scalar boson fields to the vacuum energy corresponds to  a Lagrangian $e \, \overline{\mathcal{L}}_{vac} $, where $\overline{\mathcal{L}}_{vac} $ is constant but $e$ is given by (\ref{eq8.4b}). The resulting gravitational energy-momentum (or stress-energy) tensor is
\begin{eqnarray}
T^{\mu \nu }_{vac}=2 e^{-1} \delta \left( e \overline{\mathcal{L}}_{vac} \right) / \delta g_{\mu \nu}  = g^{\mu \nu} \overline{\mathcal{L}}_{vac} \quad , \quad
\overline{\mathcal{L}}_{vac} = - \left( 8\pi \ell_{P}^2  \right)^{-1}   \Lambda 
 \label{lambda}
\end{eqnarray}
since
\begin{eqnarray}
\delta e = \frac{1}{2} \, e \,  g^{\mu \nu} \, \delta g_{\mu \nu}
 \label{lvariation}
\end{eqnarray}
and this produces a term $\Lambda g_{\mu \nu} $ on the left-hand side of the Einstein field equations, with a cosmological constant $\Lambda$. 

In (\ref{eq8.3}), the coupling of matter to gravity in $\overline{\mathcal{L}}_{matter}$ is very nearly
the same as in standard general relativity, 
but there is no factor of $e$ 
in the integrand of (\ref{eq8.3a}). This means that, for a fixed vacuum energy density due to fermions and scalar bosons, 
\begin{eqnarray}
T^{\mu \nu }_{vac}=2 e^{-1} \delta \overline{\mathcal{L}}_{vac} / \delta g_{\mu \nu}=0
\label{vac}
\end{eqnarray}
so there is no direct contribution to a cosmological constant from these fields.

The predictions of the present theory are identical to those of standard general relativity for the motion of all particles and waves in gravitational fields, and for gauge bosons acting as a source of gravity. We will now show that they are also the same for matter acting as a gravitational source. The action of (\ref{eq8.3a}) begins in the initial coordinate system. Let $\varphi$ represent any of the matter fields in (\ref{eq8.3}), with an action having the form
\begin{eqnarray}
S_{\varphi}=\int d^4 x \, \overline{\mathcal{L}}_{\varphi} \quad , \quad 
\overline{\mathcal{L}}_{\varphi} = \varphi^{\dag} A_{\varphi} \varphi \; .
 \label{varphi}
\end{eqnarray}
The contribution of $\varphi$ to the gravitational energy-momentum tensor in the present theory is then
\begin{align}
T^{\mu \nu }_{\varphi} =  2e^{-1}  \frac{\delta \overline{\mathcal{L}}_{\varphi}} {\delta g_{\mu \nu }} = 2 \, e^{-1} \varphi^{\dag} \frac{\delta A_{\varphi} } {\delta g_{\mu \nu} } \varphi 
= 2 \, \overline{\varphi}^{\dag} \frac{\delta A_{\varphi} } {\delta g_{\mu \nu} }  \overline{\varphi} \quad , \quad \overline{\varphi} = e^{-1/2}\varphi \; .
\label{var1}
\end{align}

The usual energy-momentum tensor, with
\begin{eqnarray}
S_{\varphi'}=\int d^4 x \, \mathcal{L}_{\varphi'} \quad . \quad \mathcal{L}_{\varphi'} = e \overline{\mathcal{L}}_{\varphi'} \quad , \quad 
\overline{\mathcal{L}}_{\varphi'} = \varphi'^{\dag} A_{\varphi'} \varphi'
\label{usual}
\end{eqnarray}
is
\begin{eqnarray}
 T^{\mu \nu }_{\varphi'}=
2 \, e^{-1} \frac{\delta \left( e \overline{\mathcal{L}}_{\varphi'} \right) }{ \delta g_{\mu \nu } }= 2 \, \overline{\mathcal{L}}_{\varphi'} \, e^{-1} \frac{\delta e }{\delta g_{\mu \nu }} +
2 \frac{\delta \overline{\mathcal{L}}_{\varphi'} }{ \delta g_{\mu \nu }} = 2 \frac{\delta \overline{\mathcal{L}}_{\varphi'}}{ \delta g_{\mu \nu } } 
= 2 \, \varphi'^{\dag} \frac{\delta A_{\varphi'} } {\delta g_{\mu \nu} }  \varphi'  
\label{eq8.7x}
\end{eqnarray}
since the bilinear forms of (\ref{eq8.3}) -- or similar forms including mass and interaction terms, or for composite objects like protons or planets -- imply that $\overline{\mathcal{L}}_{\varphi'} = 0 $ if the classical equations of motion are satisfied. 
Classical in the present context means that particles or composite bodies remain on the mass shell, and the energy-momentum tensor here corresponds to the energy and momentum of quantum fields, particles, or composite objects satisfying their quantum equations of motion, which yield classical trajectories according to Ehrenfest's theorem. 

There is agreement between (\ref{var1}) and (\ref{eq8.7x}) if $\overline{\varphi}$ can be identified with $\varphi'$. According to (\ref{varphi}), (\ref{var1}), and (\ref{usual}), this means that 
\begin{align}
e  \overline{\varphi}^{\dag} A_{\varphi}  \overline{\varphi} = \varphi^{\dag} A_{\varphi} \varphi \quad \mathrm{or} \quad A_{\varphi}  e^{-1/2} \varphi = e^{-1/2} A_{\varphi}  \varphi \; .
\label{var3}
\end{align}
The only part of an operator $A_{\varphi} $ in (\ref{eq8.3}) that does not immediately commute with $ e^{-1/2}$ is the covariant derivative $D_{\mu}$. But the covariant derivative of a function of only the vierbein is zero, so
\begin{align}
D_{\mu} e^{-1/2} \varphi = [D_{\mu} e^{-1/2} ] \varphi +  e^{-1/2} D_{\mu} \varphi = e^{-1/2} D_{\mu} \varphi 
\label{var4}
\end{align}
and (\ref{var3}) is satisfied.

If the $\overline{\varphi} $ are renamed and called $\varphi $, then (\ref{eq8.3}) holds with  (\ref{eq8.3a}) modified to
\begin{equation}
 S_{matter}  = \int d^{4}x\, e \, \overline{\mathcal{L}}_{matter} 
 \label{eq8.3-mod}
 \end{equation}
 but the stress-energy tensor must be calculated from 
 \begin{align}
T^{\mu \nu }_{\varphi} =  2 \, \varphi^{\dag} \frac{\delta A_{\varphi} } {\delta g_{\mu \nu} }  \varphi  \quad \mathrm{or} \quad 
T^{\varphi }_{\mu \nu } =  -2 \, \varphi^{\dag} \frac{\delta A_{\varphi} } {\delta g^{\mu \nu} }  \varphi 
\label{var-final}
\end{align}
and \textit{not} from $T^{\mu \nu }_{\varphi} = 2 \, e^{-1} \delta \left( e \overline{\mathcal{L}}_{\varphi} \right) / \delta g_{\mu \nu }$ (unless it is recognized that $\varphi$ now contains a hidden factor of $e^{-1/2} $). As shown above, the two formulas give the same result for particles satisfying their classical equations of motion, but not in general, and not for the zero-point (vacuum) action of fermions and scalar bosons (if used in the conventional way). Eq. (\ref{var-final}) is, in fact, a natural definition of the energy-momentum tensor in a quantum description, and more familiar classical forms can be obtained by using the classical equation of motion and integration by parts.

We might note that the shift from the initial fields of Section~\ref{sec:sec4} to the redefined fields of (\ref{eq8.3-mod}) (i.e. the fields of (\ref{eq8.3}) after reinterpretation), could have been made immediately after (\ref{eq4.43}), with the understanding that the effect of the $e^{-1/2}$ factors absorbed into the fields cancels that of the external factor of $e$ when $\delta / \delta g_{\mu \nu}$ is applied to the action.

If the metric tensor $g_{\mu \nu}$ is taken to be fixed -- e.g. if all other fields are taken to evolve on a fixed classical gravitational background -- then these fields can be treated via path-integral quantization in the usual way, with only the measure changed by the extra factor of $e^{-1/2}$ in the fermion and scalar boson fields.

The predictions of general relativity are thus unchanged in the present picture, except that the cosmological constant of (\ref{vac}) is zero if the vacuum Lagrangian density $\overline{\mathcal{L}}_{vac}$ is taken to be fixed.

On the other hand, $\overline{\mathcal{L}}_{vac}$ 
is not really fixed, since the fields in the vacuum will respond to variations in
the gauge potentials of (\ref{eq5.23}) and the vierbein of (\ref{eq4.40}) 
(or metric tensor of (\ref{eq8.3z})). 
In Appendix \ref{sec:appG} the Einstein-Hilbert action of gravity
\begin{eqnarray}
\mathcal{L}_{G}= \left( 16\pi \ell_{P}^2 \right)^{-1} e \, R \; ,
\label{eq8.7}
\end{eqnarray}
where $\ell _{P}^{2}= G$, is obtained as the vacuum response to the curvature of the vierbein (or metric tensor).

Similarly, we conjecture that the 
``diamagnetic'' response of  the vacuum fermion and scalar-boson fields to gauge curvature gives rise to the Maxwell-Yang-Mills action, with 
\begin{eqnarray}
\mathcal{L}_{g}=-\frac{1}{4}g_{0}^{-2} \, e \, 
g^{\mu \rho }g^{\nu \sigma } \,F_{\mu \nu }^{i}F_{\rho \sigma}^{i} 
\label{eq8.8}
\end{eqnarray}
where $g_{0}$ is the coupling constant for the fundamental gauge group. This form is analogous to the shift in the free energy
\begin{align}
\Omega \left( B \right) - \Omega \left( 0 \right) =  \frac{e^2 A}{24 \pi m c^2} B^2 \; ,
\end{align}
when electrons in a metal (with area $A$) respond to an applied magnetic field $\overrightarrow{B}$, exhibiting Landau diamagnetism. 

It appears that it is nontrivial to obtain (\ref{eq8.8}) from a proper calculation, and that a detailed knowledge of the vacuum states is required. However, (i) those states will surely be modified when they are perturbed by the curvature of gauge fields (as are the states of electrons in a metal), (ii)  (\ref{eq8.8}) has the simplest form consistent with the symmetries of the vacuum (including invariance under coordinate, Lorentz, and gauge transformations), and (iii) within the present picture (\ref{eq8.8}) can originate only from the response of the vacuum to external gauge fields. With this interpretation, $\mathcal{L}_{g}$ must necessarily vanish when these fields vanish -- i.e., in the vacuum itself:
\begin{eqnarray}
\langle \mathcal{L}_{g} \rangle_{vac} = 0 \; .
\label{eq8.9x}
\end{eqnarray}
This means that when (\ref{eq8.8}) is quantized, the field operators must be normal-ordered. It follows that there is no cosmological constant resulting from the gauge fields. On the other hand, virtual processes will still be affected by a change in their boundary conditions; a detailed treatment of this aspect, and of the observed Casimir effect~\cite{Jaffe,casimir1,casimir2},  would be inappropriately long  here, but see the discussion of this point in Section \ref{sec:sec1a}.

\subsection{\label{sec:BH}Bekenstein-Hawking entropy of black holes}

There have been many attempts to understand the Bekenstein-Hawking entropy $S_{BH}$ of black holes in terms of microscopic degrees of freedom, but none has provided a convincing explanation for even the simplest of physical black holes. On the other hand, Gibbons and Hawking~\cite{Gibbons-Hawking} have shown that, for a static black hole, an expression equal to the Euclidean action has the form required of $S_{BH}$. In the present picture, every fundamental Lorentzian action $S_L$ can alternatively be interpreted as the negative of a Boltzmann entropy $S$, determined by counting microstates (of dits) as in Sections \ref{sec:sec1a} and \ref{sec:sec2}. The Lorentzian action of a general system has the form $S_L = \int dt \, (T-V)$, and the Eucidean action the form $S_E = \int dt \, (T+V)$, where $T$ includes the time derivatives. For a static system, we have $T=0$ and the first term vanishes in both expressions. It follows that
\begin{eqnarray}
S_E = -S_L = S \quad \mathrm{for \ a \ static \ black \ hole} \; .
\label{Euclid}
\end{eqnarray}
 The entropy $S$ originates from the microstates of the gravitational field configuration called a black hole, so it has the same status as the entropy of any static thermodynamic system.
 
 The action of a rotating black hole contains an additional contribution from the angular momentum, with the same expression for the entropy.

\subsection{\label{sec:DM2}Dark matter}

There are a vast number of hypothetical dark matter
candidates, most of which do not have well-defined masses or couplings, and
many of which have already been ruled out by experiment -- or at least found to be subdominant species in a multicomponent scenario. For example, the simplest susy models which have ``natural'' values for the
parameters, and which are also compatible with limits from the LHC, are
found to be in disagreement with both the abundance of dark matter and the
limits from direct-detection 
experiments~\cite{Baer-Barger-2016,Baer-Barger-2018,Roszkowski-2018,Baer-Barger-2020,Tata-2020} -- if the lightest supersymmetric particle (LSP) is assumed to be the dominant constituent. But, as mentioned in Section \ref{sec:sec1a}, the present picture requires susy at some energy scale, and the LSP (as a subdominant component) can stably coexist with the present dark matter candidate (the lightest higgson).

Fortunately the enthusiasm of experimentalists searching for new physics has not been notably diminished by the past lack of success, and work on new facilities and capabilities has persisted even though the pandemic. For example, LZ, XENONnT, and PandaX-4T should all soon be taking data with much larger detectors than those of past experiments. They should ultimately be able to detect a dark matter WIMP with a mass of $\sim 50$ GeV/c$^2$ if the Xe collision cross-section is larger than about $1.4 \times 10^{-48}$ cm$^2$~\cite{LZ,XENON,PandaX}, within roughly the next 5 years. For the WIMP predicted by the present theory -- the lowest-mass higgson of Section \ref{sec:DM} -- this cross-section is slightly below  $10^{-47}$ cm$^2$, with a mass of about $72$ GeV/c$^2$.

More generally, this dark matter candidate is consistent with all current experiments, and observable in the near or foreseeable future through a wide variety of direct, indirect, and collider detection experiments. To review the conclusions of Refs.~\cite{DM2021a} and \cite{DM2021b}: This particle is unique in that it has (i) precisely defined couplings and (ii) a well-defined mass of about 72 GeV/c$^2$, providing specific cross-sections and other experimental signatures as targets for clean experimental tests. It has not yet been detected because it has no interactions other than second-order gauge couplings, to $W$ and $Z$ bosons. However, these weak couplings are still sufficient to enable observation by direct detection experiments which should be fully functional within the next few years, including XENONnT, LZ, and PandaX. The cross-section for collider detection at LHC energies is small -- roughly 1 femtobarn -- but observation may ultimately be achievable at the high-luminosity LHC, and should certainly be within reach of the even more powerful colliders now being planned. It is possible that the present dark matter candidate has already been observed via indirect detection: Several analyses of gamma rays from the Galactic center, observed by Fermi-LAT, and of antiprotons, observed by AMS-02, have shown consistency with the interpretation that these result from annihilation of dark matter particles having approximately the same mass and annihilation cross-section as the present candidate. Finally, there is consistency with the observations of Planck, which have ruled out many possible candidates with larger masses.

\section{\label{sec:sec9}Conclusion}

Starting with the simplest imaginable picture, and interpreting our universe as the product of two spaces with topological singularities, we obtain the following results: 
4-dimensional spacetime 
with one time coordinate; spin 1/2 fermion and spin zero boson fields defined on this spacetime; path-integral quantization of these fields; gauge fields and a fundamental gauge theory which is necessarily $SO(N)$; correct couplings of matter fields to the gauge fields; a gravitational vierbein; correct couplings of matter fields to gravity; Lorentz invariance; supersymmetry at some energy scale; elimination of the usual enormous cosmological constant; the Einstein-Hilbert action for gravity; the Bekenstein-Hawking entropy of black holes; and a new set of particles, including a new dark matter WIMP which should be detectable in the near future.

\appendix


\section{\label{sec:appA}The internal space}

The internal space of Section \ref{sec:sec5} is $(D-4)$-dimensional, with 
an $SO(D-4)$ (or more precisely $Spin(D-4)$) rotation group 
and its vector, spinor, etc. representations -- for example, the 
$\mathbf{10}$ and $\mathbf{16}$ representations when $D-4=10$. It may be
helpful to begin with an analogy, however, in which external spacetime is
replaced by the $z$-axis. The 
internal space is replaced by an $xy$-plane, with internal states described
by $2$-dimensional vector fields (rather than the higher-dimension vector
and spinor fields considered below). One of these states is occupied by the
condensate, and is represented by a vector $\boldsymbol{v}_{1}$ which
points radially outward from the origin at all points in the
$xy$-plane when $z=0$. 
The other state is an additional basis function, represented
by a vector $\boldsymbol{v}_{2}$ which is everywhere perpendicular to 
$\boldsymbol{v}_{1}$. But $\boldsymbol{v}_{1}$ is allowed to rotate
as a function of $z$, so it has both radial and tangential components after
a displacement along the $z$-axis. Then $\boldsymbol{v}_{2}$ is forced
to rotate with $\boldsymbol{v}_{1}$ -- i.e., the condensate -- in order
to preserve orthogonality.

Now let us turn to the actual internal space, first considering a
set of $(D-4)$-dimensional vector fields $\widetilde{\psi }_{vec}^{r}$. Let 
$\widetilde{\psi }_{vec}^{0}$ represent the state occupied by a bosonic
condensate. In the simplest picture, and at some fixed $x_{0}^{\mu }$, only
the $r\,$th component of the field $\widetilde{\psi }_{vec}^{r}$ is nonzero
along some radial direction in the internal space, making the fields
trivially orthogonal in that direction. Then, with $x^{\mu }$ still fixed, 
$\widetilde{\psi }_{vec}^{r}\left( x^{m}\right) $ in all other radial
directions is obtained from the original 
$\widetilde{\psi }_{vec}^{r}\left( x_{0}^{m}\right) $ by rotating it 
to $ x^{m}$. 
In other words, the field at each point in the internal space is identical
to the field that would be obtained at that point 
if the original field $\widetilde{\psi } _{vec}^{r}\left( x_{0}^{m}\right) $ 
were subjected to a rotation about the origin. This produces
an isotropic configuration for the condensate and each basis function. 
As in (\ref{eq4.14}) we can write
\begin{eqnarray}
\widetilde{\psi }_{vec}^{r}\left( x^{m}\right) =U_{vec}\left(
x^{m},x_{0}^{m}\right) \widetilde{\psi }_{vec}^{r}\left( x_{0}^{m}\right) \;.
\label{eq12.1}
\end{eqnarray}
Just as in the analogy, a field that is radial at 
$x_{0}^{m}$ will also be radial at all other points $x^{m}$. However, 
a general $\widetilde{\psi }_{vec}^{r}\left( x_{0}^{m}\right)$ 
permits a general vortex-like configuration of the condensate. 

Also as in the analogy, the state $\widetilde{\psi }_{vec}^{0}$ of the
condensate is allowed to rotate as a function of $x^{\mu }$ (because such 
a rotation does not alter the internal action). Since the other
basis functions $\widetilde{\psi }_{vec}^{r}$ are required to remain
orthogonal to $\widetilde{\psi }_{vec}^{0}$ and each other, they are
required to rotate with the condensate. Then (\ref{eq12.1}) becomes more 
generally
\begin{eqnarray}
\widetilde{\psi }_{vec}^{r}\left( x^{m},x^{\mu }\right) =U_{vec}
\left( x^{m},x_{0}^{m};x^{\mu },x_{0}^{\mu }\right) \widetilde{\psi }
_{vec}^{r}\left( x_{0}^{m},x_{0}^{\mu }\right) 
\label{eq12.2}
\end{eqnarray}
with
\begin{eqnarray}
\widetilde{\psi }_{vec}^{r\,\dag }\left( x^{m},x^{\mu }\right) 
\,\widetilde{\psi }_{vec}^{r^{\prime }}\left( x^{m},x^{\mu }\right) 
=\widetilde{\psi }
_{vec}^{r\,\dag }\left( x_{0}^{m},x_{0}^{\mu }\right) \,\widetilde{\psi }
_{vec}^{r^{\prime }}\left( x_{0}^{m},x_{0}^{\mu }\right) 
= \delta_{r r^{\prime }}
\label{eq12.2a}
\end{eqnarray}
since
\begin{eqnarray}
U_{vec}^{\dag }\left( x^{m},x_{0}^{m};x^{\mu },x_{0}^{\mu }\right)
U_{vec}\left( x^{m},x_{0}^{m};x^{\mu },x_{0}^{\mu }\right) =1 \; .
\label{eq12.2b}
\end{eqnarray}

In general (with $x^{\mu }$ fixed), let 
$\widetilde{\psi }\left( \boldsymbol{x}\right)$ 
represent a multicomponent basis function with
angular momentum $j$ at a point $\boldsymbol{x}$ in the $(D-4)$-dimensional
internal space. After a rotation about the origin 
specified by the $(D-4)\times (D-4)$ matrix $\boldsymbol{R}$, it is transformed to  
\begin{eqnarray}
\widetilde{\psi }^{\prime }\left( \boldsymbol{x}\right) = 
\mathcal{R}\left( \boldsymbol{R}\right) \,\widetilde{\psi }
\left( \boldsymbol{R}^{-1}\boldsymbol{x}\right) 
\label{eq12.1d}
\end{eqnarray}
where $\mathcal{R}\left( \boldsymbol{R}\right) $ belongs to the
appropriate representation 
of the group $Spin\left( D-4\right) $. However, we require that the field be
isotropic, so that it is left unchanged after a rotation:
\begin{eqnarray}
\widetilde{\psi }^{\prime }\left( \boldsymbol{x}\right) 
=\widetilde{\psi }\left( \boldsymbol{x}\right) \;.
\label{eq12.1e}
\end{eqnarray}
Then we can define $\widetilde{\psi }\left( \boldsymbol{x}\right) $ 
at each value of the radial coordinate $r$ by starting with
a $\widetilde{\psi }\left( \boldsymbol{x}_{0}\right) $ and
requiring that  
\begin{eqnarray}
\widetilde{\psi }\left( \boldsymbol{x}\right) =\mathcal{R}
\left( \boldsymbol{R}\right) \,\widetilde{\psi }\left( \boldsymbol{x}
_{0}\right) \quad ,\quad \boldsymbol{x=R\,x}_{0}\;.
\label{eq12.1f}
\end{eqnarray}
With this definition, $\widetilde{\psi }\left( \boldsymbol{x}\right)$
is a single-valued function of the coordinates only if $j$
is an integer. If $j=1/2$, e.g., $\widetilde{\psi }\left( \boldsymbol{x}
\right) $ acquires a minus sign after a rotation of $2\pi $,
but it is single-valued on the $Spin\left( D-4\right) $ group manifold. 

Multivalued functions are 
well-known in other similar contexts, such as the behavior of the
phase of an ordinary superfluid order parameter 
$\psi _{s}=e^{i\theta _{s}}n_{s}^{1/2}$ around a
vortex, which becomes discontinuous if it is required to be a 
single-valued function of the coordinates~\cite{kleinert}. In the same
way, $z^{1/2}$ exhibits a discontinuity across a branch cut if it
is required to be a single-valued function and $z$ is restricted to a 
single complex plane. I.e., 
$z^{1/2} = \left| z \right|^{1/2} e^{i\phi /2}$ gives 
$+\left| z \right|^{1/2}$ for $\phi=0$ and 
$-\left| z \right|^{1/2}$ for $\phi=2 \pi$. But when defined on a pair
of Riemann sheets, $z^{1/2}$ is a continuous function, and 
the same is true of 
$\widetilde{\psi }\left( \boldsymbol{x}\right) $ as we
have defined it above, on the group manifold. The key idea in either 
case is to extend the manifold over which the function is defined, 
so that there are no artificial discontinuities.
A similar principle holds in condensed matter physics, where a spinor 
can be a multivalued function of position (but with physical expectation 
values single-valued).

A vectorial condensate and vectorial basis functions 
are appropriate for the simplest Higgs-like fields and their
superpartners. Similarly, spinorial fields $\widetilde{\psi }_{sp}^{r}$ are
appropriate for ordinary fermions, sfermions, and a possible primordial
condensate occupying a state $\widetilde{\psi }_{sp}^{0}$. (In the present
context, of course, ``vector'' and
``spinor'' refer only to properties in the
internal space.) Again, let $\widetilde{\psi }_{sp}^{r}\left(
x_{0}^{m}\right) $ represent a field along some radial direction in the
internal space at some fixed  $x_{0}^{\mu }$. Then the field configuration 
for every point $x^{m}$ is obtained by taking 
$\widetilde{\psi }_{sp}^{r}\left( x^{m}\right) $ 
to be identical to the field that would be obtained at that point if 
$\widetilde{\psi }_{sp}^{r}\left( x_{0}^{m}\right) $ were subjected 
to a rotation, with 
\begin{eqnarray}
\widetilde{\psi }_{sp}^{r}\left( x^{m}\right) =U_{sp}\left(
x^{m},x_{0}^{m}\right) \widetilde{\psi }_{sp}^{r}\left( x_{0}^{m}\right) 
\label{eq12.3}
\end{eqnarray}
as in (\ref{eq12.1f}).

Again, the state $\widetilde{\psi }_{sp}^{0}$ of the condensate is allowed
to rotate as a function of $x^{\mu }$, and since the other basis functions 
$\widetilde{\psi }_{sp}^{r}$ must remain orthogonal to 
$\widetilde{\psi }_{sp}^{0}$ they are required to rotate 
with the condensate. The general version of (\ref{eq12.3}) is then
\begin{eqnarray}
\widetilde{\psi }_{sp}^{r}\left( x^{m},x^{\mu }\right) =U_{sp}\left(
x^{m},x_{0}^{m};x^{\mu },x_{0}^{\mu }\right) \widetilde{\psi }
_{sp}^{r}\left( x_{0}^{m},x_{0}^{\mu }\right) \;. 
\label{eq12.4}
\end{eqnarray}

The same reasoning applies to each irreducible representation, and thus to
the combined set of fields $\widetilde{\psi }_{int}^{r}\left( x^{m},x^{\mu
}\right) $:
\begin{eqnarray}
\widetilde{\psi }_{int}^{r}\left( x^{\prime \, m},x^{\prime \, \mu }\right)
=U_{int}\left( x^{\prime \, m},x^{m};x^{\prime \, \mu },x^{\mu }\right) 
\widetilde{\psi }_{int}^{r}\left( x^{m},x^{\mu }\right) 
\label{eq12.5}
\end{eqnarray}
with
\begin{eqnarray}
\widetilde{\psi }_{int}^{r\,\dag }\left( x^{\prime \,m},x^{\prime \,\mu
}\right) \,\widetilde{\psi }_{int}^{r^{\prime }}
\left( x^{\prime \,m},x^{\prime \,\mu
}\right) =\widetilde{\psi }_{int}^{r\,\dag }\left( x^{m},x^{\mu }\right) \,
\widetilde{\psi }_{int}^{r^{\prime }}\left( x^{m},x^{\mu }\right) 
= \delta_{r r^{\prime }} \; .
\label{eq12.6}
\end{eqnarray}

So that the internal action will be unaffected as $x^{\mu}
\rightarrow x^{\prime \,\mu }$, we require that the order parameter
experience a uniform rotation, described by a matrix $\overline{\mathcal{R}}
_{int}$ which is independent of $x^{m}$. Then $U_{int}$ has the form
\begin{eqnarray}
U_{int}\left( x^{\prime \, m},x^{m};x^{\prime \, \mu },x^{\mu }\right)  =
\overline{\mathcal{R}}_{int}\left( x^{\prime \,\mu },x^{\mu }\right) \,
\mathcal{R}_{int}\left( x^{\prime \,m},x^{m}\right) \;.
\label{eq12.7}
\end{eqnarray}
(Notice that (\ref{eq12.7}) is to be distinguished from a rotation about 
the origin, which is given by (\ref{eq12.1d}), and which according to 
(\ref{eq12.1e}) would leave $\widetilde{\psi }\left( x^{m}\right) $ 
unchanged rather than rotated at each point $x^{m}$.) It follows that 
\begin{eqnarray}
\widetilde{\psi }_{int}^{r}\left( x^{m},x^{\mu }\right) =\overline{
\mathcal{R}}_{int}\left( x^{\mu },x_{0}^{\mu }\right) \,\widetilde{\psi 
}_{int}^{r}\left( x^{m},x_{0}^{\mu }\right) \;.
\label{eq12.8}
\end{eqnarray}
At each fixed $x^m$, the order parameter has been rotated as $x_0^{\mu} \rightarrow x^{\mu}$.
We define the parameters $\delta \overline{\phi }_{i}$ by 
\begin{eqnarray}
\overline{\mathcal{R}}_{int}\left( x^{\mu}+\delta x^{\mu },x_{0}^{\mu}
\right) = \overline{\mathcal{R}}_{int}\left( x^{\mu}, x_{0}^{\mu } \right) 
\left( 1-i\,\delta \overline{\phi }_{i}\,J_{i}  \right)
\label{eq12.8b}
\end{eqnarray}
or
\begin{eqnarray}
\delta \widetilde{\psi }_{int}^{r}\left( x^{m}\right) &=& -i\,\delta 
\overline{\phi }_{i}\,\overline{J}_{i}\,\widetilde{\psi }_{int}^{r}
\left( x^{m}\right) \quad \text{as}\quad 
x^{\,\mu }\rightarrow x^{\,\mu }+\delta x^{\,\mu }
\label{eq12.9} \\
\overline{J}_{i} &=& 
\overline{\mathcal{R}}_{int}\left( x^{\mu}, x_{0}^{\mu } \right) J_{i} 
\,\overline{\mathcal{R}}_{int}^{-1}\left( x^{\mu}, x_{0}^{\mu } \right)
\label{eq12.9a} 
\end{eqnarray}
where the matrices $J_{i}$ are the generators in the reducible representation 
of $Spin(D-4)$ corresponding to $\widetilde{\psi }_{int}^{r}$.  
The matrix elements of $\overline{J}_{i}$ are independent of
$x^{\mu}$:
\begin{eqnarray}
\hspace{-1.6cm} \int d^{D-4}x\,\widetilde{\psi }_{int}^{r\dagger } 
\left( x^{m},x^{\mu }\right) \overline{J}_{i} \,
\widetilde{\psi }_{int}^{r^{\prime}} \left( x^{m},x^{\mu }\right)
= \int d^{D-4}x\,\widetilde{\psi }_{int}^{r\dagger } 
\left( x^{m},x_{0}^{\mu }\right) J_{i} \,
\widetilde{\psi }_{int}^{r^{\prime}} \left( x^{m},x_{0}^{\mu }\right) 
\; .
\label{eq12.9b}
\end{eqnarray}
The primordial condensate is in a specific representation, but the
basis functions in other representations are chosen to rotate with it 
according to (\ref{eq12.8}) and (\ref{eq12.9}). 

It may be helpful to illustrate the above ideas by returning to the
$2$-dimensional analogy. Equation (\ref{eq12.1f}) becomes  
\begin{eqnarray}
\hspace{-1.5cm}
\boldsymbol{v}\left( \boldsymbol{x}\right) =\mathcal{R}_{vec}\,
\boldsymbol{v}\left( \boldsymbol{x}_{0}\right) \;,\;\mathcal{R}_{vec}
=\left( 
\begin{array}{cc}
\cos \phi  & -\sin \phi  \\ 
\sin \phi  & \cos \phi 
\end{array}
\;\right) \;,\;\boldsymbol{v}\left( \boldsymbol{x}_{0}\right) =\left( 
\begin{array}{c}
R\left( r\right)  \\ 
0
\end{array}
\right) \;\mathrm{or}\;\left( 
\begin{array}{c}
0 \\ 
R\left( r\right) 
\end{array}
\right) 
\label{eq12.10}
\end{eqnarray}
for the vector representation and 
\begin{eqnarray}
s\left( \boldsymbol{x}\right) =\mathcal{R}_{sp}\,s\left( \boldsymbol{x}
_{0}\right) \;,\;\mathcal{R}_{sp}=e^{-i\sigma _{3}\phi /2}\;,\;s\left( 
\boldsymbol{x}_{0}\right) =\left( 
\begin{array}{c}
R\left( r\right)  \\ 
0
\end{array}
\right) \;\mathrm{or}\;\left( 
\begin{array}{c}
0 \\ 
R\left( r\right) 
\end{array}
\right) 
\label{eq12.11}
\end{eqnarray}
for the spinor representation. The matrices corresponding to the 
$J_{i}$ are 
\begin{eqnarray}
J_{vec}=\left( 
\begin{array}{cc}
0 & -i \\ 
i & 0
\end{array}
\;\right) \quad \text{and }\quad J_{sp}=\frac{\sigma _{3}}{2}=\frac{1}{2}
\left( 
\begin{array}{cc}
1 & 0 \\ 
0 & -1
\end{array}
\;\right) \;.
\label{eq12.13}
\end{eqnarray}

Notice that $\phi_{i}$ is an angular coordinate in the internal space,
whereas $\overline{\phi }_{i}$ is a parameter specifying the rotation of 
$\widetilde{\psi }_{int}^{r}$ at fixed $x^{m}$ as $x^{\mu }$ is varied.

\section{\label{sec:appB}Solutions in the internal space}

Our goal in this appendix is merely to show that there are solutions with
the form required in Appendix \ref{sec:appA}, so we will look first for 
solutions with the higher-derivative 
terms in (\ref{eq5.3}) and (\ref{eq5.4}) neglected, and with $\Psi _{int}$
sufficiently small that $V_{0} \left( x^{m}\right) $ 
can also be neglected. Then (\ref{eq5.3}) and (\ref{eq5.4}) become  
\begin{eqnarray}
\left( -\frac{1}{2m_{0}}\partial _{m}\partial _{m}-\mu _{int}
\right)\Psi _{int}\left( x^{m},x^{\mu }\right) 
= 0 \\
\left( -\frac{1}{2m_{0}}
\partial _{m}\partial _{m}-\mu _{int}\right) \widetilde{\psi }
_{int}^{r}\left( x^{m},x^{\mu }\right) =0 \; .
\label{eq11.1}
\end{eqnarray}

For simplicity of notation, let $\widetilde{\psi }_{int}^{r}\left(
x^{m},x^{\mu }\right) $ again be represented by 
$\widetilde{\psi }\left( \boldsymbol{x} \right) $, with components 
$\widetilde{\psi }_{p}\left( \boldsymbol{x} \right) $. Each component 
varies with position in the way specified by (\ref{eq12.1f}) (together
with the radial dependence of 
$\widetilde{\psi }\left( \boldsymbol{x} _{0}\right) $). 
It therefore has a kinetic energy given by 
$-\left( 2m_{0} \right)^{-1} \partial_{m} \partial_{m}
\widetilde{\psi }_{p}\left( \boldsymbol{x}\right) $, and an orbital
angular momentum given by the usual orbital angular momentum 
operators $\widehat{J}_{i}$ in $\bar{d}$ 
dimensions~\cite{narumi,gallup,louck,herrick,dong,lee}, which essentially
measure how rapidly 
$\widetilde{\psi }_{p}\left( \boldsymbol{x} \right) $ 
varies as a function of the angles $\phi _{i}$.

The Laplacian $\partial _{m}\partial _{m}$ can be rewritten in terms of 
radial derivatives and the usual $\widehat{J}^{2}$, 
giving~\cite{narumi,gallup,louck}  
\begin{eqnarray}
\left( -\frac{1}{r^{2K}}\frac{\partial}{\partial r}\left( r^{2K}
\frac{\partial}{\partial r}\right) +\frac{
\widehat{J}^{2}}{r^{2}}-1\right) \widetilde{\psi }_{p}\left( \boldsymbol{x}
\right) =0\quad ,\quad K=\frac{\bar{d}-1}{2}
\label{eq11.3}
\end{eqnarray}
after rescaling of the radial coordinate $r$, where
\begin{eqnarray}
\bar{d}=D-4 \; .
\label{equ11.3a}
\end{eqnarray}
 In addition, it is shown in
Narumi and Nakau~\cite{narumi}, Gallup~\cite{gallup}, and Louck~\cite{louck} that
\begin{eqnarray}
\widehat{J}^{2}\widetilde{\psi }_{p}\left( \boldsymbol{x}\right) 
=j\left( j+\bar{d}-2\right) \widetilde{\psi }_{p}\left( \boldsymbol{x}\right)
\label{eq11.4}
\end{eqnarray}
where $j$ is the orbital angular momentum quantum number, as defined on 
p. 677 of Gallup, but with this definition extended to 
half-integer values of $m_{\alpha}$ and $j$. Normally, of course, only 
integer values of these orbital quantum numbers are permitted. 
However, the functions $\widetilde{\psi }_{p}\left( \boldsymbol{x}\right) $
as defined in \ref{sec:appA} can have $j=1/2$ etc. (in which case 
they are multivalued functions of the coordinates but single-valued
functions on the group manifold, as discussed below (\ref{eq12.1f})). 
Also, the demonstration of (\ref{eq11.4}) of Gallup~\cite{gallup} 
can be extended in the present context to half-integer 
$j$, because it employs raising and lowering operators. 
(At each $\boldsymbol{x}$, $\widetilde{\psi }_{p}$ is a linear combination 
of states with different values of $m_{\alpha}$, but (\ref{eq11.4})
still holds.) For each 
$\widetilde{\psi }_{p}\left( \boldsymbol{x}\right) $
the radial wavefunction then satisfies 
\begin{eqnarray}
\left[ -\frac{1}{r^{2K}}\frac{d}{dr}\left( r^{2K}\frac{d}{dr}\right) 
+\frac{j\left( j+\bar{d}-2\right) }{r^{2}}-1\right] R\left( r\right) =0\;.
\label{eq11.5}
\end{eqnarray}

It may be helpful once again to consider the $2$-dimensional analogy 
of \ref{sec:appA}, where the orbital angular momentum operator 
is
\begin{eqnarray}
\widehat{J}=-i \partial / \partial \phi \; .
\label{eq12.12}
\end{eqnarray}
For the vector representation, (\ref{eq12.10}) implies that
the kinetic energy is given by  
\begin{eqnarray}
\partial _{m}\partial _{m}\boldsymbol{v}\left( \boldsymbol{x}\right)  &=&
\left[ \frac{1}{r}\frac{\partial}{\partial r}
\left( r\frac{\partial}{\partial r}\right) + 
\frac{1}{r^{2}}\frac{\partial ^{2}}{\partial \phi ^{2}}\right] \left( 
\begin{array}{cc}
\cos \phi  & -\sin \phi  \\ 
\sin \phi  & \cos \phi 
\end{array}
\;\right) \boldsymbol{v}\left( \boldsymbol{x}_{0}\right)  \\
&=&\left[ \frac{1}{r}\frac{\partial}{\partial r}
\left( r\frac{\partial}{\partial r}\right) - 
\frac{1}{r^{2}}\right] \boldsymbol{v}\left( \boldsymbol{x}\right) 
\label{eq11.5b}
\end{eqnarray}
in agreement with (\ref{eq11.5}) for $j=1$. For the spinor representation, 
(\ref{eq12.11}) gives
\begin{eqnarray}
\partial _{m}\partial _{m}s\left( \boldsymbol{x}\right)  &=&
\left[ \frac{1}{r}\frac{\partial}{\partial r}
\left( r\frac{\partial}{\partial r}\right) + 
\frac{1}{r^{2}}\frac{\partial ^{2}}{\partial \phi ^{2}}\right] e^{-i\sigma
_{3}\phi /2} \, s \left( \boldsymbol{x}_{0}\right)  \\
&=&\left[ \frac{1}{r}\frac{\partial}{\partial r}
\left( r\frac{\partial}{\partial r}\right) - 
\frac{1/4}{r^{2}}\right] s\left( \boldsymbol{x}\right) 
\label{eq11.5d}
\end{eqnarray}
in agreement with (\ref{eq11.5}) for $j=1/2$. 

Equation (\ref{eq11.5}) can be further reduced to~\cite{herrick,lee} 
\begin{eqnarray}
\left[ -\frac{d^{2}}{dr^{2}}+\frac{k\left( k-1\right) }{r^{2}}-1\right] \chi
\left( r\right) =0\quad ,\quad k=j+K=j+\frac{\bar{d}-1}{2}\;
\label{eq11.6}
\end{eqnarray}
where $\chi \left( r\right) \equiv r^{K}R\left( r\right) \;.$
It is then easy to show that 
\begin{eqnarray}
\chi \left( r\right) \propto r^{k}\quad \mathrm{as}\;r\rightarrow 0\quad
,\quad \chi \left( r\right) \propto \sin \left( r + \delta \right)
\quad \mathrm{as}\;r\rightarrow
\infty 
\label{eq11.7}
\end{eqnarray}
where $\delta$ is a phase. 

The higher derivatives in the full internal wave equation (\ref{eq5.4})
permit exponentially decaying solutions which are then normalizable  
and have finite action. Suppose that the above equation at large $r$ is
modified  to
\begin{eqnarray}
\left[ \alpha ^{2}\frac{d^{4}}{dr^{4}}-\frac{d^{2}}{dr^{2}}-1\right] \chi
\left( r\right) =0 \; .
\label{equa11.1}
\end{eqnarray}
The solutions are 
\begin{eqnarray}
\chi \left( r\right) \propto e^{iq\,r}\quad ,\quad q^{2}=-\frac{1}{2\alpha
^{2}}\pm \frac{\sqrt{1+4\alpha ^{2}}}{2\alpha ^{2}} \; .
\label{equa11.2}
\end{eqnarray}
There is then an exponentially decaying solution with the form
$q=i/\bar{\alpha} $ and
\begin{eqnarray}
\chi \left( r\right) \propto e^{-\,r/\bar{\alpha} }
\label{equa11.4}
\end{eqnarray}
so both the order parameter and the basis functions fall to zero as $r \rightarrow \infty$.

\section{\label{sec:appC}Euclidean and Lorentzian Propagators}

For Weyl fermions, the Euclidean 2-point function is
\begin{eqnarray}
&&G_{f}\left( x_{1},x_{2}\right)  
= \left\langle \psi _{f}\left( x_{1}\right)
\psi _{f}^{\dag }\left( x_{2}\right) \right\rangle  
=\frac{\int \mathcal{D}\,\psi _{f}^{\dagger }\,\mathcal{D}\,\psi
_{f}\,\psi _{f}\left( x_{1}\right) \psi _{f}^{\dag }\left( x_{2}\right)
\,e^{-S_{f}}}{\int \mathcal{D}\,\psi _{f}^{\dagger }\,\mathcal{D}\,\psi
_{f}\;e^{-S_{f}}} \label{eq13.1}\\
\hspace{-1.6cm} &=&\frac{\prod\nolimits_{s}
\int d \overline{\psi }_{f}^{\, \ast } \left(
s\right) \int d\overline{\psi }_{f}\left( s\right) \,e^{-\overline{\psi }
_{f}^{\, \ast }\left( s\right) a\left( s\right) \overline{\psi }_{f}\left(
s\right) }\sum\nolimits_{s_{1},s_{2}}\overline{\psi }_{f}\,\left(
s_{1}\right)  \overline{\psi }_{f}^{\, \ast }\left( s_{2}\right) U\left(
x_{1},s_{1}\right) \,U^{\dag }\left( x_{2},s_{2}\right) }{
\prod\nolimits_{s}\int d \overline{\psi }_{f}^{\, \ast }\left( s\right) \int 
d \overline{\psi }_{f}\left( s\right) 
\,e^{-\overline{\psi }_{f}^{\, \ast
}\left( s\right) a\left( s\right) \overline{\psi }_{f}\left( s\right) }} \nonumber
\end{eqnarray}
where (\ref{eq6.2}) and (\ref{eq6.3}) have been used. In a term with  
$s_{2}\neq s_{1}$, the numerator contains the factor
\begin{eqnarray}
\int d \overline{\psi }_{f}^{\, \ast }\left( s_{1}\right) \int d\overline{\psi 
}_{f}\left( s_{1}\right) \,e^{- \overline{\psi }_{f}^{\, \ast }\left(
s_{1}\right) a\left( s_{1}\right) \overline{\psi }_{f}\left( s_{1}\right) }
\overline{\psi }_{f}\left( s_{1}\right) =0
\label{eq13.3}
\end{eqnarray}
according to the rules for Berezin integration. But a term with $s_{2}=s_{1}$
contributes
\begin{eqnarray}
\hspace{-1.0cm} \frac{\int d \overline{\psi }_{f}^{\, \ast }\left( s_{1}\right) \int 
d \overline{\psi }_{f}\left( s_{1}\right) 
\,e^{-\overline{\psi }_{f}^{\, \ast
}\left( s_{1}\right) a\left( s_{1}\right) \overline{\psi }_{f}\left(
s_{1}\right) }\overline{\psi }_{f}\left( s_{1}\right) \overline{\psi }
_{f}^{\, \ast }\left( s_{1}\right) }{\int d\overline{\psi}_{f}^{\, \ast
}\left( s_{1}\right) \int d\overline{\psi }_{f}\left( s_{1}\right) \,e^{-
\overline{\psi }_{f}^{\, \ast }\left( s_{1}\right) a\left( s_{1}\right) 
\overline{\psi }_{f}\left( s_{1}\right) }} \; U\left( x_{1},s_{1}\right)
\,U^{\dag }\left( x_{2},s_{1}\right)  \nonumber \\
=a\left( s_{1}\right) ^{-1}U\left( x_{1},s_{1}\right) \,U^{\dag }\left(
x_{2},s_{1}\right) 
\label{eq13.4}
\end{eqnarray}
so
\begin{eqnarray}
G_{f}\left( x_{1},x_{2}\right) =\sum\nolimits_{s}\overline{G}_{f}\left(
s\right) U\left( x_{1},s\right) \,U^{\dag }\left( x_{2},s\right) 
\quad , \quad 
\overline{G}_{f}\left( s\right) =a\left( s\right) ^{-1} \; .
\label{eq13.5}
\end{eqnarray}
If the $U\left( x,s\right)$ used to represent 
$\psi _{f}\left( x \right)$ are a complete set, the
propagator $G_{f}\left( x,x' \right)$ is a true Green's function: 
\begin{eqnarray}
L_{f} \left( x \right) U\left( x,s\right) = a \left( s\right)
U\left( x,s\right) \quad , \quad 
\psi _{f}\left( x\right) = \sum_{s} U\left( x,s\right) \,
\overline{\psi }_{f}\left( s\right)  
\label{eq13.6a}
\end{eqnarray}
and $\sum\nolimits_{s}
U\left( x,s\right) \,U^{\dag }\left( x',s \right) =
\delta \left( x-x' \right) $ imply that
\begin{eqnarray}
L_{f} \left( x \right) G_{f}\left( x,x' \right) = 
\delta \left( x-x' \right) 
\label{eq13.7}
\end{eqnarray}
as usual.

The treatment for scalar bosons is similar:
\begin{eqnarray}
\hspace{-1.6cm}&&G_{b}\left( x_{1},x_{2}\right)  =\left\langle \phi _{b}\left( x_{1}\right)
\phi _{b}^{\dag }\left( x_{2}\right) \right\rangle  
=\frac{\int \mathcal{D}\,\phi _{b}^{\dagger }\,\,\mathcal{D}\,\phi
_{b}\,\phi _{b}\left( x_{1}\right) \phi _{b}^{\dagger }\,\left( x_{2}\right)
\,e^{-S_{f}}}{\int \mathcal{D}\,\phi _{b}^{\dagger }\,\mathcal{D}\,\phi
_{b}\;e^{-S_{f}}} 
\label{eq13.8} \\
&=&\frac{\prod\nolimits_{s}\int\nolimits_{-\infty }^{\infty }
d \, \mathrm{Re} \, 
\overline{\phi }_{b}\left( s\right) \int\nolimits_{-\infty }^{\infty }
d \, \mathrm{Im} \, 
\overline{\phi }_{b}\left( s\right) \,e^{-\widetilde{a}\left(
s\right) \left[ \left( \mathrm{Re} \,  \overline{\phi }_{b}
\left( s\right) \right)^{2}+\left(\mathrm{Im} \, \overline{\phi }_{b}
\left( s\right) \right) ^{2}\right] }
\sum\nolimits_{s_{1},s_{2}}\overline{\phi }_{b}\left( s_{1}\right) 
\overline{\phi }_{b}^{\, \ast }\,\left( s_{2}\right) }
{\prod\nolimits_{s}\int\nolimits_{- \infty }^{\infty }
d \, \mathrm{Re} \, \overline{\phi }_{b}\left( s\right)
\int\nolimits_{-\infty }^{\infty }
d \, \mathrm{Im} \, \overline{\phi }_{b}\left(
s\right) \,e^{-\widetilde{a}\left( s\right) \left[ 
\left( \mathrm{Re} \, \overline{
\phi }_{b}\left( s\right) \right) ^{2}+
\left( \mathrm{Im} \, \overline{\phi }
_{b}\left( s\right) \right) ^{2}\right] }} \nonumber \\
&& \hspace{8cm}
\times U_{b}\left( x_{1},s_{1}\right)\,  
U_{b}^{\dag }\left( x_{2},s_{2}\right) 
\label{eq13.9}
\end{eqnarray}
where
\begin{eqnarray}
L_{b} \left( x \right) U_{b}\left( x,s\right) = \widetilde{a}\left( s\right)
U_{b}\left( x,s\right) \quad , \quad 
\phi _{b}\left( x\right) = \sum_{s} U_{b}\left( x,s\right) \,
\overline{\phi }_{b}\left( s\right)    \; .
\label{eq13.9a}
\end{eqnarray}

In a term with $s_{2}\neq s_{1}$, the numerator contains the factor
\begin{eqnarray}
\int\nolimits_{-\infty }^{\infty }
d \, \mathrm{Re} \, \overline{\phi }_{b}\left(
s_{1}\right) \int\nolimits_{-\infty }^{\infty }
d \, \mathrm{Im} \, \overline{\phi }
_{b}\left( s_{1}\right) \,e^{-\widetilde{a}\left( s_{1}\right) \left[ \left( 
\mathrm{Re} \, \overline{\phi }_{b}\left( s_{1}\right) \right) ^{2}
+\left( \mathrm{Im} \, 
\overline{\phi }_{b}\left( s_{1}\right) \right) ^{2}\right] }
\left[\mathrm{Re} \, \overline{\phi }_{b}\left( s_{1}\right) 
+i \, \mathrm{Im} \, \overline{\phi }_{b}\left( s_{1}\right) \right] 
\nonumber \\
 =0 \hspace{1cm}
 \label{eq13.10}
\end{eqnarray}
since the integrand is odd. But a term with $s_{2}=s_{1}$ contains
the factor
\begin{eqnarray}
\hspace{-1cm}
\frac{\int\nolimits_{-\infty }^{\infty }
d \, \mathrm{Re} \, \overline{\phi }
_{b}\left( s_{1}\right) e^{-\widetilde{a}\left( s_{1}\right) 
\left( \mathrm{Re} \, 
\overline{\phi }_{b}\left( s_{1}\right) \right) ^{2}}
\left( \mathrm{Re} \, \overline{
\phi }_{b}\left( s_{1}\right) \right) ^{2}}{\int\nolimits_{-\infty }^{\infty}
d \, \mathrm{Re} \, \overline{\phi }_{b}\left( s_{1}\right) 
\,e^{-\widetilde{a}\left( s_{1}\right) 
\left( \mathrm{Re} \, \overline{\phi }_{b}\left( s_{1}\right) \right) ^{2}}}
+\frac{\int\nolimits_{-\infty }^{\infty }
d \, \mathrm{Im} \, \overline{\phi }
_{b}\left( s_{1}\right) e^{-\widetilde{a}\left( s_{1}\right) 
\left( \mathrm{Im} \, 
\overline{\phi }_{b}\left( s_{1}\right) \right) ^{2}}
\left( \mathrm{Im} \, \overline{
\phi }_{b}\left( s_{1}\right) \right) ^{2}}{\int\nolimits_{-\infty }^{\infty}
d \, \mathrm{Im} \, \overline{\phi }_{b}\left( s_{1}\right) 
\,e^{-\widetilde{a}\left( s_{1}\right) 
\left( \mathrm{Im} \, \overline{\phi }_{b}\left( s_{1}\right) \right) ^{2}
}} \nonumber \\
= \widetilde{a}\left( s_{1}\right) ^{-1} \hspace{1cm}
\label{eq13.11}
\end{eqnarray}
so
\begin{eqnarray}
G_{b}\left( x_{1},x_{2}\right) =\sum\nolimits_{s}\overline{G}_{b}
\left( s\right) U_{b}\left( x_{1},s\right) \, U^{\dag }_{b}
\left( x_{2},s\right) 
\quad , \quad 
\overline{G}_{b}\left( s\right) =\widetilde{a}\left( s\right) ^{-1}  \; .
\label{eq13.12} 
\end{eqnarray}

As usual, $a\left( s\right) $ and $\widetilde{a}\left( s\right) $ contain a 
$ +i\epsilon $ which is associated with a convergence factor in the path
integral (and which gives a well-defined inverse). 

The above are the propagators in the Euclidean formulation. The Lorentzian
propagators are obtained through the same procedure with 
$a\left( s\right) \rightarrow -ia\left(
s\right) $ and $\widetilde{a}\left( s\right) \rightarrow -i\widetilde{a}
\left( s\right) $:
\begin{eqnarray}
\overline{G}_{f}^{L}\left( s\right)  = ia\left( s\right) ^{-1} 
\quad , \quad 
\overline{G}_{b}^{L}\left( s\right)  = i\widetilde{a}\left( s\right)
^{-1} \; .
\label{eq13.15}
\end{eqnarray}
The propagators in the Euclidean and Lorentzian formulations thus differ by
only a factor of $i$. More generally, in the present picture, the action,
fields, operators, classical equations of motion, quantum transition
probabilities, propagation of particles, and meaning of time are the same in
both formulations.

For a single noninteracting bosonic field with a mass $m_{b}$, the basis
functions are
\begin{eqnarray}
U_{b}\left( x,p\right) =\mathcal{V}^{-1/2}e^{ip\cdot x}=\mathcal{V}
^{-1/2}e^{-i\omega t}e^{i\overrightarrow{p}\cdot \overrightarrow{x}}
\label{eq13.16}
\end{eqnarray}
so with $s\rightarrow p$ we have
\begin{eqnarray}
\widetilde{a}\left( p\right) =\omega ^{2}-\left\vert \overrightarrow{p}
\right\vert ^{2}-m_{b}^{2}+i\epsilon 
\label{eq13.17}
\end{eqnarray}
and
\begin{eqnarray}
\overline{G}_{b}\left( p\right)  &=&\frac{1}{\omega ^{2}-\left\vert 
\overrightarrow{p}\right\vert ^{2}-m_{b}^{2}+i\epsilon }\; 
\label{eq13.18} \\
\overline{G}_{b}^{L}\left( p\right)  &=&\frac{i}{\omega ^{2}-\left\vert 
\overrightarrow{p}\right\vert ^{2}-m_{b}^{2}+i\epsilon } \; .
\label{eq13.19}
\end{eqnarray}
Notice that (\ref{eq13.5}) and (\ref{eq13.12}) hold even when the basis 
functions in (\ref{eq13.6a}) or (\ref{eq13.9a}) are not a complete set.

\section{\label{sec:appD}Negative-action modes}

According to (\ref{eq7.32b}), there is a negative contribution
\begin{equation}
S^{\phi}_{-}=-\displaystyle\sideset{}{'}\sum\limits_{s<0}\overline{\phi }_{b}^{\,\ast }\left(
s^{\prime }\right) \left\vert \widetilde{a}\left( s\right) \right\vert 
\overline{\phi }_{b}\left( s^{\prime }\right) 
\end{equation}
to the action for what will become scalar boson fields. Here
it will be assumed that each primitive variable $x_{s}=\mathrm{Re} \; \overline{\phi }
_{b}\left( s^{\prime }\right) $ or $\mathrm{Im} $ $\overline{\phi }_{b}\left(
s^{\prime }\right) $ makes a quartic contribution $\frac{1}{2}
b_{s}x_{s}^{4}$ to the action, stabilizing the vacuum, so that its 
total action is
\begin{equation}
-\left\vert \widetilde{a}\left( s\right) \right\vert x_{s}^{2}+\frac{1}{2}
b_{s}x_{s}^{4}\;.
\end{equation}
Minimization gives the value
\begin{equation}
\overline{x}_{s\,}^{2}=\left\vert \widetilde{a}\left( s\right) \right\vert
/b_{s}\;.
\end{equation}
An excitation out of the vacuum can be represented by $x_{s}=\overline{x}
_{s\,}+\delta x_{s}$, and the lowest-order additional action is
\begin{eqnarray}
\delta S^{\phi}_{-} &=&\left[ \frac{\partial S^{\phi}_{-}}{\partial x_{s}}\right] _{
\overline{x}_{s\,}}\delta x_{s}+\frac{1}{2}\left[ \frac{\partial ^{2}S^{\phi}_{-}}{
\partial x_{s}^{2}}\right] _{\overline{x}_{s\,}}\left( \delta x_{s}\right) ^{2} \\
&=&0+\left[ - \left\vert \widetilde{a}\left( s\right) \right\vert +3b_{s}
\overline{x}_s^2 \right] \left( \delta x\right) ^{2} \\
&=&2\left\vert \widetilde{a}\left( s\right) \right\vert \left( \delta
x\right) ^{2}\;.
\end{eqnarray}
Notice that the quartic coefficient $b_{s}$ is not present in this result.
For an excitation $\delta \overline{\phi }_{b}\left(s^{\prime }\right) =\mathrm{Re} \;
\delta \overline{\phi }_{b}\left( s^{\prime }\right) +i \, \mathrm{Im} \; \delta \overline{\phi }_{b}\left( s^{\prime }\right) $, therefore, the action is 
\begin{eqnarray}
\delta S^{\phi}_{-} &=&2\left\vert \widetilde{a}\left( s\right) \right\vert \left[
\left( \mathrm{Re} \; \delta \overline{\phi }_{b}\left( s^{\prime }\right) \right)
^{2}+\left( \mathrm{Im} \; \delta \overline{\phi }_{b}\left( s^{\prime }\right) \right)
^{2}\right]  \\
&=&2\left\vert \widetilde{a}\left( s\right) \right\vert \delta \overline{
\phi }_{b}^{\,\ast }\left( s^{\prime }\right) \delta \overline{\phi }
_{b}\left( s^{\prime }\right)  \\
&=&\left\vert \widetilde{a}\left( s\right) \right\vert \overline{\phi }
_{b}^{\prime \ast }\left( s^{\prime }\right) \overline{\phi }_{b}^{\prime
}\left( s^{\prime }\right) 
\end{eqnarray}
where
\begin{equation}
\overline{\phi }_{b}^{\prime }\left( s^{\prime }\right) = \sqrt{2} \, \delta \overline{
\phi }_{b}\left( s^{\prime }\right) \;.
\end{equation}
When $\overline{\phi }_{b}^{\prime }\left( s^{\prime }\right) $ is renamed $
\overline{\phi }_{b}\left( s^{\prime }\right) $, as in (\ref{eq7.32a}), the excitations of
negative-action modes have the same form for the action as those of
positive-action modes. Of course, there still remains a negative-action
``condensate density'' buried in the vacuum (and there is also a change in the measure within the path integral). 

\section{\label{sec:appG}Einstein-Hilbert action}

Since (\ref{eq7.32b}) has the same form as (\ref{eq7.32a}), all the subsequent steps for positive-action modes can be repeated for the negative-action modes in the second term of (\ref{minus}), with a result that differs by an overall minus sign. In the present context, the relevant part of the final action is 
\begin{eqnarray}
\mathcal{L}_{G} = \Phi_- ^{\dag }\left( x\right) \left( + \frac{1}{4} R \right) \Phi _- \left( x\right) 
\label{G1}
\end{eqnarray}
where $\Phi _- $ contains these negative-action modes, which all have the same interaction with $R$. (Other states in the vacuum will make contributions with the opposite sign, but we assume the modes contributing to (\ref{G1}) are dominate in this context. A quantitative calculation would have to include all bosonic vacuum contributions.) There is a very large ``condensate density'' for these modes, with the dimension of inverse length squared (in natural units), so we can define a very small length $\ell_P$ by
\begin{eqnarray}
\ell _{P} = \left( 4\pi n_-  \right) ^{-1/2} \quad , \quad n_- = \Phi_- ^{\dag } \Phi _- 
\label{G2}
\end{eqnarray}
and the action corresponding to  (\ref{G1})  is
\begin{equation}
S_{G}=\frac{1}{16\pi }\ell _{P}^{-2}\int d^{4}x \, R \; .
\label{G3}
\end{equation}

The above treatment is in a locally inertial coordinate system with coordinates $x^{\mu}$. 
(It is assumed that the parameters determining $n_-$ are themselves determined in an inertial frame.) 
Let $d^{4}x^{\prime}$ be the volume element in a general coordinate system, which is held fixed as the metric tensor varies. 
Since $d^{4}x^{\prime} \, e = d^{4} x $, where $e=\left( -\det \, g_{\mu \nu} \right) ^{1/2}$, the volume element in the locally inertial frame varies in proportion to $e$, and in the fixed general coordinate system $\int d^{4}x \, R$ is replaced by $\int d^{4}x^{\prime} \,e \, R$.

With the general coordinates renamed $x^{\mu}$, we have
\begin{equation}
S_{G}=\frac{1}{16\pi }\ell _{P}^{-2}\int d^{4}x \, e \, R \; .
\label{cond2}
\end{equation}
$S_{G}$ is interpreted as the Einstein-Hilbert action, $\ell _{P}$ as the Planck length, and $G=\ell _{P}^2$ as the gravitational constant.

\end{document}